\documentclass{ian}

\usepackage{dimer}
\usepackage{amssymb}
\usepackage{dsfont}
\usepackage{graphicx}
\usepackage{largearray}
\usepackage{constants}

\def\define#1{{\bf #1}}

\begin{document}

\hbox{}
\hfil{\bf\LARGE
Nematic liquid crystal phase\par
\vskip10pt
\hfil  in a system of interacting dimers and monomers
}
\vfill

\hfil{\bf\large Ian Jauslin}\par
\hfil{\it School of Mathematics, Institute for Advanced Study}\par
\vskip20pt

\hfil{\bf\large Elliott H. Lieb}\par
\hfil{\it Departments of Mathematics and Physics, Princeton University}\par

\vfill

\hfil {\bf Abstract}\par
\medskip
We consider a monomer-dimer system with a strong attractive dimer-dimer interaction that favors alignment. In 1979, Heilmann and Lieb conjectured that this model should exhibit a nematic liquid crystal phase, in which the dimers are mostly aligned, but do not manifest any translational order. We prove this conjecture for large dimer activity and strong interactions. The proof follows a Pirogov-Sinai scheme, in which we map the dimer model to a system of hard-core polymers whose partition function is computed using a convergent cluster expansion.
\vfill

\tableofcontents

\vfill

\hfil{\footnotesize \copyright\ 2017 by the authors. This paper may be reproduced, in its entirety, for non-commercial purposes.}
\smallskip

{\footnotesize\hfil e-mail: {\tt jauslin@ias.edu}, {\tt lieb@princeton.edu}}

\eject

\setcounter{page}1
\pagestyle{plain}

\section{Introduction}
\indent In a 1979 paper, O.J.\-~Heilmann and one of us \cite{HL79} attempted to construct a simple statistical mechanical lattice model of a liquid crystal phase transition. Such a model would have to have the property that the constituent `molecules' would have to show no long-range order at high temperature and, at low temperature, have a transition to a phase in which there is long-range rotational order of the molecules, but {\it no} long-range translational order. In other words, the molecules are nearly parallel, but their centers show no long-range correlations. Such a model had not been constructed before then, although there was the 1949 heuristic ultra-thin, ultra-long molecule model of L.\-~Onsager \cite{On49}.

\indent In the model considered in\-~\cite{HL79} the molecules are represented by interacting dimers or fourmers on a square or cubic lattice. It was shown, by reflection positivity and chessboard estimates, that, for several different models, the system exhibits long-range orientational order at low temperature. Thus, if we specify the orientation of one dimer somewhere in the lattice, any other dimer is oriented in the same way with large probability. It was not proved, however, that this rotational order is {\it not} accompanied by translational order, that is, it was not proved that fixing a dimer somewhere on the lattice does not induce correlations in the position of distant dimers, even though it does induce a preference for their orientation.

\indent Since then, there have been many new developments in the field, though a complete proof of the lack of translational order for any of the models considered in\-~\cite{HL79} was, until now, still lacking. In\-~\cite{AH80}, a new three-dimensional model was added to the list by extending one of the two-dimensional models in\-~\cite{HL79}. In\-~\cite{AZ82,Za96}, the result was extended to a model of elongated molecules on a lattice admitting {\it continuous} orientations, with short- (in three dimensions) and long- (in two dimensions) range attractive interactions. A liquid crystalline (also called {\it nematic}) phase was later proved to exist\-~\cite{BKL84} (that is, both orientational order and a lack of translational order are shown) in a model of infinitely thin long molecules in two dimensions admitting a {\it finite} number of orientations (although the discussion in\-~\cite{BKL84} is limited to a remark in the concluding section of the paper). This behavior was also shown to occur in an integrable lattice model of rods admitting two orientations and of varying length\-~\cite{IVZ06} or of a fixed, long length\-~\cite{DG13}. Finally, in\-~\cite{ACM14}, a mean-field interacting dimer model was introduced and solved.

\indent There has also been some progress towards proving the conjecture in\-~\cite{HL79}. Most efforts have focused on one of the models in\-~\cite{HL79}, model\-~I (see figure\-~\ref{fig:interaction}), which is two-dimensional, and involves an interaction between collinear, neighboring dimers. In\-~\cite{Al16}, D.\-~Alberici tweaked this model by making the activity of horizontal and vertical dimers different, thus favoring one orientation, and showed the emergence of a liquid crystalline phase. There have also been numerical results\-~\cite{PCF14} supporting the conjecture.

\indent In this paper, we shall prove the conjecture in\-~\cite{HL79} that there is no long-range translational order in model\-~I. There is little doubt that similar proofs could be devised for the other models and other dimensions for which orientational order was proved in\-~\cite{HL79}.
\bigskip

\indent Let us describe the model in more precise terms. It is a monomer-dimer system on the square lattice, in which a dimer is an object that covers exactly two neighboring vertices, and a monomer covers a single vertex. No two objects are permitted to cover the same vertex. Monomers are to be thought of, in this context, as empty sites, whereas dimers represent molecules. The dimer-activity $z$ is large, which favors dimers heavily, but the presence of monomers is crucial. In addition we introduce a strong attractive force that favors alignment. Without this interaction, as was shown in\-~\cite{HL72}, the monomer-dimer model would not have phase transitions at positive temperature, and thus, would exhibit no liquid crystalline ordering.

\indent The attractive interaction assigns a negative energy $-J$ to every pair of dimers that are adjacent and aligned, that is, that are on the same row or the same column, see figure\-~\ref{fig:interaction}. We offer two interpretations of this model. One is of polar molecules of length 1, represented by individual dimers; the other is of molecules of varying length, modeled by chains of adjacent and aligned dimers.

\indent We choose boundary conditions that favors vertical dimers, and focus on the parameter regime $J\gg z\gg 1$. We first prove that horizontal dimers are unlikely in the bulk, in accordance with the result of\-~\cite{HL79}. The method of proof is completely different from that in\-~\cite{HL79}; in particular we do not use reflection positivity. We further show that the probability of finding a vertical dimer on a given edge is, in the thermodynamic limit, independent of the position of the edge. Furthermore, the joint probability of finding a dimer at an edge $e$ and another at $e'$, up to a constant, decays exponentially in the distance between $e$ and $e'$ with a rate $\gtrsim e^{-\frac32 J}z^{-\frac12}$. This proves the absence of translational order.

\indent The proof follows a Pirogov-Sinai\-~\cite{PS75} scheme, which is an extension of the Peierls argument. The main idea is to map the interacting dimer model to a system of hard-core polymers, and show that the effective activity of these polymers decays {\it sufficiently} fast in their size. We then use a cluster expansion to compute the partition function of the model in terms of an absolutely convergent series, and estimate the one- and two-point correlation functions.
\bigskip

\indent This paper is organized as follows. In section\-~\ref{sec:model}, we define the model in precise terms, state our main theorem and provide a detailed sketch of the proof. Section\-~\ref{sec:1d} describes the solution to an ancillary model in which only one dimer orientation is allowed, which plays an important role in the rest of the proof. In section\-~\ref{sec:polymer_model}, we map the interacting dimer model to the polymer model. In section\-~\ref{sec:cluster} we prove bounds on the polymer activity and entropy, and compute the partition function of the polymer model in terms of an absolutely convergent cluster expansion. Finally, the proof of the main theorem is concluded in section\-~\ref{sec:nematic}.
\bigskip

\delimtitle{\bf Acknowledgements}
We thank Alessandro Giuliani, Diego Alberici and Daniel Ueltschi for helpful discussions about this work. The work of E.H.L. was partially supported by U.S. National Science Foundation grant PHY\-~1265118. The work of I.J. was supported by The Giorgio and Elena Petronio Fellowship Fund and The Giorgio and Elena Petronio Fellowship Fund II.
\enddelim

\section{The Model}\label{sec:model}
\subsection{Definition of the model}
\begin{figure}
  \hfil\includegraphics[width=8cm]{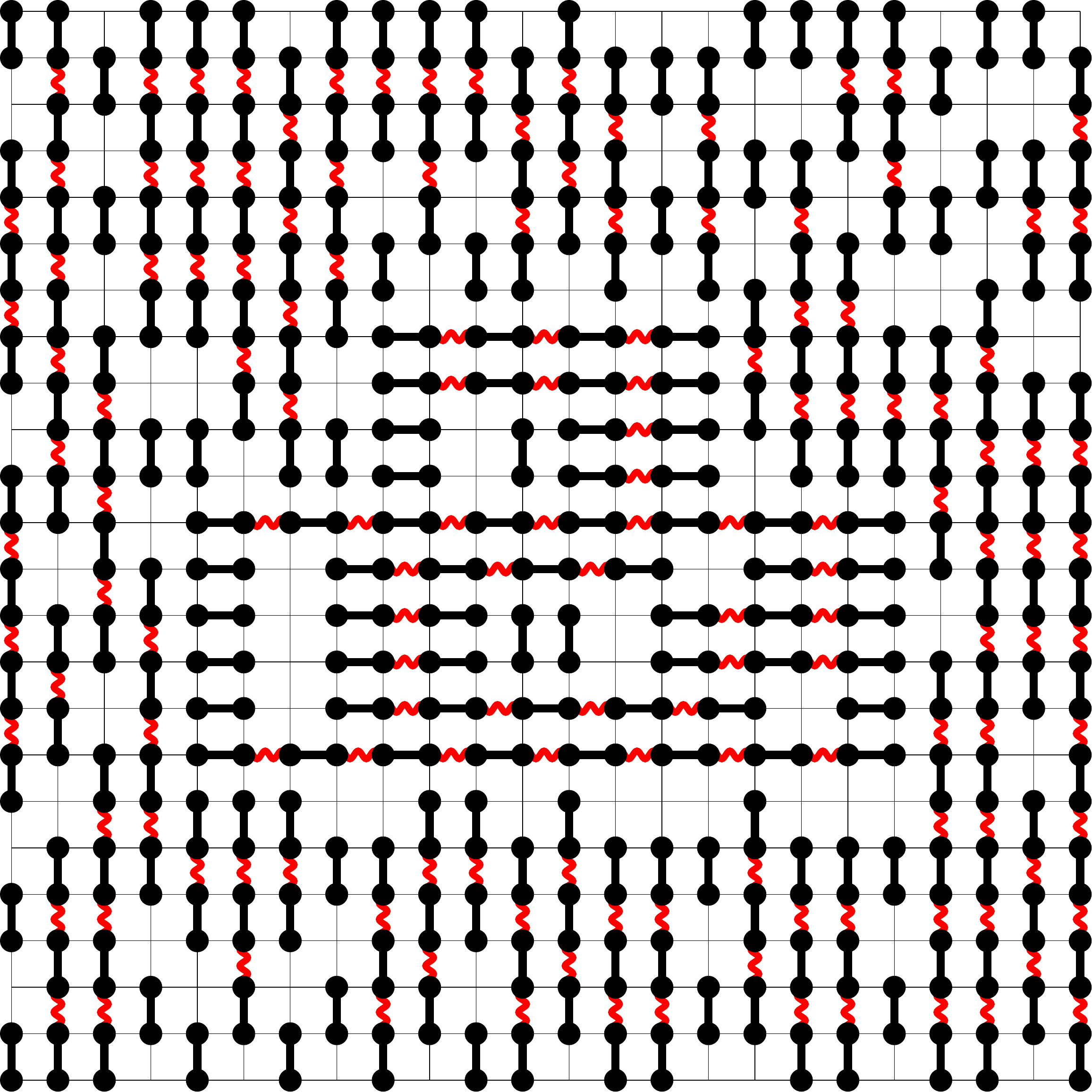}
  \caption{An example of a dimer configuration. Interacting dimers are depicted as connected by a red (color online) wavy line.}
  \label{fig:interaction}
\end{figure}

\indent A dimer configuration is a collection of non-overlapping edges of $\mathbb Z^2$. In order to define these formally, we denote the set of edges of a subset $\Lambda\subset\mathbb Z^2$ by
\begin{equation}
  \mathcal E(\Lambda):=\{\{x,x'\},\ x,x'\in\Lambda,\ \|x-x'\|=1\}
  \label{mcE}
\end{equation}
in which $\|\cdot\|$ denotes the Euclidean distance. The edges of $\Lambda$ are either horizontal ($\mathrm h$-edges) or vertical ($\mathrm v$-edges), and given a set of edges $E\subset\mathcal E(\Lambda)$ and $q\in\{\mathrm{h},\mathrm{v}\}$, we denote the set of \define{$q$-edges} in $E$ by $\mathbb D_q(E)$. We then define the set of \define{dimer configurations} in $\Lambda$ as
\begin{equation}
  \Omega(\Lambda):=\{E\subset\mathcal E(\Lambda),\quad \forall e\neq e'\in E,\ e\cap e'=\emptyset\}
\end{equation}
(see figure\-~\ref{fig:interaction} for an example).
\bigskip

\point{\bf Interaction.} We introduce a strong interaction between dimers, that favors configurations in which dimers are aligned, collinear and neighbors (see figure\-~\ref{fig:interaction}). Every such pair of dimers contributes $-J$ to the energy of the configuration, and $J$ will be taken to be large.
\bigskip

\point{\bf Boundary condition.} We choose the boundary condition in such a way that either vertical or horizontal dimers are favored. To determine which it is, we introduce a variable $q\in\{\mathrm v,\mathrm h\}$ which is set to $\mathrm v$ if vertical dimers are favored and $\mathrm h$ if horizontal ones are. In addition, we define $-q$ as the {\it opposite} of $q$, that is, if $q=\mathrm h$, then $-q=\mathrm v$ and vice-versa. The boundary condition consists of two forces: first, $-q$-dimers are not allowed to be too close to the boundary, and $q$-dimers may be attracted by certain parts of the boundary.

\indent Note that we could consider different boundary conditions, as long as they favor horizontal or vertical dimers. It would require an extra computation, which we have chosen not to carry out, as we have achieved our goal of showing that there are two extremal Gibbs states in which the rotational symmetry is broken, but the translational one is not.

\begin{figure}
  \hfil\includegraphics[width=8cm]{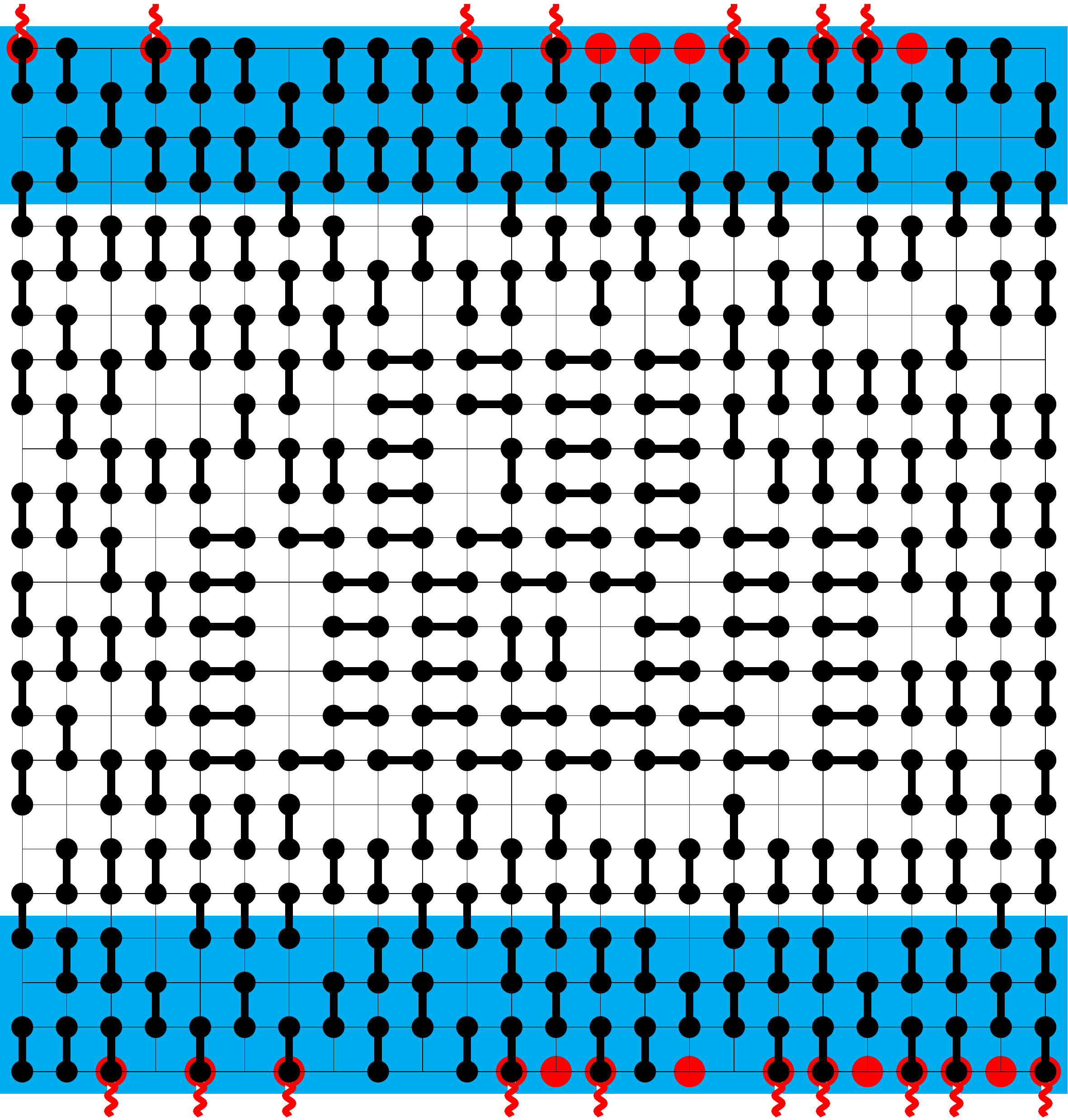}
  \caption{The boundary condition. Here, we have taken $q=\mathrm v$ and $\ell_0=4$. There can be no horizontal dimers in the cyan (color online) region. Some of the vertices on the boundary, depicted as large red (color online) discs, are {\it magnetized}: they contribute $-J$ to the energy when they are occupied.}
  \label{fig:boundary}
\end{figure}

\indent In order to define the boundary condition precisely, let us first define the \define{boundary} of a bounded subset $\Lambda\subset\mathbb Z^2$, denoted by $\partial\Lambda$, as the set of edges $\{x,x'\}\in\mathcal E(\mathbb Z^2)$ with $x\in\Lambda$ and $x'\in\mathbb Z^2\setminus\Lambda$. In addition, we define the \define{$q$-distance} between two points $x,x'\in\mathbb Z^2$, denoted by $\mathfrak d_q(x,x')\in\mathbb R\cup\{\infty\}$, in the following way. If $x$ and $x'$ are in the same $q$-line (a v-line is a vertical line and an h-line is a horizontal one), then $\mathfrak d_q(x,x')=\|x-x'\|$, and if they are not, then $\mathfrak d_q(x,x')=\infty$. We can now define the boundary condition, which, we recall, consists of two forces (see figure\-~\ref{fig:boundary} for an example).
\begin{itemize}
  \item We fix a length scale $\ell_0>1$ and require that every $-q$-dimer in $\Lambda$ be separated from the boundary of $\Lambda$ by a $q$-distance of at least $\ell_0$. We denote the set of dimer configurations satisfying this condition by
  \begin{equation}
    \Omega_{q,\ell_0}(\Lambda):=\{\underline\delta\in\Omega(\Lambda),\quad\mathfrak d_q(\mathbb D_{-q}(\underline\delta),\partial\Lambda)\geqslant\ell_0\}.
  \end{equation}
  (In this paper, we will use the convention that the distance between two sets is the smallest distance between the elements of the set. Furthermore, we will use this convention recursively to define the distance between sets of sets, and so forth...)

  \item In addition to this condition, we will allow part or all of $\partial\Lambda$ to be {\it magnetized}, by which we mean that parts of the boundary may attract $q$-dimers, as if there were $q$-dimers right outside it. Formally, we introduce a subset $\varrho\subset\partial\Lambda$ of edges on the boundary which are magnetized, and, given a dimer configuration $\underline\delta\in\Omega_{q,\ell_0}(\Lambda)$, we define the set of dimers that are bound to the boundary as
  \begin{equation}
    \mathbb B_q(\underline\delta,\varrho):=
    \{d\in\underline\delta,\quad \mathfrak d_q(d,\varrho)=0\}.
  \end{equation}
  Every dimer in $\mathbb B_q(\underline\delta,\varrho)$ contributes $-J$ to the interaction, as if every such dimer interacted with a phantom dimer outside $\Lambda$.
\end{itemize}

The boundary condition is thus specified by the triplet $(q,\varrho,\ell_0)\equiv\mathbf q$, and we will use the shorthand $\Omega_{\mathbf q}\equiv\Omega_{q,\ell_0}$ and $\mathbb B_{\mathbf q}(\underline\delta)\equiv\mathbb B_q(\underline\delta,\varrho)$.
\bigskip

\point{\bf Observables.} In this paper, we will compute the {\it grand-canonical partition function} of the system, defined as
\begin{equation}
  Z(\Lambda|\mathbf q)=\sum_{\underline\delta\in\Omega_{\mathbf q}(\Lambda)}z^{|\underline\delta|}e^{-W_0(\underline\delta)}\mathfrak B_{\mathbf q}(\underline\delta)
  \label{Zmodel}
\end{equation}
with
\begin{equation}
  e^{-W_0(\underline\delta)}:=\prod_{d_1,d_2\in\underline\delta}e^{\frac12J\mathds1(d_1\sim d_2)}
  ,\quad
  \mathfrak B_{\mathbf q}(\underline\delta):=
  \prod_{d\in\mathbb B_{\mathbf q}(\underline\delta)}e^J
  \label{eW0}
\end{equation}
in which
\begin{itemize}
  \item $z>0$ is the {\it dimer activity},
  \item $J>0$ is the {\it interaction strength}, (the factor $\frac12$ accounts for the fact that each pair is counted twice)
  \item $|\underline\delta|$ denotes the number of dimers in $\underline\delta$,
  \item $\mathds1(d_1\sim d_2)\in\{0,1\}$ identifies which pairs of dimers interact: it is equal to 1 if and only if $\exists q'\in\{\mathrm v,\mathrm h\}$ such that $d_1$ and $d_2$ are both $q'$-dimers and are at $q'$-distance $1$ from one another.
\end{itemize}
\bigskip

\indent In addition, we will compute the $\mathfrak n$-point correlation functions, defined as follows. We fix a set of edges $\Upsilon\equiv\{\upsilon_1,\cdots,\upsilon_{\mathfrak n}\}\subset\mathcal E(\Lambda)$, and define
\begin{equation}
  \left<\mathds 1_{\upsilon_1}\cdots\mathds 1_{\upsilon_{\mathfrak n}}\right>_{\Lambda,\mathbf q}
  :=
  \frac1{Z(\Lambda|\mathbf q)}
  \sum_{\displaystyle\mathop{\scriptstyle\underline\delta\in\Omega_{\mathbf q}(\Lambda)}_{\delta\supset\Upsilon}}z^{|\underline\delta|}e^{-W_0(\underline\delta)}\mathfrak B_{\mathbf q}(\underline\delta).
\end{equation}
The infinite-volume limit of this correlation function is defined by considering a square $L\times L$ box $\Lambda_L$ and taking the $L\to\infty$ limit
\begin{equation}
  \left<\mathds 1_{\upsilon_1}\cdots\mathds 1_{\upsilon_{\mathfrak n}}\right>_{\mathbf q}
  :=\lim_{L\to\infty}
  \left<\mathds 1_{\upsilon_1}\cdots\mathds 1_{\upsilon_{\mathfrak n}}\right>_{\Lambda_L,\mathbf q}
  .
\end{equation}
We will assume that the different $\upsilon_i$ are at a distance of at least $\ell_0$ from each other (this assumption is merely a technical requirement). Note that the partition function of dimers in $\Lambda$ that contain $\Upsilon$ can, equivalently, be viewed as the partition function on $\Lambda\setminus(\bigcup_{\upsilon\in\Upsilon}\upsilon)$ with a special boundary condition. Namely, the endpoints of $\upsilon$ are {\it magnetized}, in the sense discussed above, but, unlike the boundary of $\Lambda$, which excludes $-q$-dimers at a distance $\ell_0$, the boundary of $\upsilon$ does not exclude any dimers. Formally, defining
\begin{equation}
  \Lambda^{(\Upsilon)}:=\Lambda\setminus\left({\textstyle\bigcup_{\upsilon\in\Upsilon}\upsilon}\right)
\end{equation}
and
\begin{equation}
  \Omega_{q,\ell_0}^{(\Upsilon)}(\Lambda):=\{\underline\delta\in\Omega_{q,\ell_0}(\Lambda),\quad\forall\delta\in\underline\delta,\ \delta\subset\Lambda^{(\Upsilon)}\}
\end{equation}
we have (recall that we assume that different $\upsilon_i$'s are not neighbors (so that sources do not interact directly))
\begin{equation}
  \left<\mathds 1_{\upsilon_1}\cdots\mathds 1_{\upsilon_{\mathfrak n}}\right>_{\Lambda,\mathbf q}
  =
  \frac{Z^{(\Upsilon)}(\Lambda|\mathbf q)}{Z(\Lambda|\mathbf q)}
  ,\quad
  Z^{(\Upsilon)}(\Lambda|\mathbf q)
  :=
  z^{\mathfrak n}
  \sum_{\scriptstyle\underline\delta\in\Omega_{\mathbf q}^{(\Upsilon)}(\Lambda)}z^{|\underline\delta|}e^{-W_0(\underline\delta)}\mathfrak B_{\mathbf q}^{(\Upsilon)}(\underline\delta)
\end{equation}
in which $\mathfrak B_{\mathbf q}^{(\Upsilon)}(\underline\delta)$ includes the interactions with the sources:
\begin{equation}
  \mathfrak B_{\mathbf q}^{(\Upsilon)}(\underline\delta):=
  \mathfrak B_{\mathbf q}(\underline\delta)
  \left(\prod_{\upsilon\in \mathbb D_q(\Upsilon)}\mathfrak B_{(q,\partial_q\upsilon,\ell_0)}(\underline\delta)\right)
  \left(\prod_{\upsilon\in \mathbb D_{-q}(\Upsilon)}\mathfrak B_{(-q,\partial_{-q}\upsilon,\ell_0)}(\underline\delta)\right)
  \label{frakB_Upsilon}
\end{equation}
in which $\partial_q\equiv\mathbb D_q(\partial)$ (by which we mean that for any set $X$, $\partial_q X\equiv\mathbb D_q(\partial X)$) (note that the index $\ell_0$ is redundant; we have kept it in in order not to have to introduce yet another notation for the boundary condition $\mathbf q\equiv(q,\varrho,\ell_0)$).
\bigskip

\point{\bf Oriented dimer model.} As was shown in \cite{HL79}, when the interaction strength is sufficiently large, the probability of horizontal and vertical dimers coexisting is low. In fact, the main idea is to compute how much the partition function of the model with $q$-boundary conditions differs from that of a similar model in which there are {\it only} $q$-dimers and monomers, and to show that, in a sense to be made precise, this difference is small. We first formally define the {\it oriented dimer model}, in which only one of the two dimer orientations is allowed: let $\Theta_q(\Lambda)\subset\Omega(\Lambda)$ denote the set of $q$-dimer configurations on $\Lambda$:
\begin{equation}
  \Theta_q(\Lambda):=\{\underline\delta\in\Omega_q(\Lambda),\ \delta\in\mathbb D_q(\mathcal E(\Lambda))\}
\end{equation}
in terms of which the partition function of the $q$-dimer model is
\begin{equation}
  \mathfrak Z_{\mathbf q}^{(\Upsilon)}(\Lambda)=z^{\mathfrak n}\sum_{\underline\delta\in\Theta_q(\Lambda^{(\Upsilon)})}z^{|\underline\delta|}e^{-W_0(\underline\delta)}\mathfrak B_{\mathbf q}^{(\Upsilon)}(\underline\delta)
  .
  \label{Zv}
\end{equation}
In order to compare $Z^{(\Upsilon)}(\Lambda|\mathbf q)$ and $\mathfrak Z_{\mathbf q}^{(\Upsilon)}(\Lambda)$, we will compute the ratio
\begin{equation}
  \frac{Z^{(\Upsilon)}(\Lambda|\mathbf q)}{\mathfrak Z_{\mathbf q}^{(\Upsilon)}(\Lambda)}.
  \label{goal}
\end{equation}
Note that, in the oriented dimer model, since different columns of vertical dimers and different rows of horizontal dimers do not interact, in order to compute $\mathfrak Z_{\mathbf q}^{(\Upsilon)}(\Lambda)$, it suffices to compute the partition function of dimers on a one-dimensional chain.
\bigskip

\subsection{Result}
\indent Our main result is that, at large activities and yet larger interaction strengths, this model exhibits {\it nematic} order, that is, it exhibits long-range orientational order, yet no long-range translational order. This is stated precisely in the following theorem.
\bigskip

\theoname{Theorem}{nematic phase}\label{theo:nematic}
  Let $\mathbf v\equiv(\mathrm v,\emptyset,\ell_0)$, which corresponds to open boundary conditions coupled with the condition that no horizontal dimers come within a distance $\ell_0$ of the boundary. There exist {\it large} constants (which, in principle, can be worked out) $\cst C{cst:ineqJ},\cst C{cst:ineqz}>0$ such that, if
  \begin{equation}
    J>\cst C{cst:ineqJ} z
    \quad\mathrm{and}\quad
    z>\cst C{cst:ineqz}
  \end{equation}
  then, taking $\ell_0=\cst C{cst:ell0}e^{\frac32J}\sqrt z$ for some constant $\cst C{cst:ell0}>0$ ($\ell_0$ is of the order of the correlation length of the oriented dimer model), the following statements hold.
  \begin{itemize}
    \item Let $e_{\mathrm v}\in \mathbb D_{\mathrm v}(\mathcal E(\mathbb Z^2))$ be a vertical edge, $\left<\mathds 1_{e_{\mathrm v}}\right>_{\mathbf v}$ is {\it independent} of the position of $e_{\mathrm v}$, and
    \begin{equation}
      \left<\mathds 1_{e_{\mathrm v}}\right>_{\mathbf v}=\frac12\left(1+O\left({\textstyle\frac1{\sqrt{ze^J}}}\right)\right).
      \label{bound_1v}
    \end{equation}
    In other words, the probability of finding a dimer at a given edge is independent of the position of that edge, and most vertices are occupied by a vertical dimer (if the lattice were fully packed, then half the edges are occupied).

    \item Let $e_{\mathrm h}\in \mathbb D_{\mathrm h}(\mathcal E(\mathbb Z^2))$ be a horizontal edge,
    \begin{equation}
      \left<\mathds 1_{e_{\mathrm h}}\right>_{\mathbf v}=O(e^{-3J}).
      \label{bound_1h}
    \end{equation}
    Thus, horizontal dimers are unlikely. This implies orientational order (in particular, this implies that $\left<\mathds 1_{e_{\mathrm h}}\mathds 1_{e_{\mathrm v}}\right>_{\mathrm v}=O(e^{-3J})$.

    \item For any pair of edges $e,e'\in \mathcal E(\mathbb Z^2)$ which are at a distance of at least $\ell_0$,
    \begin{equation}
      \left<\mathds 1_e\mathds 1_{e'}\right>_{\mathbf v}
      -
      \left<\mathds 1_e\right>_{\mathbf v}
      \left<\mathds 1_{e'}\right>_{\mathbf v}
      =
      O(e^{-\cst C{cst:exp1vv}\mathrm{dist}_{\mathrm{HL}}(e,e')})
      \label{bound_1vv}
    \end{equation}
    for some constant $\cst C{cst:exp1vv}>0$, in which the distance $\mathrm{dist}_{\mathrm{HL}}$ is that induced by the norm
    \begin{equation}
      \|(x,y)\|_{\mathrm{HL}}:=J|x|+\ell_0^{-1}|y|
      .
    \end{equation}
    This means that the probability of placing two dimers at $e$ and $e'$ is equal to a term that does not depend on the position of the edges plus a term that decays exponentially with the distance between them. There is, thus, no long-range translational order. The decay rate is of order $J\gg 1$ in the horizontal direction and $e^{-\frac32J}z^{-\frac12}\ll 1$ in the vertical.
  \nopagebreakafteritemize
  \end{itemize}
\restorepagebreakafteritemize
\endtheo

\subsection{Sketch of the proof}
\indent Before discussing the proof that is carried out in this paper, let us mention two simpler approaches we have tried which have failed.

\indent In\-~\cite{HL79}, orientational order was proved using reflection positivity and chessboard estimates. The main difficulty with extending this method to prove the lack of translational order is that, as can be seen from theorem\-~\ref{theo:nematic}, the correlation length of the system is very large: $\ell_0\approx e^{\frac32J}\sqrt z$, and the lack of order is only visible on that scale, and seems difficult to see using only reflection positivity.

\indent Another natural approach to the problem is to integrate out the vertical dimers and manipulate the resulting effective horizontal dimer model. The idea being that, if vertical dimers are favored on the boundary, then they should dominate, so the horizontal dimer model would be a rarefied gas, which could be treated by standard cluster expansion methods. However, since horizontal dimers are subjected to a surface tension, they tend to bunch together into large {\it swarms}. In order for this approach to be successful, the swarms would have to pay an energetic price proportional to their volume, in order to counterbalance their entropy. Unfortunately, they do not do so. Note, however, that if we made the activity of horizontal dimers slightly smaller than that of vertical ones, as in\-~\cite{Al16}, then the horizontal swarms would have a sufficiently large volume cost, and this approach would be successful.

\indent Instead, we opted for a Pirogov-Sinai argument.
\bigskip

\indent The main idea of the proof is to estimate how much the partition function of the full dimer model differs from that of the oriented dimer model, which is integrable, and to show that the dominant contribution to the observables in theorem\-~\ref{theo:nematic} come from the oriented dimer model. The oriented dimer model is integrable, and one easily shows that the local dimer density is invariant under translations and satisfies\-~(\ref{bound_1v}). In addition, pair correlations decay in the vertical direction with a rate $\ell_0^{-1}\approx e^{-\frac32J}z^{-\frac12}$, and are identically zero in the horizontal direction. Therefore, (\ref{bound_1vv}) holds in the oriented dimer model, with the improvement that the decay rate in the horizontal direction is infinite, rather than of order $J$. The full model does have horizontal correlations, mediated by horizontal dimers. In order to bound the difference in the partition functions of the oriented dimer model and the full one, we will compute the ratio of the dimer partition function to the oriented dimer partition function\-~(\ref{goal}) in terms of absolutely convergent series.
\bigskip

\indent Obviously, the difference between the full and the oriented dimer models is that there are both horizontal and vertical dimers in the former. With that in mind, we consider dimer configurations in terms of {\it horizontal} and {\it vertical phases} and {\it defects} (see figure\-~\ref{fig:dimer_contour}). A vertical phase is a region of $\mathbb Z^2$ that is occupied only by vertical dimers (and monomers); similarly, a horizontal phase is occupied by horizontal dimers. The interface between a vertical and a horizontal phase is a {\it defect}. This point of view is similar to the {\it Peierls argument} for the ferromagnetic Ising model, in which one can consider a spin configuration as a collection of contours which delineate regions containing only $+$ or $-$ spins. Unlike the Ising model, the configuration in a uniform phase is not unique (because they can contain monomers), but, since the oriented dimer model is integrable, we can compute the partition function in these regions (this is reminiscent of the models considered in\-~\cite{BKL84,BKL85}). In addition, given that we are computing the ratio\-~(\ref{goal}), the partition functions in uniform phases appearing in the numerator are {\it approximately} canceled out by the oriented dimer partition function in the denominator, leaving an effective weight for the defects.

\indent The dominant contribution to the weight of a defect comes from the fact that {\it most} dimers in the denominator of\-~(\ref{goal}) interact with a neighboring dimer (because the dimer activity is large), which means that almost every other vertical edge (we choose $q=\mathrm v$) contributes a factor $e^J$. We can keep track of these factors by assigning a weight $e^{\frac J2}$ to each endpoint of a dimer. On the other hand, the dimers on either side of a defect have different orientations, and, therefore, do not interact. By cutting these interactions, a defect of length $|l|$ contributes a factor $\approx e^{-\frac J2|l|}$ (see figure\-~\ref{fig:dimer_contour}). This is encouraging: in the language of Pirogov-Sinai theory\-~\cite{PS75}, this would indicate that the system satisfies the {\it Peierls condition} with a {\it large decay rate} $\frac J2$, which is a sufficient condition for general Pirogov-Sinai constructions\-~\cite{KP84,BKL84} to apply.

\indent There is, however, one important complication. As was mentioned earlier, the partition functions in the uniform phases only approximately cancel. Indeed, in the numerator, one has a product of oriented partition functions over a partition of $\Lambda$, whereas, in the denominator, there is only one oriented partition function over all of $\Lambda$. However, the partition function in a region depends on its geometry. In addition, while correlations in the oriented dimer model decay exponentially, they have a large correlation length $\ell_0\approx e^{\frac32J}\sqrt z$. There are, therefore, two length scales at play in this system: the microscopic size of a dimer, and the mesoscopic correlation length of the oriented dimer model. Therefore, the dependence of the oriented dimer partition function on the geometry of the region which it describes is strong when the diameter of the region does not exceed $\ell_0$. As a consequence, defects interact with each other, with an exponentially decaying interaction that has a very small decay rate $\ell_0^{-1}$. In order to deal with this interaction, we use the Mayer trick\-~\cite{Ur27,Ma37} (that is, we write the pair interaction $e^{-W}$ as $(e^{-W}-1)+1$ and expand) and split defect configurations into isolated bunches of interacting defects, called {\it polymers}, which interact only via a hard-core repulsion. We represent polymers graphically as a collection of defects connected to each other by lines representing the interaction (see figure\-~\ref{fig:polymers}). The effective activity of the polymers can then be shown to be $\approx e^{-\frac J2|l|-\ell_0^{-1}|\sigma|}$ where $|l|$ is the total length of the defects and $|\sigma|$ is the total length of the interaction lines. This looks much worse than $e^{-\frac J2|l|}$: the decay rate is now $\approx\ell_0^{-1}$ which is extremely small and may not, a priori, suffice to control the entropy of the polymers: in a model of arbitrary polymers with activity $e^{-\ell_0^{-1}|l|}$, it would be likely to find polymers, whereas we need them to be rare.

\indent The key ingredient to overcome this difficulty is that the interaction is one-dimensional: it comes from the oriented dimer model, and takes place over vertical or horizontal lines, so the contribution to the entropy of a polymer from its interactions is only a one-dimensional sum. In addition, interaction lines are always connected to a defect, which has a very small weight. In fact, the smallest possible defect is of length $6$, so the largest possible weight for a defect is $e^{-3J}$. On the other hand, the sum over the length of the interaction lengths yields $\sum_\ell e^{-\ell_0^{-1}\ell}\approx\ell_0$. Now, since every new interaction line must connect to a new defect, the overall contribution of the interaction line along with the defect to which it is connected, is, at most, $\ell_0e^{-3J}\ll 1$. This allows us to control the entropy of the polymers, even though the decay rate of the interaction lines is small. Having done so, we use a cluster expansion\-~\cite{Ru99,GBG04,KP86,BZ00} to compute the partition function of the polymer model and\-~(\ref{goal}).

\indent There are some more technical complications that arise in the proof. One of these is standard in Pirogov-Sinai theory: unlike the Ising model, the partition function of the oriented dimer model may take different values for vertical and horizontal boundary conditions, which prevents us from using a straight Peierls argument. In order to avoid long-range interactions in the defect model, we must {\it flip} the boundary condition inside each defect back to the vertical, and, in doing so, introduce an extra factor in the activity of the defect that depends on the partition function of the full dimer model inside the defect with both boundary conditions. We then show that this term is, at most, exponentially large in the size of the defect with a rate that is much smaller than $\frac J2$, and thereby causes no trouble. To do so, we must bound the partition function inside the defect from above and below, which we do by induction, and is the main reason why we compute the ratio\-~(\ref{goal}) instead of merely bounding it.

\indent In addition, we have found it necessary to avoid interaction lines of length $<\ell_0$. This is due to the fact that the polymer model we have constructed contains {\it trivial polymers}, which do not contain any defect and consist of a single interaction line going all the way through $\Lambda$. Whenever such lines are of length $<\ell_0$ (which may occur since, in order to carry out the inductive argument mentioned above, we cannot restrict our attention to $\Lambda$'s of large volume), their activity can be close to $\pm 1$. This causes a number of issues, which we have opted to remedy by ensuring that no short trivial polymers may arise. This can be accomplished by grouping defects that are closer than $\ell_0$ from each other into bunches, called {\it contours} (see figure\-~\ref{fig:contours}).

\indent Finally, the introduction of sources to compute correlation functions comes with its share of pesky complications, which we will not comment on here. In fact, readers who are not interested in the fine details of the proof are invited to consider only the case $\Upsilon=\emptyset$, and skip the source-specific paragraphs on a first reading.

\section{Solution of the one-dimensional problem}\label{sec:1d}
\indent In this section, we compute the partition function of the oriented dimer model on a finite, connected chain $\{1,\cdots,\ell\}\subset\mathbb Z$, with various boundary conditions.
\bigskip

\indent In order to specify the boundary condition, we introduce the following notation. We introduce a real vector space,
\begin{equation}
  \Delta:=\mathrm{span}\{\left|\rightdimer\right>,\left|\leftdimer\right>,\left|\times\right>\}
\end{equation}
Given a pair of vectors $\omega\equiv(\omega_1,\omega_\ell)\in\Delta^2$, we define the partition function $\Psi^{(\omega)}(\ell)$ in the following way.
\begin{itemize}
  \item If $\omega_1=\left|\rightdimer\right>$, $\left|\leftdimer\right>$ or $\left|\times\right>$, then the first vertex must be covered by, respectively, a half-dimer pointing right, a half-dimer pointing left or a monomer;
  \item For symmetry reasons, $\omega_\ell$ is defined the other way around (this notation may seem slightly counter-intuitive, but it will be useful in the following): if $\omega_\ell=\left|\rightdimer\right>$, $\left|\leftdimer\right>$ or $\left|\times\right>$, then the last vertex must be covered by, respectively, a half-dimer pointing left, a half-dimer pointing right or a monomer.
  \item $\Psi^{(\omega)}(\ell)$ is bilinear in $\omega$. 
\end{itemize}
For example, the partition function with open boundary conditions is obtained by taking $\omega=(\left|\rightdimer\right>+\left|\times\right>,\left|\rightdimer\right>+\left|\times\right>)$.
\bigskip

\theo{Lemma}\label{lemma:1d}
  For every $\ell\geqslant 1$, we have, for $\omega\equiv(\omega_1,\omega_\ell)\in\Delta^2$,
  \begin{equation}
    \Psi^{(\omega)}(\ell)=\nu_+(\omega_1)\nu_+(\omega_\ell)b_+\lambda_+^{\ell}+\nu_-(\omega_1)\nu_-(\omega_\ell)b_-\lambda_-^{\ell}+\nu_0(\omega_1)\nu_0(\omega_\ell)b_0\lambda_0^{\ell}
    \label{Zv0}
  \end{equation}
  with
  \begin{equation}
    \begin{array}{>\displaystyle c}
      \lambda_+=\left(\sqrt{ze^J}+\frac12e^{-J}\right)(1+O(e^{-J}\epsilon^2))
      ,\quad
      \lambda_-=\left(-\sqrt{ze^J}+\frac12e^{-J}\right)(1+O(e^{-J}\epsilon^2))
      ,\\[0.3cm]
      \lambda_0=1-e^{-J}(1+\epsilon^2)+O(e^{-2J}\epsilon^2)
    \end{array}
    \label{eigenvalues}
  \end{equation}
  in which
  \begin{equation}
    \epsilon:=\frac1{\sqrt{ze^J}}
    \label{epsilon}
  \end{equation}
  and, for $i\in\{+,-,0\}$,
  \begin{equation}
    b_i:=\frac z{(2\lambda_i(\lambda_i-1)^2+z)\lambda_i}
    \label{b}
  \end{equation}
  $\nu_i(\omega_j)$ is linear in $\omega_j$, and
  \nopagebreakaftereq
  \begin{equation}
    \nu_i(\left|\rightdimer\right>)=\lambda_i-1
    ,\quad
    \nu_i(\left|\leftdimer\right>)=\frac{\lambda_i}z(\lambda_i-1)
    ,\quad
    \nu_i(\left|\times\right>)=1.
    \label{nu}
  \end{equation}
\endtheo
\restorepagebreakaftereq
\bigskip

\indent\underline{Proof}:
  We will use a {\it transfer matrix} approach.
  \bigskip
  
  \point Every vertex may be in one of three states: it is either covered by a half-dimer pointing right (\rightdimer), a half-dimer pointing left (\leftdimer), or no dimer ($\times$). One easily checks that the partition function can be written as
  \begin{equation}
    \Psi^{(\omega_1,\omega_\ell)}(\ell)=\omega_1\cdot T^{\ell-1}\omega_\ell
  \end{equation}
  where $T$ is the {\it transfer matrix}, whose expression, in the $(\left|\rightdimer\right>, \left|\leftdimer\right>, \left|\times\right>)$ basis, is
  \begin{equation}
    T:=
    \left(\begin{array}{*{3}{c}}
      0&z&0\\
      e^J&0&1\\
      1&0&1
    \end{array}\right).
  \end{equation}
  \bigskip

  \point By straightforward computation, we diagonalize $T$:
  \begin{equation}
    T=P
    \left(\begin{array}{*{3}{c}}
      \lambda_+&0&0\\
      0&\lambda_-&0\\
      0&0&\lambda_0
    \end{array}\right)
    P^{-1}
  \end{equation}
  where $\lambda_\pm$ and $\lambda_0$ satisfy\-~(\ref{eigenvalues}),
  \begin{equation}
    P=
    \left(\begin{array}{*{3}{c}}
      \lambda_+-1&\lambda_--1&\lambda_0-1\\
      \frac{\lambda_+}z(\lambda_+-1)&\frac{\lambda_-}z(\lambda_--1)&\frac{\lambda_0}z(\lambda_0-1)\\
      1&1&1
    \end{array}\right)
    \left(\begin{array}{*{3}{c}}
      \frac1{N_+}&0&0\\
      0&\frac1{N_-}&0\\
      0&0&\frac1{N_0}
    \end{array}\right).
  \end{equation}
  and
  \begin{equation}
    P^{-1}=
    \left(\begin{array}{*{3}{c}}
      \frac1{N_+}&0&0\\
      0&\frac1{N_-}&0\\
      0&0&\frac1{N_0}
    \end{array}\right)
    \left(\begin{array}{*{3}{c}}
      \frac{\lambda_+}z(\lambda_+-1)&\lambda_+-1&1\\[0.3cm]
      \frac{\lambda_-}z(\lambda_--1)&\lambda_--1&1\\[0.3cm]
      \frac{\lambda_0}z(\lambda_0-1)&\lambda_0-1&1
    \end{array}\right)
  \end{equation}
  with, for $i\in\{+,-,0\}$,
  \begin{equation}
    N_i:=\sqrt{2\frac{\lambda_i}z(\lambda_i-1)^2+1}.
  \end{equation}
  Therefore, the lemma holds with
  \begin{equation}
    b_i=\frac1{N_i^2\lambda_i}
    ,\quad
    \nu_i(\omega)=(\omega^TP)_i.
  \end{equation}
\qed

\section{Dilute hard-core polymer model}\label{sec:polymer_model}
\indent In this section, we will map the high-density dimer model to a dilute model of polymers, which only interact with each other via a hard-core repulsion. We proceed in five steps: we first map the dimer model to a {\it loop} model, then to an {\it external} contour model, for which we compute the activity and interaction of external contours, and then map the contour model to a system of {\it external} polymers, and, finally, introduce the polymer model.
\bigskip

\subsection{Loop model}\label{subsec:loops}
\indent First of all, for $c\in\{\mathrm v,\mathrm h\}$, we define the \define{$c$-support} of $\underline\delta$ as the set of vertices that are covered by $c$-dimers:
\begin{equation}
  \mathrm{supp}_{c}(\underline\delta):=\bigcup_{\{x,x'\}\in \mathbb D_{c}(\underline\delta)}\{x,x'\}
\end{equation}
(recall that $\mathbb D_c(\underline\delta)$ is the set of $c$-dimers in $\delta$).
\bigskip

\indent We construct a family $\mathcal L_q(\underline\delta)$ of {\it bounding loops} associated to $\underline\delta$. A \define{loop} is a set of edges $l$, such that there exists a {\it simply connected} (a simply connected set is a set whose complement is connected) set $\bar l\subset\mathbb Z^2$ such that $l$ is the boundary of $\bar l$: $l=\partial\bar l$. To assign a loop to $\underline\delta$, we will proceed by induction. If $\underline\delta$ only consists of $q$-dimers, then $\mathcal L_q(\underline\delta)=\emptyset$. If not, then the boundary of $\mathrm{supp}_{-q}(\underline\delta)$ is a non-empty union of disjoint loops, denoted by $\underline{\mathfrak l}\equiv\{\mathfrak l_1,\cdots,\mathfrak l_{|\underline{\mathfrak l}|}\}$. From these, we extract the {\it most external} ones $\underline{\mathfrak l}'\equiv\{\mathfrak l'_1,\cdots,\mathfrak l'_{|\underline{\mathfrak l}'|}\}\subset\underline{\mathfrak l}$ by discarding loops that lie inside other loops, that is, $\bar{\mathfrak l}'_i\cap\bar{\mathfrak l}_j\neq\emptyset$ if and only if $\mathfrak l'_i=\mathfrak l_j$ (the fact that there is no prime in $\bar{\mathfrak l}_j$ or $\mathfrak l_j$ is not a typo: the $\mathfrak l'_i$ are external to {\it all} loops).
\bigskip

\begin{figure}
  \hfil\includegraphics[width=6.667cm]{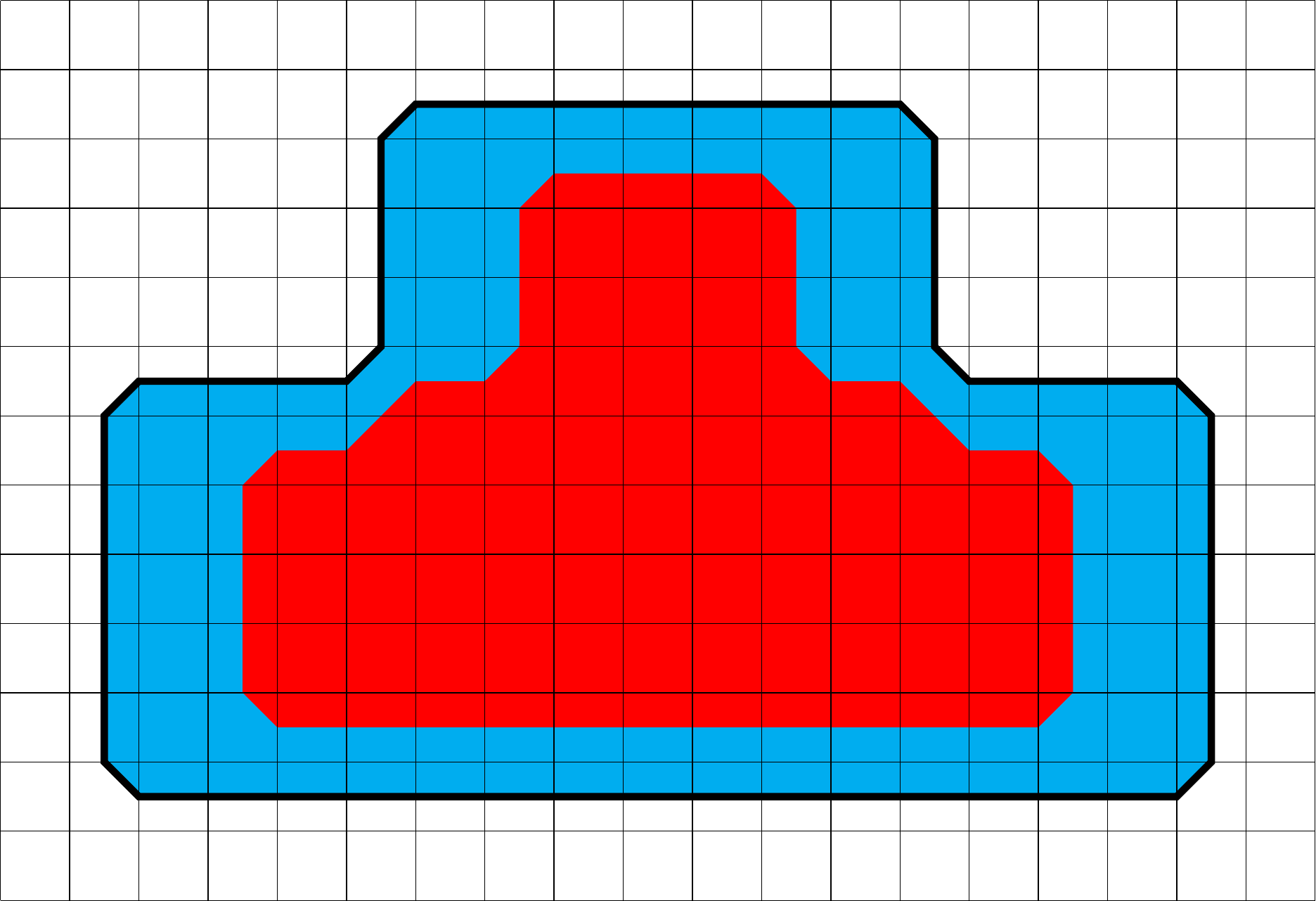}
  \caption{An $\mathrm h$-bounding loop. The loop is depicted as a thick line running through the edges that make it up. The core of the loop is colored red and its mantle is colored cyan (color online).}
  \label{fig:bounding_loop}
\end{figure}

\begin{figure}
  \hfil\includegraphics[width=8cm]{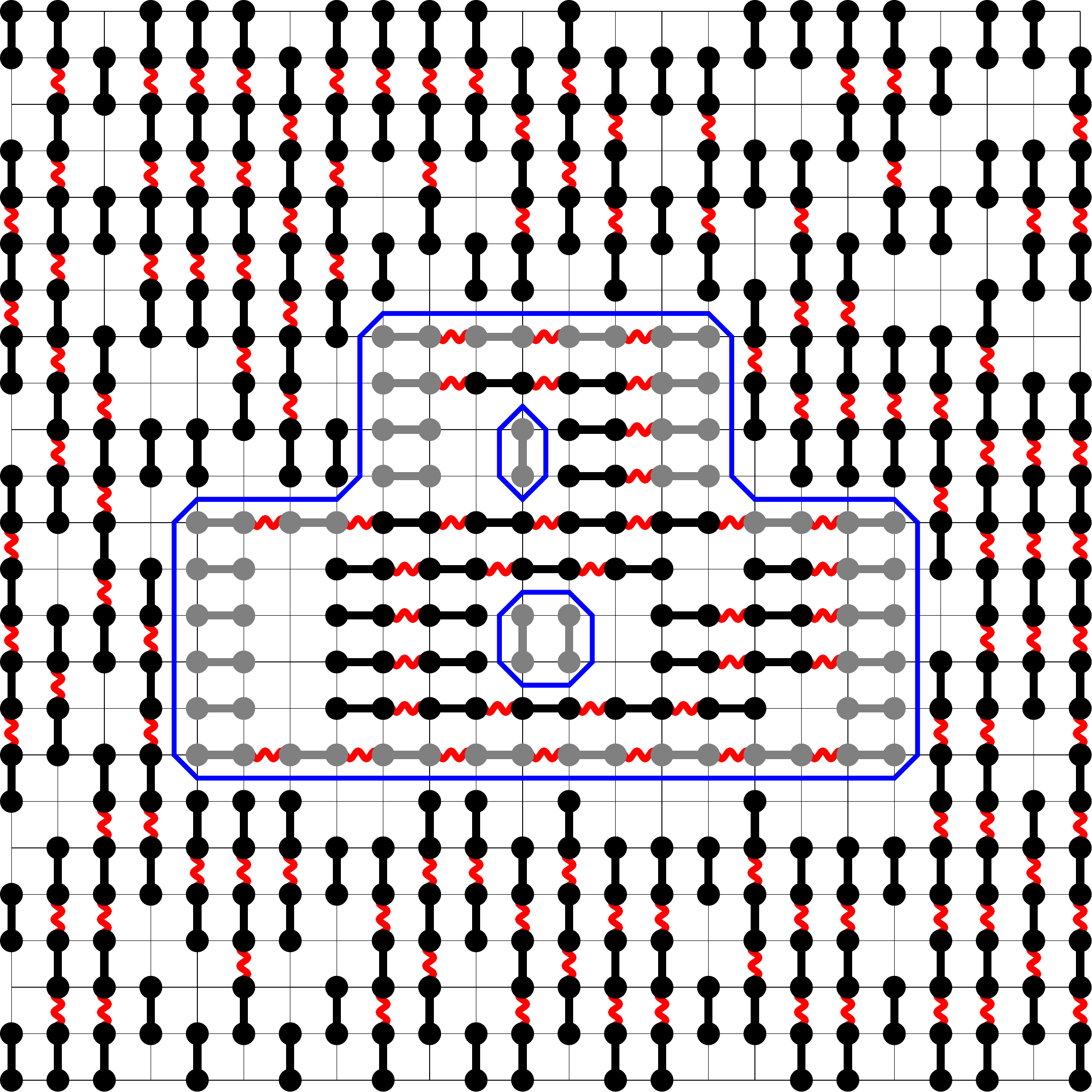}
  \caption{The loops associated to the dimer configuration in figure\-~\ref{fig:interaction}. The dimers in the mantles are gray.}
  \label{fig:dimer_loops}
\end{figure}

\indent These loops separate a $q$-phase from a $-q$ phase, which implies some geometric constraints. For one, the inside of each loop is lined with $-q$-dimers. To capture these properties, we define the notion of a {\it $c$-bounding loop} for $c\in\{\mathrm v,\mathrm h\}$. To do so, we split the interior of a loop into a region which {\it must} be covered by $c$-dimers (which we call the {\it mantle} of the loop), and the rest (the {\it core}), see figure\-~\ref{fig:bounding_loop}. Formally, the \define{core} is defined as
\begin{equation}
  \mathbb I^{(\Upsilon)}_{c}(l):=\left\{x\in\bar l\setminus\left({\textstyle\bigcup_{\upsilon\in\Upsilon}}\upsilon\right),\quad \mathfrak d_c(x,\mathbb D_c(l))\geqslant 2,\ \mathfrak d_{-c}(x,\mathbb D_{-c}(l))\geqslant 1\right\}
  \label{bbI}
\end{equation}
(recall that $\mathfrak d_c$ is the $c$-distance on $\mathbb Z^2$) (remark: we do not count the sources as being part of the interior) and its \define{mantle} as
\begin{equation}
  \mathbb O_{c}(l):=\bar l\setminus\mathbb I^{(\emptyset)}_{c}(l).
\end{equation}
A \define{$c$-bounding loop} is a loop whose mantle is disjoint from the sources $\bigcup_{\upsilon\in\Upsilon}\upsilon$, and can be completely covered by $c$-dimers (see figures\-~\ref{fig:bounding_loop} and\-~\ref{fig:dimer_loops}). We denote the set of $c$-bounding loops by $\mathfrak L_c^{(\Upsilon)}(\Lambda)$, and the set of all bounding loops by $\mathfrak L^{(\Upsilon)}(\Lambda)\equiv\mathfrak L_{\mathrm v}^{(\Upsilon)}(\Lambda)\cup\mathfrak L_{\mathrm h}^{(\Upsilon)}(\Lambda)$. (Note that some loops could be $\mathrm v$-bounding loops as well as $\mathrm h$-bounding loops. The index $c$ is meant as an extra structure, which is not a function of the geometry of the loop.)
\bigskip

\indent The loops $\mathfrak l'_i$ are $-q$-{\it bounding loops} and are disjoint. Finally, denoting the set of dimers that are contained inside a bounding loop $\mathfrak l'_j$ by $\underline\delta\cap\bar{\mathfrak l}'_j$, we define, inductively, (see figure\-~\ref{fig:dimer_loops})
\begin{equation}
  \mathcal L_q(\underline\delta):=\underline{\mathfrak l}'\cup\bigcup_{j=1}^{|\underline{\mathfrak l}'|}\mathcal L_{-q}(\underline\delta\cap\bar{\mathfrak l}'_j).
\end{equation}
The loops in $\mathcal L_q(\underline\delta)$ are disjoint, and their mantles are disjoint.
\bigskip

\begin{figure}
  \hfil\includegraphics[height=8cm]{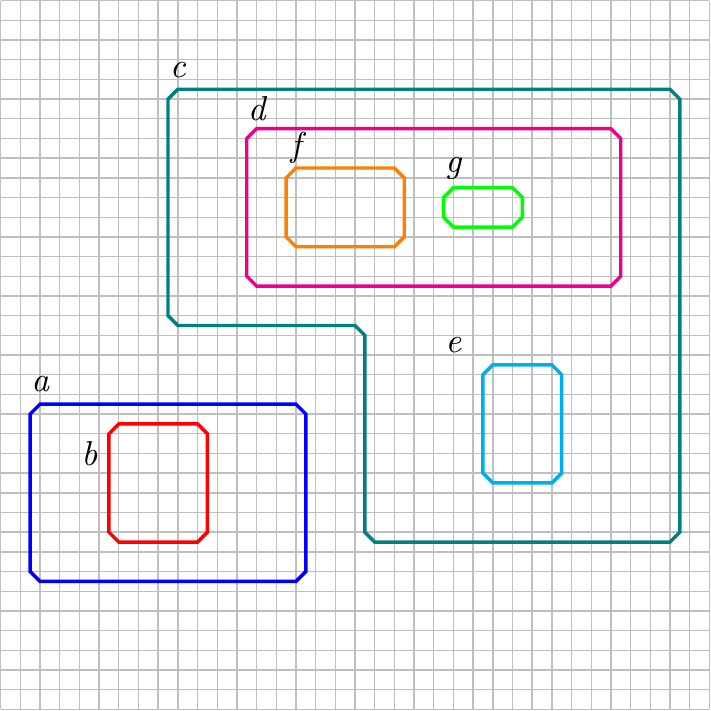}
  \hfil\raise2cm\hbox{\includegraphics[height=4cm]{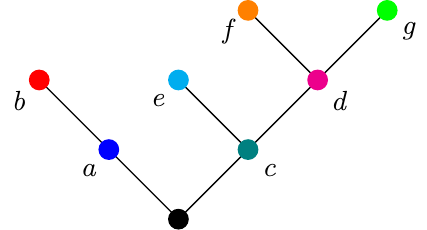}}
  \caption{Nested bounding loops and their corresponding inclusion tree. The root of the tree is drawn in black. The colors (color online) and labels of the nodes in the tree match with those of the loops. The tree is {\it alternating} because the blue, teal, orange and green ({\it a}, {\it c}, {\it f} and {\it g}) loops are $\mathrm h$-bounding loops and the red, cyan and magenta ({\it b}, {\it e} and {\it d}) ones are $\mathrm v$-bounding loops.}
  \label{fig:inclusion_tree}
\end{figure}

\indent These bounding loops are {\it alternating}, in the sense that $q$-loops may only encircle $-q$-loops. It is useful, to define this notion, to introduce {\it inclusion trees}. The \define{inclusion tree} associated to $\mathcal L_q(\underline\delta)$ is a tree $T(\mathcal L_q(\underline\delta))$ (see figure\-~\ref{fig:inclusion_tree} for an example) that is such that
\begin{itemize}
  \item $T(\mathcal L_q(\underline\delta))$ has $|\mathcal L_q(\underline\delta)|+1$ nodes. One node corresponds to $\partial\Lambda$, and is called the {\it root} of $T(\mathcal L_q(\underline\delta))$, while the $|\mathcal L_q(\underline\delta)|$ others each correspond to a loop $l\in\mathcal L_q(\underline\delta)$.
  \item For each $l\in\mathcal L_q(\underline\delta)$, the corresponding node has a unique {\it parent}, chosen in such a way that every loop containing $l$ is an {\it ancestor} of $l$ (a loop $l'$ contains a loop $l$ if $l\subset\mathcal E(\bar l')$ (recall that $\mathcal E(\bar l')$ is the set of edges in $\bar l'$)).
\end{itemize}
The inclusion tree is \define{alternating}, in the sense that the children of a $c$-bounding loop are $-c$-bounding loops. We define $c(l)$ as the orientation of the loop $l$ (that is, $l$ is a $c(l)$-bounding loop), which we will also call the \define{index} of $l$.
\bigskip

\begin{figure}
  \hfil\includegraphics[width=8cm]{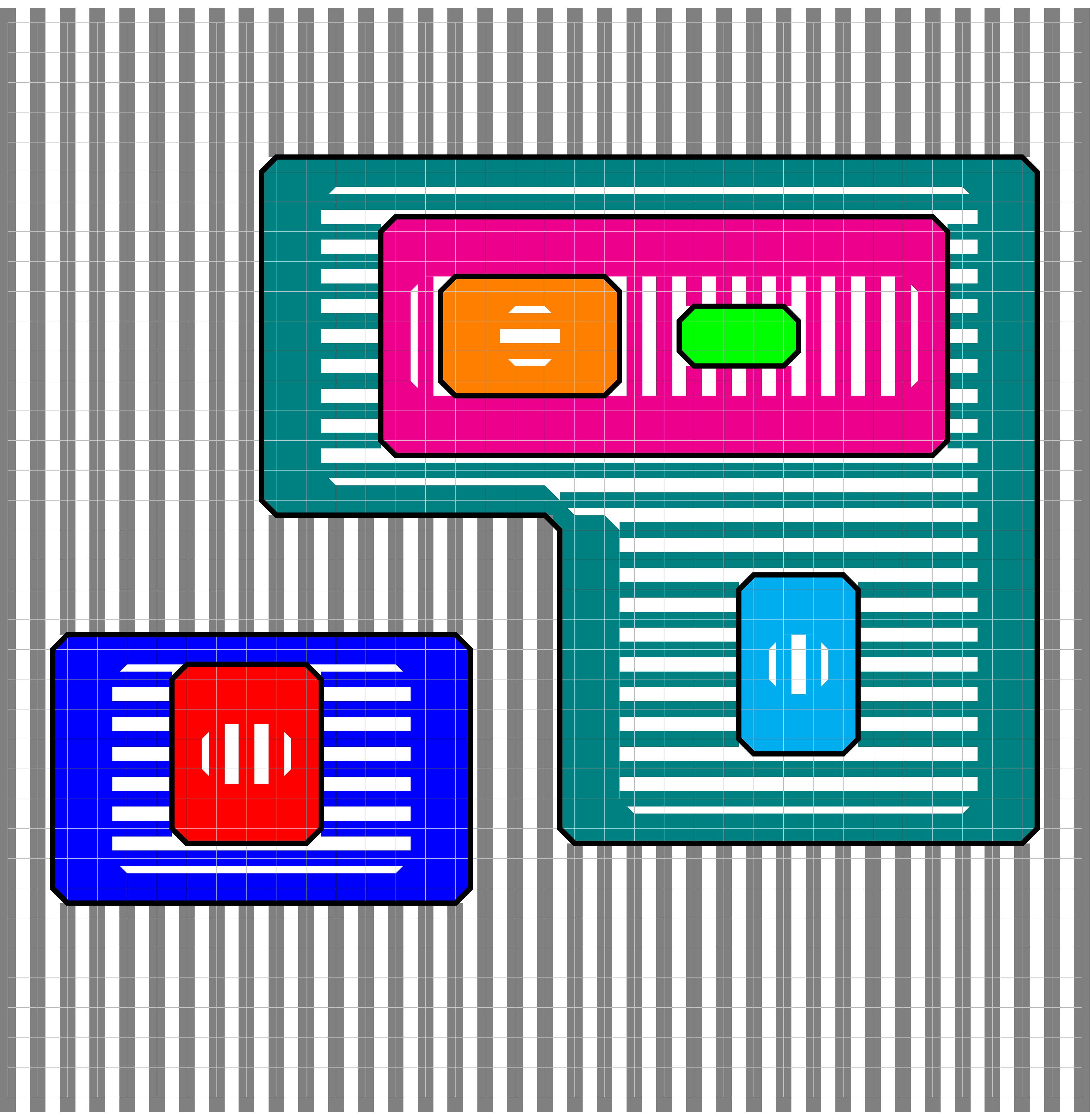}
  \caption{The segments inside and outside the loops. The sets $\iota_{c(l)}(l,\mathcal L)$ are depicted in the same color (color online) as the loop to which they correspond. The mantles of the loops are covered in black.}
  \label{fig:segments}
\end{figure}

\indent Thus, given a dimer configuration, we have constructed a set of bounding loops $\mathcal L_q(\underline\delta)$. Conversely, if we fix a family $\mathcal L_q$ of bounding loops that are disjoint and whose mantles are disjoint, whose most external loops are $q$-bounding loops, and are such that $T(\mathcal L_q)$ is alternating, then the set of dimer configurations $\underline\delta$ such that $\mathcal L_q(\underline\delta)=\mathcal L_q$ is equal to the set of configurations satisfying the following: for every loop $l\in\mathcal L_q$,
\begin{itemize}
  \item for every $-q$-dimer, there exists $l'\in\mathcal L_q$ such that the dimer is in $\bar l'$,
  \item the {\it mantle} $\mathbb O_{c(l)}(l)$ of $l$ is completely covered by $c(l)$-dimers.
\end{itemize}
Therefore, we can rewrite the dimer partition function\-~(\ref{goal}) as
\begin{equation}
  \frac{Z^{(\Upsilon)}(\Lambda|\mathbf q)}{\mathfrak Z_{\mathbf q}^{(\Upsilon)}(\Lambda)}=
  \sum_{\mathcal L\subset\mathfrak L^{(\Upsilon)}(\Lambda)}
  \left(\prod_{l\neq l'\in\mathcal L}\varphi(l,l')\right)
  \mathds 1_q(\mathcal L)
  \frac{\mathfrak Z_{\mathbf q}^{(\Upsilon)}(\Lambda\setminus\cup_{l\in\mathcal L}\bar l)}{\mathfrak Z_{\mathbf q}^{(\Upsilon)}(\Lambda)}
  \prod_{l\in\mathcal L}\left(
    \mathfrak Y_{c(l)}^{(\Upsilon)}(\mathbb O_{c(l)}(l))
    \mathfrak Z_{\mathbf c(l)}^{(\Upsilon)}(\iota^{(\Upsilon)}_{c(l)}(l,\mathcal L))
  \right)
  \label{loops}
\end{equation}
in which, roughly (these quantities are formally defined below), $\varphi_{\mathrm{ext}}(l,l')$ is a hard-core pair interaction that keeps the loops or their mantles from intersecting, $\mathds 1_q(\mathcal L)$ ensures that $T(\mathcal L)$ is alternating and that the index of its root is $q$, $\mathfrak Z_{\mathbf q}^{(\Upsilon)}(\Lambda\setminus\cup_{l\in\mathcal L}\bar l)$ is the partition function of $q$-dimers outside all loops, $\mathfrak Y_{c(l)}^{(\Upsilon)}(\mathbb O_{c(l)}(l))$ is the weight of the $c(l)$-dimers in the mantle of $l$, and $\mathfrak Z_{\mathbf c(l)}^{(\Upsilon)}(\iota_{c(l)}^{(\Upsilon)}(l,\mathcal L))$ is the partition function of $c(l)$-dimers in between loops.
\begin{itemize}
  \item $\varphi(l,l')\in\{0,1\}$ is equal to 1 if and only if $l$ and $l'$ are disjoint and their mantles are disjoint.
  \item $\mathds 1_q(\mathcal L)\in\{0,1\}$ is equal to 1 if and only if $T(\mathcal L)$ is alternating, and the index of its root is $q$.
  \item $\iota_{c}^{(\Upsilon)}(l,\mathcal L)$ is the \define{padding} of $l$, and is the space inside the loop $l$ that is external to all other loops (see figure\-~\ref{fig:segments}):
  \begin{equation}
    \iota^{(\Upsilon)}_{c(l)}(l,\mathcal L):=\mathbb I^{(\Upsilon)}_{c(l)}(l)\setminus\left({\textstyle\bigcup_{l'\in\mathcal L\setminus\{l\}}\bar l'}\right).
    \label{iota}
  \end{equation}
  (Note that one could restrict the union over $l'$ to loops {\it inside} $l$, but this is not necessary.)
  \item $\mathbf c(l)$ is the boundary condition of $\iota_{c(l)}^{(\Upsilon)}(l,\mathcal L)$. Because the $c(l)$-dimers that are in $\mathbb O_{c(l)}(l)$ can interact with those in $\iota_{c(l)}^{(\Upsilon)}(l,\mathcal L)$, the entire boundary is {\it magnetized} (see figure\-~\ref{fig:dimer_loops}): with the notation of section\-~\ref{sec:model}, $\mathbf c(l):=(c(l),\partial_{c(l)}\mathbb I_{c(l)}^{(\emptyset)}(l),\ell_0)$ (we recall that $\partial_{c(l)}\equiv\mathbb D_{c(l)}(\partial)$).
  \item $\mathfrak Y_{c(l)}^{(\Upsilon)}(\mathbb O_{c(l)}(l))$ is the weight of the dimers in $\mathbb O_{c(l)}(l)$, which is packed with dimers (see figure\-~\ref{fig:dimer_loops}):
  \begin{equation}
    \mathfrak Y_{c}(\mathbb O)
    :=z^{|\underline\delta_c(\mathbb O)|}e^{-W_0(\underline\delta_c(\mathbb O))}\mathfrak B_{\mathbf q}^{(\Upsilon)}(\underline\delta_c(\mathbb O))
    \label{frakY}
  \end{equation}
  where $\underline\delta_c(\mathbb O)$ is the unique closely-packed $c$-dimer configuration in $\mathbb O$, and $\mathfrak B_{\mathbf q}^{(\Upsilon)}$ is the boundary term at the sources (see\-~(\ref{eW0}) and\-~(\ref{frakB_Upsilon})).
\end{itemize}
\bigskip

\subsection{External contour model}\label{subsec:external_contours}
\indent These bounding loops {\it interact} with each other through the dimers in the space between them. Actually, as we will see in the following, $q$-bounding loops that are at a $-q$-distance of less than $\ell_0$ interact {\it strongly}. In order to avoid dealing with distinct objects that interact strongly, we group these loops together to form a larger object, called a {\it contour}, which is a set of loops that are all at a distance $<\ell_0$ from each other.
\bigskip

\indent The interaction is one-dimensional, and is either horizontal or vertical. To represent it graphically, it is convenient to introduce the notion of a {\it segment}. To that end, we define, for $y\in\mathbb Z^2$, the $\mathrm v$- and $\mathrm h$-lines going through $y$:
\begin{equation}
  \chi_y^{(\mathrm v)}:=\{(y,x),\ x\in\mathbb Z\}
  ,\quad
  \chi_y^{(\mathrm h)}:=\{(x,y),\ x\in\mathbb Z\}.
\end{equation}
Put simply, a $\mathrm v$-line is a vertical line and an $\mathrm h$-line is a horizontal one. In addition, we define a map $\Sigma_{c}$ that takes a bounded set $\Lambda'\subset\mathbb Z^2$ as an argument, and splits it up into $c$-segments. Formally, given $y\in\mathbb Z$, we denote the set of connected components of $\Lambda'\cap\chi_y^{(c)}$ by $\underline\sigma_{c,y}(\Lambda')$ (that is, $\underline\sigma_{c,y}(\Lambda')$ is a set whose elements are the connected components of $\Lambda'\cap\chi_y^{(c)}$), and define
\begin{equation}
  \Sigma_{c}(\Lambda'):=\bigcup_{y\in\mathbb Z}\underline\sigma_{c,y}(\Lambda').
  \label{Sigma}
\end{equation}
We then define the set of \define{segments} in between the loops in $\mathcal L_q(\underline\delta)$ (see figure\-~\ref{fig:segments}) as
\begin{equation}
  \mathcal S_\Lambda^{(\Upsilon)}(\mathcal L_q(\underline\delta)):=\left(\Sigma_q\left(\Lambda^{(\Upsilon)}\setminus\left({\textstyle\bigcup_{l\in\mathcal L}\bar l}\right)\right)\right)\cup\left(\bigcup_{l\in\mathcal L}\Sigma_{c(l)}(\iota^{(\Upsilon)}_{c(l)}(l,\mathcal L_q(\underline\delta)))\right).
  \label{calS}
\end{equation}
\bigskip

\indent We wish to gather together the loops that are separated by segments of length $<\ell_0$. To do so, we define the \define{support} of $\mathcal L_q(\underline\delta)$ as the pair
\begin{equation}
  \mathrm{supp}_{\Lambda,\ell_0}^{(\Upsilon)}(\mathcal L_q(\underline\delta)):=(\mathcal L_q(\underline\delta),\mathcal S_{\Lambda,<\ell_0}^{(\Upsilon)}(\mathcal L_q(\underline\delta)))
\end{equation}
in which $\mathcal S_{\Lambda,<\ell_0}^{(\Upsilon)}(\mathcal L_q(\underline\delta))$ is the set of segments of length $<\ell_0$:
\begin{equation}
  \mathcal S_{\Lambda,<\ell_0}^{(\Upsilon)}(\mathcal L_q(\underline\delta)):=\{\sigma\in\mathcal S_\Lambda^{(\Upsilon)}(\mathcal L_q(\underline\delta)),\ |\sigma|<\ell_0\}.
  \label{Sleell0}
\end{equation}
A {\it contour} is a subset of $\mathcal L_q(\underline\delta)$ that has a {\it connected} support, in the following sense. A $c$-segment $s$ is said to be \define{connected} to a bounding loop $l$ if $\partial_c s\cap \partial\mathbb O_{c(l)}(l)\neq\emptyset$. Similarly, two bounding loops $l,l'$ are said to be connected if $\partial\mathbb O_{c(l)}(l)\cap\partial\mathbb O_{c(l')}(l')\neq\emptyset$. Finally, given a set of segments $\mathcal S$ and a set $\mathcal L$ of disjoint loops, $(\mathcal L,\mathcal S)$ is said to be \define{connected} if, for every $x,y\in\mathcal S\cup\mathcal L$, there exists a path from $x$ to $y$, that is, there exists $\underline p\equiv(p_1,\cdots,p_{|\underline p|})$ with $p_i\in\mathcal S\cup\mathcal L$, $p_1=x$, $p_{|\underline p|}=y$, $p_i$ and $p_{i+1}$ are never both segments, and $p_i$ and $p_{i+1}$ are connected.
\bigskip

\begin{figure}
  \hfil\includegraphics[width=8cm]{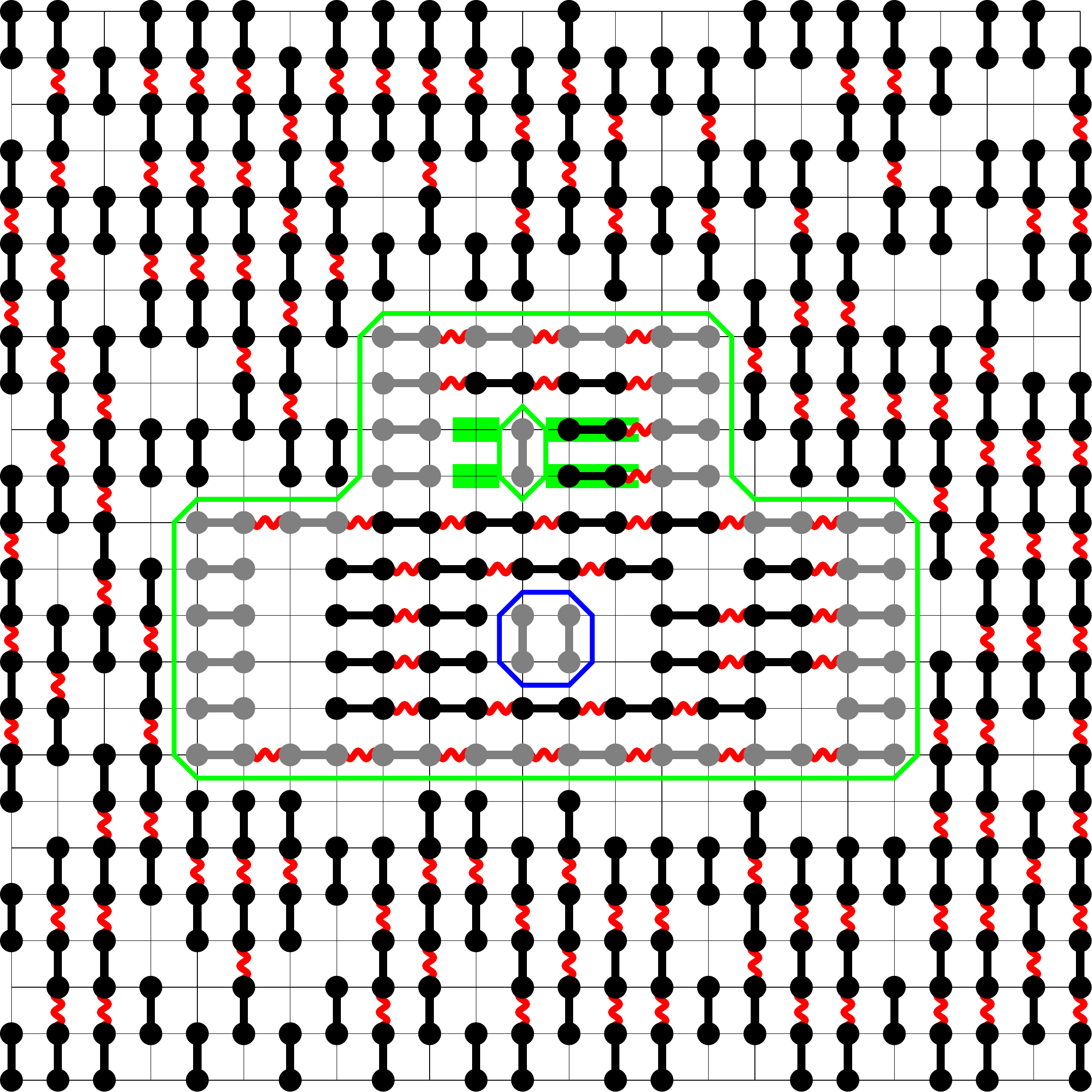}
  \caption{The contour configuration associated to the dimer configuration in figure\-~\ref{fig:interaction}. Here, we took $q=\mathrm v$ and $\ell_0=4$. There are three bounding loops. The two green (color online) bounding loops form a single contour because the segments that separate the inner loop from the mantle of the outer one (drawn in green in the figure) are of length 1 or 2, which is less than $\ell_0$. The blue (color online) bounding loop forms its own contour, but, since it is not external to the green one, it is dropped.}
  \label{fig:dimer_contour}
\end{figure}

\indent We then split $\mathcal L_q(\underline\delta)$ into {\it connected components} (see figure\-~\ref{fig:dimer_contour}), $\underline\Gamma_q(\underline\delta)\equiv\{\Gamma_1,\cdots,\Gamma_{|\underline\Gamma_q(\underline\delta)|}\}$, that is, $\Gamma_i\subset\mathcal L_q(\underline\delta)$, the support of $\Gamma_i$ is connected, $\Gamma_i\cap\Gamma_j=\emptyset$ whenever $i\neq j$, $\Gamma_1\cup\cdots\cup\Gamma_{|\underline\Gamma_q(\underline\delta)|}=\mathcal L_q(\underline\delta)$ and, finally, the support of $\Gamma_i\cup\Gamma_j$ is {\it disconnected} when $i\neq j$ (when two contours are disconnected from each other, we say they are \define{compatible}).
\bigskip

\begin{figure}
  \hfil\includegraphics[width=8cm]{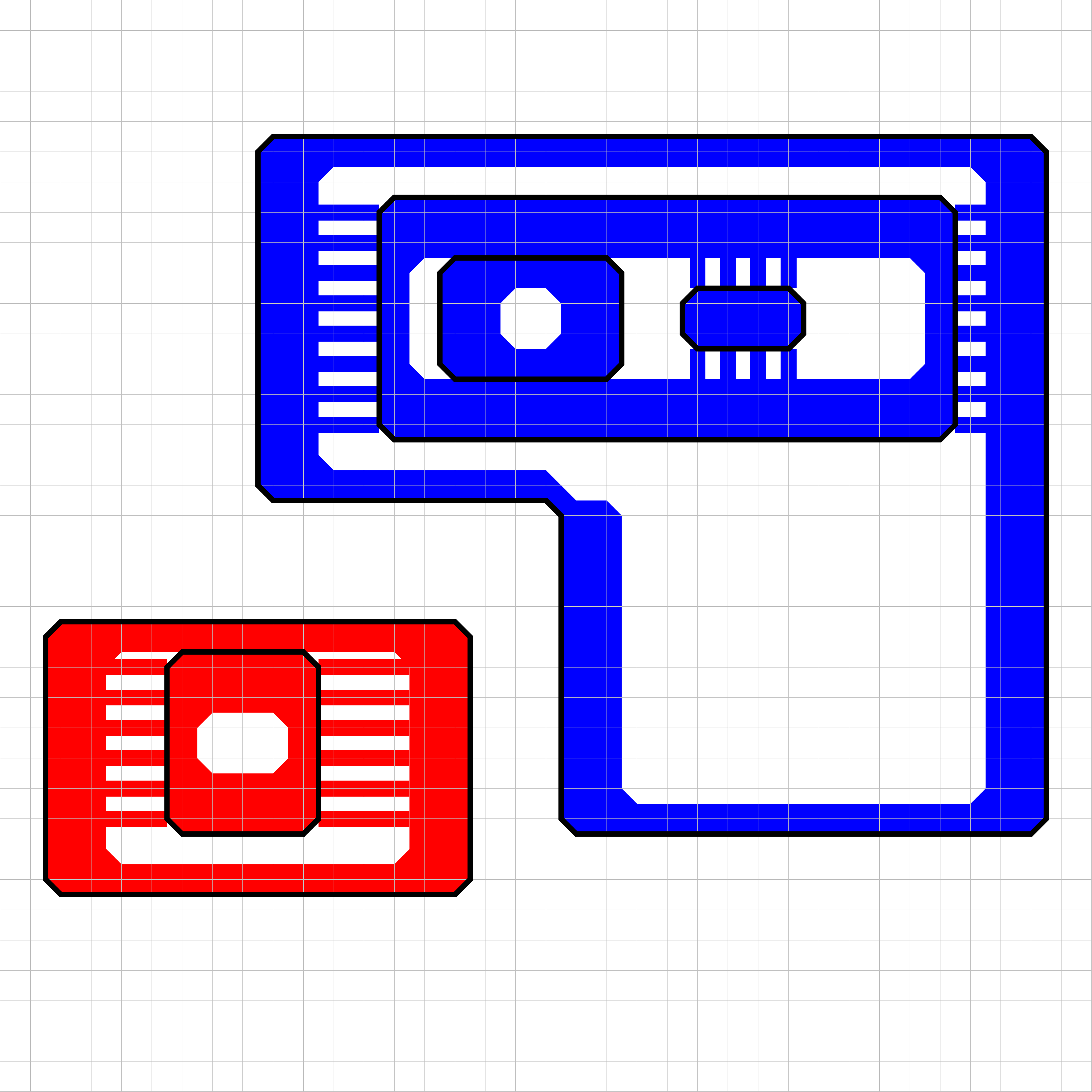}
  \caption{Two contours, depicted in different colors (color online). Here, we have taken $\ell_0=4$ and $q=\mathrm v$. The segments in the support are drawn as thick lines, and the mantles have been filled in. These two contours are compatible and external to each other. Comparing with figure\-~\ref{fig:inclusion_tree}, note that the cyan loop labeled as {\it e} is not depicted here, since it would not be external to the other contours.}
  \label{fig:contours}
\end{figure}

\indent The fact that the inclusion tree $T(\mathcal L_q(\underline\delta))$ is alternating induces a long-range interaction between contours: a $q$-bounding loop must lie inside a $-q$-bounding loop, independently of the distance that separates them. In order to avoid this, we will call upon a technique used in Pirogov-Sinai theory\-~\cite{PS75,KP84}. The first step is to focus on the contours that are the most {\it external}, in the following sense. Two contours $\Gamma,\Gamma'$ are \define{external} to each other if every loop $l\in\Gamma$ and every $l'\in\Gamma'$ are external to each other: $\bar l\cap\bar l'=\emptyset$. The set of external contours is denoted by $\underline\Gamma'_q(\underline\delta)\equiv\{\Gamma'_1,\cdots,\Gamma'_{|\underline\Gamma'_q(\underline\delta)|}\}$, and defined as the subset of $\underline\Gamma_q(\underline\delta)$ such that for every $l_i'\in\Gamma_i'$ and $l_j\in\Gamma_j$, $\bar l'_i\cap\bar l_j\neq\emptyset$ if and only if $\Gamma'_i=\Gamma_j$. (Note that, even though we dropped the contours that are not external, the contours we are left with may still contain several loops, and their inclusion tree is alternating, see figure\-~\ref{fig:contours}.)
\bigskip

\indent Grouping the loops in\-~(\ref{loops}) in this way, and dropping the contours that are not the most external, we rewrite the partition function as
\begin{equation}
  \begin{largearray}
    \frac{Z^{(\Upsilon)}(\Lambda|\mathbf q)}{\mathfrak Z_{\mathbf q}^{(\Upsilon)}(\Lambda)}=
    \sum_{\underline\Gamma\subset\mathfrak C_q^{(\Upsilon)}(\Lambda)}
    \left(\prod_{\Gamma\neq\Gamma'\in\underline\Gamma}\varphi_{\mathrm{ext}}(\Gamma,\Gamma')\right)
    \cdot\\[0.5cm]\hfill\cdot
    \frac{\mathfrak Z_{\mathbf q}^{(\Upsilon)}(\Lambda\setminus\mathcal I(\underline\Gamma))}{\mathfrak Z_{\mathbf q}^{(\Upsilon)}(\Lambda)}
    \prod_{\Gamma\in\underline\Gamma}\prod_{l\in\Gamma}\left(
      \mathfrak Y_{c(l)}^{(\Upsilon)}(\mathbb O_{c(l)}(l))
      \mathfrak Z_{\mathbf c(l)}^{(\Upsilon)}(\iota_{c(l),<\ell_0}^{(\Upsilon)}(l,\Gamma))Z^{(\Upsilon)}(\iota_{c(l),\geqslant\ell_0}^{(\Upsilon)}(l,\Gamma)|\mathbf c(l))
    \right)
  \end{largearray}
  \label{external_contour}
\end{equation}
in which, roughly (see below for a formal definition of these quantities), $\varphi_{\mathrm{ext}}(\Gamma,\Gamma')$ is a hard-core pair interaction that ensures that contours are compatible and external to each other, $\mathfrak Z^{(\Upsilon)}_{\mathbf q}(\Lambda\setminus\mathcal I(\underline\Gamma))$ is the partition function of $q$-dimers outside the contours, $\mathfrak Y_{c(l)}^{(\Upsilon)}(\mathbb O_{c(l)}(l))$ is, as before, the weight of the $c(l)$-dimers in the mantle of $l$, $\mathfrak Z_{\mathbf c(l)}^{(\Upsilon)}(\iota_{c(l),<\ell_0}^{(\Upsilon)}(l,\Gamma))$ is the partition function of $c(l)$-dimers in the segments inside the loop that are of length $<\ell_0$, and $Z^{(\Upsilon)}(\iota_{c(l),\geqslant\ell_0}^{(\Upsilon)}(l,\Gamma)|\mathbf c(l))$ is the partition function of dimers in the remainder of the loops.
\begin{itemize}
  \item $\mathfrak C_q^{(\Upsilon)}(\Lambda)$ is the set of \define{contours}, which is defined as the set of collections of loops $\Gamma$ which are pairwise disjoint and have disjoint mantles, have an alternating inclusion tree whose root label is $q$, and whose support is connected (see figure\-~\ref{fig:contours}).
  \item $\varphi_{\mathrm{ext}}(\Gamma,\Gamma')\in\{0,1\}$ is equal to 1 if and only if $\Gamma$ and $\Gamma'$ are compatible (that is, disconnected from each other) and external to each other.
  \item $\mathcal I(\underline\Gamma)$ is the union of the interiors of the loops:
  \begin{equation}
    \mathcal I(\underline\Gamma):=\bigcup_{\Gamma\in\underline\Gamma}\bigcup_{l\in\Gamma}\bar l.
    \label{mcI}
  \end{equation}
  \item $\iota_{c,<\ell_0}^{(\Upsilon)}(l,\Gamma)$ and $\iota_{c,\geqslant\ell_0}^{(\Upsilon)}(l,\Gamma)$ are the restrictions of $\iota_c^{(\Upsilon)}(l,\Gamma)$ to the parts of the segments that are of length, respectively, $<\ell_0$ and $\geqslant\ell_0$:
  \begin{equation}
    \iota_{c,<\ell_0}^{(\Upsilon)}(l,\Gamma):=\bigcup_{\displaystyle\mathop{\scriptstyle\sigma\in\Sigma_c(\iota_c^{(\Upsilon)}(l,\Gamma))}_{|\sigma|<\ell_0}}\sigma
    ,\quad
    \iota_{c,\geqslant\ell_0}^{(\Upsilon)}(l,\Gamma):=\iota_c^{(\Upsilon)}(l,\Gamma)\setminus\iota_{c,<\ell_0}^{(\Upsilon)}(l,\Gamma).
  \end{equation}
  
  \item $\mathbf c(l)$ and $\mathfrak Y_{c(l)}^{(\Upsilon)}$ were defined at the end of section\-~\ref{subsec:loops}.
\end{itemize}
\bigskip

\subsection{Effective activity and interaction of the external contour model}
\indent We will now re-organize and re-express the right side of\-~(\ref{external_contour}). First of all, by inserting trivial identities, we rewrite
\begin{equation}
  \begin{largearray}
    \frac{Z^{(\Upsilon)}(\Lambda|\mathbf q)}{\mathfrak Z_{\mathbf q}^{(\Upsilon)}(\Lambda)}=
    \sum_{\underline\Gamma\subset\mathfrak C^{(\Upsilon)}_q(\Lambda)}
    \left(\prod_{\Gamma\neq\Gamma'\in\underline\Gamma}\varphi_{\mathrm{ext}}(\Gamma,\Gamma')\right)
    \cdot\\[0.5cm]\hfill\cdot
    A_{\mathbf q}^{(\Upsilon)}(\Lambda,\underline\Gamma)
    \prod_{\Gamma\in\underline\Gamma}\prod_{l\in\Gamma}\left(B_{\mathbf q,\mathbf c(l)}^{(\Upsilon)}(\iota_{c(l),\geqslant\ell_0}^{(\Upsilon)}(l,\Gamma))\frac{\widetilde Z^{(\Upsilon)}(\iota_{c(l),\geqslant\ell_0}^{(\Upsilon)}(l,\Gamma)|\mathbf q)}{\mathfrak Z_{\mathbf q}^{(\Upsilon)}(\iota_{c(l),\geqslant\ell_0}^{(\Upsilon)}(l,\Gamma))}\right)
  \end{largearray}
  \label{ZAB}
\end{equation}
in which $\widetilde Z>0$ will be defined later (see\-~(\ref{tildeZ_external_polymers})) (since $\widetilde Z$ appears in both the numerator and denominator, it is not crucial, at this stage, to specify what it is, as long as it does not vanish, which it does not),
\begin{equation}
  B_{\mathbf q,\mathbf c}^{(\Upsilon)}(\iota):=
  \frac{Z^{(\Upsilon)}(\iota|\mathbf c)}{\mathfrak Z_{\mathbf c}^{(\Upsilon)}(\iota)}
  \frac{\mathfrak Z_{\mathbf q}^{(\Upsilon)}(\iota)}{\widetilde Z^{(\Upsilon)}(\iota|\mathbf q)}
  \label{B}
\end{equation}
and
\begin{equation}
  A_{\mathbf q}^{(\Upsilon)}(\Lambda,\underline\Gamma):=
  \frac1{\mathfrak Z^{(\Upsilon)}_{\mathbf q}(\Lambda)}
  \mathfrak Z_{\mathbf q}^{(\Upsilon)}(\Lambda\setminus\mathcal I(\underline\Gamma))
  \prod_{\Gamma\in\underline\Gamma}\prod_{l\in\Gamma}\left(\mathfrak Y_{c(l)}^{(\Upsilon)}(\mathbb O_{c(l)}(l))\mathfrak Z_{\mathbf c(l)}^{(\Upsilon)}(\iota_{c(l)}^{(\Upsilon)}(l,\Gamma))\right)
  \label{A}
\end{equation}
for which we used the fact that
\begin{equation}
  \mathfrak Z_{\mathbf c(l)}^{(\Upsilon)}(\iota_{c(l),<\ell_0}^{(\Upsilon)}(l,\Gamma))\mathfrak Z_{\mathbf c(l)}^{(\Upsilon)}(\iota_{c(l),\geqslant\ell_0}^{(\Upsilon)}(l,\Gamma))=\mathfrak Z_{\mathbf c(l)}^{(\Upsilon)}(\iota_{c(l)}^{(\Upsilon)}(l,\Gamma))
  .
\end{equation}
Recall that $\mathbf q\equiv(q,\varrho,\ell_0)$ and $\varrho$ is a subset of $\partial\Lambda$, not of $\partial\iota_{c(l),\geqslant\ell_0}^{(\Upsilon)}(l,\Gamma)$. When $\mathbf q$ appears in $\mathfrak Z_{\mathbf q}^{(\Upsilon)}(\iota)$ and the like, it is to be understood in the sense that the boundary of $\iota$ is {\it not} magnetized, that is, $\varrho$ could be replaced with $\emptyset$ (which is consistent with the notation since the elements of $\varrho$ never come in contact with the dimers inside $\iota$).
\bigskip

\indent The factors $B_{\mathbf q,\mathbf c(l)}^{(\Upsilon)}(\iota_{c(l),\geqslant\ell_0}^{(\Upsilon)})$ contribute to the {\it activity} of the contour, whereas $A_{\mathbf q}^{(\Upsilon)}(\Lambda,\underline\Gamma)$ contributes to both the activity and the interaction. In order to separate these contributions, we compute $A_{\mathbf q}^{(\Upsilon)}(\Lambda,\underline\Gamma)$ more explicitly.
\bigskip

\point Let us first compute the partition function $\mathfrak Z_{\mathbf c}^{(\Upsilon)}(\Lambda')$ of the oriented dimer model, for any bounded subset $\Lambda'\subset\mathbb Z^2$ and any boundary condition $\mathbf c\equiv(c,\varrho,\ell_0)$. Different $c$-lines are independent, so $\mathfrak Z_{\mathbf c}^{(\Upsilon)}(\Lambda')$ can be expressed as a product over $c$-lines. The boundary conditions of each line depends on $\varrho$ and $\Upsilon$. To specify them, we first introduce two 1-dimensional boundary conditions: $\omega_0$ and $\omega_1$ which correspond, respectively, to {\it open} and {\it magnetized} boundary conditions:
\begin{equation}
  \omega_0:=\left|\rightdimer\right>+\left|\times\right>
  ,\quad
  \omega_1:=e^J\left|\rightdimer\right>+\left|\times\right>.
  \label{omegabeta}
\end{equation}
We then define the boundary condition of a $c$-line $\sigma$, $\varpi_{\varrho,\Upsilon}(\sigma)\equiv(\varpi^{(0)}_{\varrho,\Upsilon}(\sigma),\varpi^{(1)}_{\varrho,\Upsilon}(\sigma))$, as follows. Let $x_0(\sigma)$ denote the lower-left-most vertex of $\sigma$, and $x_1(\sigma)$ the upper-right-most. For $j\in\{0,1\}$,
\begin{equation}
  \varpi^{(j)}_{\varrho,\Upsilon}(\sigma):=
  \left\{\begin{array}l
    \omega_1\mathrm{\ if\ }\exists e\in\varrho\cup \partial^{(\mathrm{mag})}\Upsilon,\ e\ni x_j(\sigma)\\
    \omega_0\mathrm{\ if\ not}
  \end{array}\right.
  \label{varpi}
\end{equation}
in which $\partial^{(\mathrm{mag})}\Upsilon$ is the {\it magnetized} portion of the boundaries of the sources:
\begin{equation}
  \partial^{(\mathrm{mag})}\Upsilon:=\bigcup_{\upsilon\in \mathbb D_q(\Upsilon)}\mathbb D_q(\partial\upsilon)\cup\bigcup_{\upsilon\in \mathbb D_{-q}(\Upsilon)}\mathbb D_{-q}(\partial\upsilon)
  .
\end{equation}
\bigskip

\indent We now reexpress $\mathfrak Z_{\mathbf c}^{(\Upsilon)}(\Lambda')$:
\begin{equation}
  \mathfrak Z_{\mathbf c}^{(\Upsilon)}(\Lambda')=z^{|\Upsilon\cap\Lambda'|}\prod_{\sigma\in\Sigma_c(\Lambda'{}^{(\Upsilon)})}\Psi^{(\varpi_{\varrho,\Upsilon}(\sigma))}(|\sigma|)
\end{equation}
in which $\Sigma_c$ was defined in\-~(\ref{Sigma}), and $\Psi$ is the partition function of the one-dimensional dimer model, computed in section\-~\ref{sec:1d}. Now, by\-~(\ref{Zv0}),
\begin{equation}
  \mathfrak Z_{\mathbf c}^{(\Upsilon)}(\Lambda')=
  z^{|\Upsilon\cap\Lambda'|}
  \prod_{\sigma\in\Sigma_c(\Lambda')}\left(\nu_+(\varpi^{(0)}_{\varrho,\Upsilon}(\sigma))\nu_+(\varpi^{(1)}_{\varrho,\Upsilon}(\sigma))b_+\lambda_+^{|\sigma|}e^{-W_{\varpi_{\varrho,\Upsilon}(\sigma)}(|\sigma|)}\right)
  \label{idfrakZ}
\end{equation}
where, for $\omega\equiv(\omega^{(0)},\omega^{(1)})\in\Delta^2$,
\begin{equation}
  e^{-W_{\omega}(|\sigma|)}:=
  1
  +\frac{\nu_-(\omega^{(0)})\nu_-(\omega^{(1)})b_-}{\nu_+(\omega^{(0)})\nu_+(\omega^{(1)})b_+}
  \left(\frac{\lambda_-}{\lambda_+}\right)^{|\sigma|}
  +\frac{\nu_0(\omega^{(0)})\nu_0(\omega^{(1)})b_0}{\nu_+(\omega^{(0)})\nu_+(\omega^{(1)})b_+}
  \left(\frac{\lambda_0}{\lambda_+}\right)^{|\sigma|}
  \label{eW}
\end{equation}
and $\lambda$ and $\nu$ were defined in lemma\-~\ref{lemma:1d}.
\bigskip

\point In addition, $\mathfrak Y^{(\Upsilon)}_c(\mathbb O_c)$, which, we recall, is the partition function of close-packed $c$-dimers in $\mathbb O_c$ (see\-~(\ref{frakY})), is equal to
\begin{equation}
  \mathfrak Y^{(\Upsilon)}_c(\mathbb O_c)=(ze^J)^{\frac12|\mathbb O_c|}e^{-\frac12J|\partial_c\mathbb O_c|}
  \exp\left(J|\{e\in \partial_c\mathbb O_c,\ e\cap\partial_c\mathbb D_c(\Upsilon)\neq\emptyset\}|\right)
  \label{idfrakY}
\end{equation}
with $\partial_c\mathbb D_c(\Upsilon):=\bigcup_{\upsilon\in\mathbb D_c(\Upsilon)}\partial_c\upsilon$.
\bigskip

\point We now plug\-~(\ref{idfrakZ}) and\-~(\ref{idfrakY}) into\-~(\ref{A}) to compute $A_{\mathbf q}(\Lambda,\underline\Gamma)$. We split the resulting terms into three contributions as follows.
\bigskip

\subpoint First, we focus on the terms involving $\lambda_+$. By definition, for any bounded $\Lambda'\subset\mathbb Z^2$ and $c\in\{\mathrm v,\mathrm h\}$,
\begin{equation}
  \sum_{\sigma\in\Sigma_c(\Lambda')}|\sigma|=|\Lambda'|
\end{equation}
so (recall\-~(\ref{iota}) and\-~(\ref{mcI}))
\begin{equation}
  \frac{
    \prod_{\sigma\in\Sigma_q(\Lambda^{(\Upsilon)}\setminus\mathcal I(\underline\Gamma))}\lambda_+^{|\sigma|}
  }{
    \prod_{\sigma\in\Sigma_q(\Lambda^{(\Upsilon)})}\lambda_+^{|\sigma|}
  }
  \left(\prod_{\Gamma\in\underline\Gamma}\prod_{l\in\Gamma}\left((ze^J)^{\frac12|\mathbb O_{c(l)}(l)|}\kern-10pt\prod_{\sigma\in\Sigma_{c(l)}(\iota_{c(l)}^{(\Upsilon)}(l,\Gamma))}\kern-10pt\lambda_+^{|\sigma|}\right)\right)
  =\left(\frac{\sqrt{ze^J}}{\lambda_+}\right)^{\sum_{\Gamma\in\underline\Gamma}\sum_{l\in\Gamma}|\mathbb O_{c(l)}(l)|}.
  \label{idlambda}
\end{equation}
\bigskip

\begin{figure}
  \hfil\includegraphics[width=6cm]{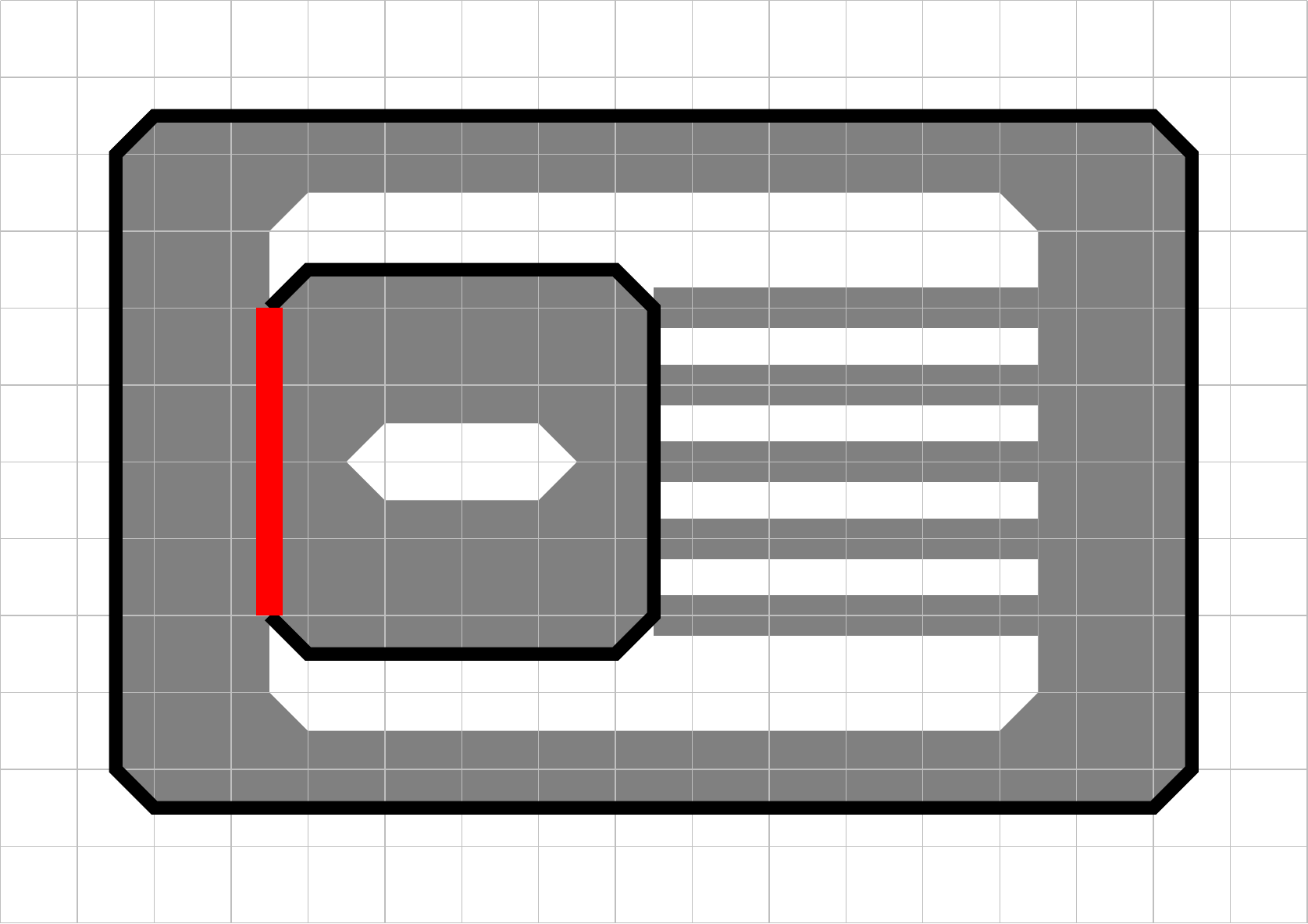}
  \caption{In this example, one of the loops in the contour touches the mantle of the loop that contains it. The set of edges at which this occurs is denoted by $\mathbb X(\Gamma)$. In the figure, $\mathbb X(\Gamma)$ is rendered as a thick red (color online) line.}
  \label{fig:mbbX}
\end{figure}

\subpoint We turn, now, to the terms involving $b_+$ and $\nu_+$. For the moment, we will ignore the sources. The factors in
\begin{equation}
  \begin{largearray}
    \mathfrak N(\underline\Gamma,\Lambda):=\frac{
      \prod_{\sigma\in\Sigma_q(\Lambda\setminus\mathcal I(\underline\Gamma))}\left(\nu_+(\varpi^{(0)}_{\varrho,\emptyset}(\sigma))\nu_+(\varpi^{(1)}_{\varrho,\emptyset}(\sigma))b_+\right)
    }{
      \prod_{\sigma\in\Sigma_q(\Lambda)}\left(\nu_+(\varpi^{(0)}_{\varrho,\emptyset}(\sigma))\nu_+(\varpi^{(1)}_{\varrho,\emptyset}(\sigma))b_+\right)
    }
    \cdot\\[1cm]\hfill\cdot
    \prod_{\Gamma\in\underline\Gamma}\prod_{l\in\Gamma}\left(e^{-\frac12J|\partial_{c(l)}\mathbb O_{c(l)}(l)|}\prod_{\sigma\in\Sigma_{c(l)}(\iota_{c(l)}^{(\emptyset)}(l,\Gamma))}\left(\nu_+(\varpi^{(0)}_{\varrho(l),\emptyset}(\sigma))\nu_+(\varpi^{(1)}_{\varrho(l),\emptyset}(\sigma))b_+\right)\right)
  \end{largearray}
  \label{idnu_expr}
\end{equation}
with $\varrho(l):=\partial_{c(l)}\mathbb I_{c(l)}^{(\emptyset)}(l)$ (the inside of the loops is entirely magnetized), regard the boundaries of the lines $\sigma$ (see figure\-~\ref{fig:segments}). They fall in one of the following categories.
\begin{itemize}
  \item The terms that are attached to $\partial_q\Lambda$ appear in the numerator and the denominator and cancel each other out.
  \item In addition, there are factors attached to the contours. These come in two flavors: those coming from {\it outside} the contour, which have open boundary conditions, and those coming from {\it inside},which have magnetized boundary conditions. Thus, there is a factor $\sqrt{b_+}\nu_+(\omega_0)$ associated with each edge of the outer boundary of the mantle of a loop, {\it provided} that edge comes in contact with a segment. And there is a factor $\sqrt{b_+}\nu_+(\omega_1)$ associated with each edge of the inner boundary of the mantle, again, {\it provided} that edge comes in contact with a segment. The latter caveat is not innocuous: there are cases (see figure\-~\ref{fig:mbbX}) in which a loop in the contour comes in contact with the inner boundary of the mantle of another loop, in which case there are no such terms. To keep track of these events, we introduce the set
  \begin{equation}
    \mathbb X(\Gamma):=\bigcup_{l\neq l'\in\Gamma}l'\cap\partial_{c(l)}\mathbb O_{c(l)}(l)
    .
  \end{equation}
\end{itemize}
All in all,
\begin{equation}
  \mathfrak N(\underline\Gamma,\Lambda)
  =
  \prod_{\Gamma\in\underline\Gamma}
  \prod_{l\in\Gamma}
  \left(
    e^{-\frac12J|l|}
    \left(\sqrt{b_+}\nu_+(\omega_0)e^{\frac12J}\right)^{|\mathbb D_{-c(l)}(l\setminus \mathbb X(\Gamma))|}
    \left(\sqrt{b_+}\nu_+(\omega_1)e^{-\frac12J}\right)^{|\partial_{c(l)}\mathbb I_{c(l)}^{(\emptyset)}(l)\setminus\mathbb X(\Gamma)|}
  \right)
  \label{idnu}
\end{equation}
in which we used the identities
\begin{equation}
  \partial_{c(l)}\mathbb O_{c(l)}(l)=\mathbb D_{c(l)}(l)\cup\partial_{c(l)}\mathbb I_{c(l)}^{(\emptyset)}(l)
  ,\quad
  l=\mathbb D_{c(l)}(l)\cup \mathbb D_{-c(l)}(l)
\end{equation}
(and both unions are disjoint unions) and (since the edges in $\mathbb X$ appear on the inside of the mantle of a loop and the outside of another loop ($\mathbb X$ is the interface between a mantle and a loop that touches the mantle), see figure\-~\ref{fig:mbbX})
\begin{equation}
  \sum_{l\in\Gamma}\left|\mathbb D_{-c(l)}(l\cap\mathbb X(\Gamma))\right|
  =
  \sum_{l\in\Gamma}\left|\partial_{c(l)}\mathbb I_{c(l)}^{(\emptyset)}(l)\cap\mathbb X(\Gamma)\right|
\end{equation}
which means that the $e^{\frac12J}$ and $e^{-\frac12J}$ in\-~(\ref{idnu}) that come from edges in $\mathbb X$ cancel each other out.
\bigskip

\subpoint Let us now take the sources into account. Sources break up the segments, and, in doing so, contribute their own boundary terms. The main contribution comes from the sources $\upsilon$ that are surrounded by an odd number of loops: indeed, in this case, $\upsilon$ contributes to the denominator $\mathfrak Z_{\mathbf q}^{(\Upsilon)}(\Lambda)$ through the boundary terms on its $q$-boundary $\partial_q\upsilon$, whereas, in the numerator, if it is surrounded by an odd number of loops, then it contributes boundary terms on its $-q$-boundary $\partial_{-q}\upsilon$. In addition, in cases where sources come in contact with contours, the boundary of the contour may be erased at the source. Finally, the sources may interact directly with the dimers in the mantle of a contour. We denote the product of all of these factors by $\mathfrak u_q^{(\Upsilon)}(\Gamma)$. To define $\mathfrak u$ formally, we will introduce the notion of {\it contact points}. Given a source $\upsilon\in\Upsilon$ and a contour $\Gamma\in\mathfrak C_q(\Lambda)$ the \define{contact points} of $\upsilon$ and $\Gamma$ is the set of edges $e$ that both intersect $\upsilon$ and neighbor the mantle of a loop of $\Gamma$. Formally, given $c\in\{\mathrm v,\mathrm h\}$, we define the set of interior and exterior $c$-contact points of $\upsilon$ and $\Gamma$ as
\begin{equation}
  \mathfrak K_c^{(\mathrm{int})}(\upsilon,\Gamma)
  :=
  \bigcup_{\displaystyle\mathop{\scriptstyle l\in\Gamma}_{\upsilon\in\bar l}}\partial_c\upsilon\cap\partial_c \mathbb O_{c(l)}(l)
  ,\quad
  \mathfrak K_c^{(\mathrm{ext})}(\upsilon,\Gamma)
  :=
  \bigcup_{\displaystyle\mathop{\scriptstyle l\in\Gamma}_{\upsilon\not\in\bar l}}\partial_c\upsilon\cap\partial_c \mathbb O_{c(l)}(l)
  \label{frakK}
\end{equation}
and the set of $c$-contact points as
\begin{equation}
  \mathfrak K_c(\upsilon,\Gamma)
  :=
  \mathfrak K_c^{(\mathrm{int})}(\upsilon,\Gamma)
  \cup
  \mathfrak K_c^{(\mathrm{ext})}(\upsilon,\Gamma)
  .
\end{equation}
In addition, the \define{background} $\mathfrak g(\upsilon,\Gamma)\in\{\mathrm v,\mathrm h\}$ of a source $\upsilon$ is defined in the following way: for each $l\in\Gamma$, if $\upsilon\in\mathcal E(\iota_{c(l)}^{(\emptyset)}(l,\Gamma))$, then $\mathfrak g(\upsilon,\Gamma)$ is set to $c(l)$. If there is no such loop, then $\mathfrak g(\upsilon,\Gamma)$ is set to $q$. Correspondingly, we split the set of sources into sources in a vertical and horizontal background:
\begin{equation}
  \Upsilon=\Upsilon_{\mathrm v}\cup\Upsilon_{\mathrm h}
  ,\quad
  \Upsilon_c:=\{\upsilon\in\Upsilon,\ \mathfrak g(\upsilon,\Gamma)=c\}
  .
\end{equation}
This allows us to express $\mathfrak u$, following the description given above:
\begin{equation}
  \begin{array}{r@{\ }>\displaystyle l}
    \mathfrak u_q^{(\Upsilon)}(\Gamma)
    =&
    \left(\frac{\nu_+^4(\omega_0)b_+^2}{\nu_+^2(\omega_1)b_+}\right)^{|\mathbb D_q(\Upsilon_{-q})|-|\mathbb D_{-q}(\Upsilon_{-q})|}
    \left(
      \prod_{\upsilon\in\Upsilon}
      b_+^{-|\mathfrak K_{\mathfrak g(\upsilon,\Gamma)}(\upsilon,\Gamma)|}
    \right)
    \\[0.75cm]
    &
    \left(
      \prod_{\upsilon\in \mathbb D_q(\Upsilon_q)}
      \nu_+(\omega_1)^{-|\mathfrak K_q(\upsilon,\Gamma)|}
      \nu_+(\omega_0)^{-|\mathfrak K_q^{(\mathrm{ext})}(\upsilon,\Gamma)|}
      \nu_+(\omega_1)^{-|\mathfrak K_q^{(\mathrm{int})}(\upsilon,\Gamma)|}
      e^{J|\mathfrak K_q^{(\mathrm{int})}(\upsilon,\Gamma)|}
    \right)
    \\[0.75cm]
    &
    \left(
      \prod_{\upsilon\in \mathbb D_q(\Upsilon_{-q})}
      \nu_+(\omega_0)^{-|\mathfrak K_{-q}(\upsilon,\Gamma)|}
      \nu_+(\omega_0)^{-|\mathfrak K_{-q}^{(\mathrm{ext})}(\upsilon,\Gamma)|}
      \nu_+(\omega_1)^{-|\mathfrak K_{-q}^{(\mathrm{int})}(\upsilon,\Gamma)|}
      e^{J|\mathfrak K_q^{(\mathrm{ext})}(\upsilon,\Gamma)|}
    \right)
    \\[0.75cm]
    &
    \left(
      \prod_{\upsilon\in \mathbb D_{-q}(\Upsilon_q)}
      \nu_+(\omega_0)^{-|\mathfrak K_q(\upsilon,\Gamma)|}
      \nu_+(\omega_0)^{-|\mathfrak K_q^{(\mathrm{ext})}(\upsilon,\Gamma)|}
      \nu_+(\omega_1)^{-|\mathfrak K_q^{(\mathrm{int})}(\upsilon,\Gamma)|}
      e^{J|\mathfrak K_{-q}^{(\mathrm{ext})}(\upsilon,\Gamma)|}
    \right)
    \\[0.75cm]
    &
    \left(
      \prod_{\upsilon\in \mathbb D_{-q}(\Upsilon_{-q})}
      \nu_+(\omega_1)^{-|\mathfrak K_{-q}(\upsilon,\Gamma)|}
      \nu_+(\omega_0)^{-|\mathfrak K_{-q}^{(\mathrm{ext})}(\upsilon,\Gamma)|}
      \nu_+(\omega_1)^{-|\mathfrak K_{-q}^{(\mathrm{int})}(\upsilon,\Gamma)|}
      e^{J|\mathfrak K_{-q}^{(\mathrm{int})}(\upsilon,\Gamma)|}
    \right)
    .
  \end{array}
  \label{fraku}
\end{equation}
\bigskip

\indent Thus, the actual contribution of the terms involving $b_+$ and $\nu_+$ is
\begin{equation}
  \mathfrak N^{(\Upsilon)}(\underline\Gamma,\Lambda)=
  \mathfrak N(\underline\Gamma,\Lambda)
  \prod_{\Gamma\in\underline\Gamma}
  \mathfrak u_q^{(\Upsilon)}(\Gamma)
  .
  \label{idnu_sources}
\end{equation}
\bigskip

\subpoint We now put things together: by plugging\-~(\ref{idlambda}), (\ref{idnu}) and\-~(\ref{idnu_sources}) into\-~(\ref{A}), keeping track of the $e^{-W}$ terms, and noting that the $z^{|\Upsilon\cap\Lambda'|}$ factors in\-~(\ref{idfrakZ}) cancel out, we find
\begin{equation}
  A_{\mathbf q}(\Lambda,\underline\Gamma)=
  \mathfrak y_{\mathbf q}^{(\Upsilon)}(\Lambda)
  e^{-\mathfrak W_{\mathbf q}^{(\Upsilon)}(\underline\Gamma)}
  \prod_{\Gamma\in\underline\Gamma}\left(
    \mathfrak x_{\mathbf q}^{(\Upsilon)}(\Gamma)
    \mathfrak u_q^{(\Upsilon)}(\Gamma)
    \prod_{l\in\Gamma}\frac{e^{-\frac12J|l|}}{\mathfrak y_{\mathbf c(l)}^{(\Upsilon)}(\iota_{c(l),\geqslant\ell_0}^{(\Upsilon)}(l,\Gamma))}
  \right)
  \label{Acomp}
\end{equation}
where
\begin{equation}
  \mathfrak y_{\mathbf q}^{(\Upsilon)}(\Lambda):=
  \prod_{\sigma\in\Sigma_{\mathbf q}(\Lambda)}\frac1{e^{-W^{(\varpi_{\varrho,\Upsilon}(\sigma))}(|\sigma|)}}
  \label{fraky}
\end{equation}
(recall\-~(\ref{eW})), and, if $\bar\varrho\equiv\varrho\cup\bigcup_{l\in\Gamma}\partial\mathbb I_{c(l)}^{(\emptyset)}(l)$, (recall that $\mathbf q\equiv(q,\varrho,\ell_0)$ is the boundary condition) then
\begin{equation}
  \begin{largearray}
    \mathfrak x_{\mathbf q}^{(\Upsilon)}(\Gamma):=
    \left(\prod_{\sigma\in\mathcal S_{\Lambda,<\ell_0}^{(\Upsilon)}(\Gamma)}e^{-W^{(\varpi_{\bar\varrho,\Upsilon}(\sigma))}(|\sigma|)}\right)
    \cdot\\[0.75cm]\hfill\cdot
    \prod_{l\in\Gamma}\left(
      \left(\sqrt{b_+}\nu_+(\omega_0)e^{\frac12J}\right)^{|\mathbb D_{-c(l)}(l\setminus\mathbb X(\Gamma))|}
      \left(\sqrt{b_+}\nu_+(\omega_1)e^{-\frac12J}\right)^{|\partial_{c(l)}\mathbb I_{c(l)}^{(\emptyset)}(l)\setminus\mathbb X(\Gamma)|}
      \left(\frac{\sqrt{ze^J}}{\lambda_+}\right)^{|\mathbb O_{c(l)}(l)|}
    \right)
  \end{largearray}
  \label{frakx}
\end{equation}
and $\mathfrak W^{(\Upsilon)}_{\mathbf q}$ is the effective interaction:
\begin{equation}
  e^{-\mathfrak W_{\mathbf q}^{(\Upsilon)}(\underline\Gamma)}:=
  \prod_{\displaystyle\mathop{\scriptstyle\sigma\in\Sigma_q(\Lambda^{(\Upsilon)}\setminus\mathcal I(\underline\Gamma))}_{|\sigma|\geqslant\ell_0}}e^{-W^{(\varpi_{\varrho,\Upsilon}(\sigma))}(|\sigma|)}.
  \label{efrakW}
\end{equation}
\bigskip

\point Finally, we are in a position to write the contour model in terms of an effective activity and interaction: by inserting\-~(\ref{Acomp}) into\-~(\ref{ZAB}), and multiplying and dividing $Z/\mathfrak Z_{\mathbf q}$ by $\mathfrak y_{\mathbf q}^{(\Upsilon)}$, we find
\begin{equation}
  \frac{Z^{(\Upsilon)}(\Lambda|\mathbf q)}{\widetilde{\mathfrak Z}^{(\Upsilon)}_{\mathbf q}(\Lambda)}=
  \sum_{\underline\Gamma\subset\mathfrak C^{(\Upsilon)}_q(\Lambda)}
  \left(\prod_{\Gamma\neq\Gamma'\in\underline\Gamma}\varphi_{\mathrm{ext}}(\Gamma,\Gamma')\right)
  e^{-\mathfrak W^{(\Upsilon)}_{\mathbf q}(\underline\Gamma)}
  \prod_{\Gamma\in\underline\Gamma}\left(
    \eta^{(\Upsilon)}_{\mathbf q}(\Gamma)
    \prod_{l\in\Gamma}
    \frac{\widetilde Z^{(\Upsilon)}(\iota_{c(l),\geqslant\ell_0}^{(\Upsilon)}(l,\Gamma)|\mathbf q)}{\widetilde{\mathfrak Z}^{(\Upsilon)}_{\mathbf q}(\iota_{c(l),\geqslant\ell_0}^{(\Upsilon)}(l,\Gamma))}
  \right)
  \label{Z_contour}
\end{equation}
in which
\begin{equation}
  \widetilde{\mathfrak Z}^{(\Upsilon)}_{\mathbf q}(\Lambda'):=
  \mathfrak Z^{(\Upsilon)}_{\mathbf q}(\Lambda')
  \mathfrak y^{(\Upsilon)}_{\mathbf q}(\Lambda')
  \label{tildefrakZ}
\end{equation}
and
\begin{equation}
  \eta^{(\Upsilon)}_{\mathbf q}(\Gamma):=
  \mathfrak x_{\mathbf q}^{(\Upsilon)}(\Gamma)
  \mathfrak u_q^{(\Upsilon)}(\Gamma)
  \prod_{l\in\Gamma}\left(
    K_{\mathbf q,\mathbf c(l)}^{(\Upsilon)}(\iota_{c(l),\geqslant\ell_0}^{(\Upsilon)}(l,\Gamma))
    e^{-\frac J2|l|}
  \right)
  \label{eta}
\end{equation}
with
\begin{equation}
  K^{(\Upsilon)}_{\mathbf q,\mathbf c}(\Lambda'):=
  \frac{Z^{(\Upsilon)}(\Lambda'|\mathbf c)}{\widetilde{\mathfrak Z}^{(\Upsilon)}_{\mathbf c}(\Lambda')}
  \frac{\widetilde{\mathfrak Z}^{(\Upsilon)}_{\mathbf q}(\Lambda')}{\widetilde Z^{(\Upsilon)}(\Lambda'|\mathbf q)}
  .
  \label{K}
\end{equation}
The factor $\eta^{(\Upsilon)}_{\mathbf q}(\Gamma)$ is the {\it effective activity} of $\Gamma$, $\varphi_{\mathrm{ext}}$ is a hard-core pair interaction between the contours, and $\mathfrak W^{(\Upsilon)}_{\mathbf q}(\underline\Gamma)$ is a many-body, short-range {\it effective interaction}, arising from the 1-dimensional partition functions of the dimer configurations separating them.
\bigskip

\subsection{External polymer model}\label{subsec:external_polymers}
\begin{figure}
  \hfil\includegraphics[width=8cm]{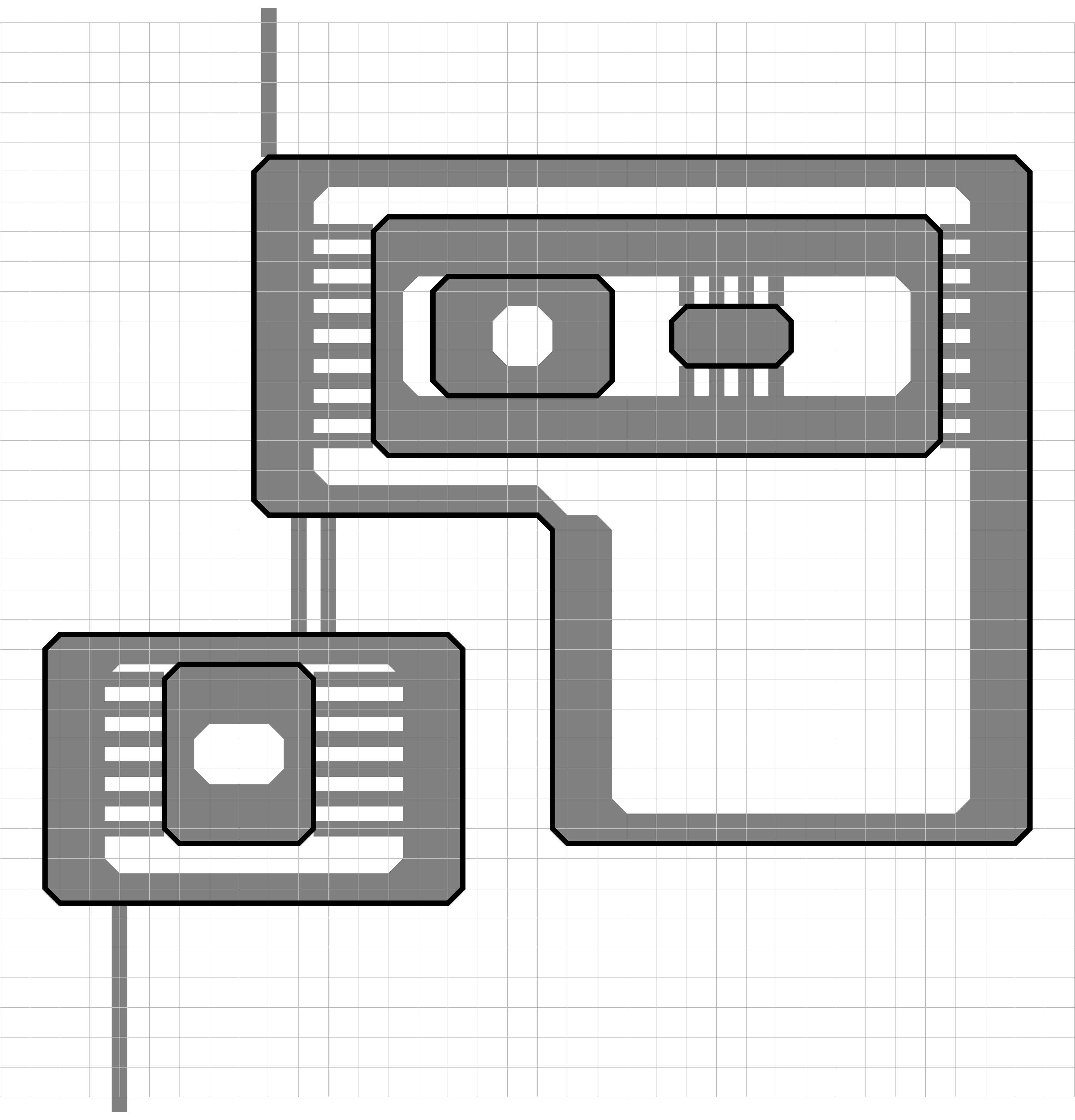}
  \caption{An external polymer. It is obtained by joining the two contours in figure\-~\ref{fig:contours} by segments of length $\geqslant\ell_0$. This polymer is connected to $\partial\Lambda$.}
  \label{fig:polymers}
\end{figure}

\indent We have mapped the dimer model to a contour model with hard-core and short-range (exponentially decaying) interactions. The next step is to dispense with the short-range interactions. To that end, we re-sum the interaction $e^{-\mathfrak W^{(\Upsilon)}_q}$ by inserting trivial identities into\-~(\ref{efrakW}):
\begin{equation}
  e^{-\mathfrak W^{(\Upsilon)}_{\mathbf q}(\underline\Gamma)}=
  \prod_{\displaystyle\mathop{\scriptstyle\sigma\in\Sigma_q(\Lambda^{(\Upsilon)}\setminus\mathcal I(\underline\Gamma))}_{|\sigma|\geqslant\ell_0}}(w_{\varpi_{\varrho,\Upsilon}(\sigma)}(|\sigma|)+1)
\end{equation}
where
\begin{equation}
  w_{\varpi_{\varrho,\Upsilon}(\sigma)}(|\sigma|):=
  e^{-W^{(\varpi_{\varrho,\Upsilon}(\sigma))}(|\sigma|)}-1.
  \label{w}
\end{equation}
and expand $e^{-\mathfrak W}$:
\begin{equation}
  e^{-\mathfrak W^{(\Upsilon)}_{\mathbf q}(\underline\Gamma)}:=
  \sum_{S\subset\{\sigma\in\Sigma_q(\Lambda^{(\Upsilon)}\setminus\mathcal I(\underline\Gamma)),\ |\sigma|\geqslant\ell_0\}}\ \prod_{\sigma\in S}w_{\varpi_{\varrho,\Upsilon}(\sigma)}(|\sigma|).
\end{equation}
Each term in the sum over $S$ gives rise to a new object, called an {\it external polymer}, which consists of contours joined together by segments in $S$ (see figure\-~\ref{fig:polymers}). Formally, an \define{external polymer} is a couple $\xi\equiv(\underline\Gamma(\xi),\underline\sigma(\xi))$ with
\begin{itemize}
  \item $\underline\Gamma(\xi)\equiv\{\Gamma_1,\cdots,\Gamma_{|\underline\Gamma(\xi)|}\}$ is a (possibly empty) set of contours $\Gamma_i\in\mathfrak C_q^{(\Upsilon)}(\Lambda)$, that are pairwise {\it compatible} and {\it external} to each other,
  \item $\underline\sigma(\xi)\equiv\{\sigma_1(\xi),\cdots,\sigma_{|\underline\sigma(\xi)|}\}$ is a (possibly empty) set of $q$-segments
  \begin{equation}
    \textstyle
    \sigma_i\in\left\{
      \sigma\in\Sigma_q\left(\Lambda^{(\Upsilon)}\setminus\left(\bigcup_{\Gamma\in\underline\Gamma(\xi)}\bigcup_{l\in\Gamma}\bar l\right)\right)
      ,\quad
      |\sigma|\geqslant\ell_0
    \right\}
  \end{equation}
\end{itemize}
satisfying the following conditions.
\begin{itemize}
  \item $\underline\Gamma(\xi)$ and $\underline\sigma(\xi)$ cannot both be empty.
  \item We define the \define{support} of $\xi$ as
  \begin{equation}
    \mathrm{supp}_{\Lambda,\ell_0}^{(\Upsilon)}(\xi):=
    \left(
      {\textstyle\bigcup_{\Gamma\in\underline\Gamma(\xi)}\Gamma}
      ,
      \mathcal S_{\Lambda}^{(\Upsilon)}(\xi)
    \right)
    ,\quad
    \mathcal S_{\Lambda}^{(\Upsilon)}(\xi):=
    \underline\sigma(\xi)\cup
    \left({\textstyle\bigcup_{\Gamma\in\underline\Gamma(\xi)}\mathcal S_{\Lambda,<\ell_0}^{(\Upsilon)}(\Gamma)}\right)
    .
    \label{supp_polymer}
  \end{equation}
  The support of $\xi$ is required to be {\it connected}.
\end{itemize}
Denoting the set of external polymers in $\Lambda$ by $\mathfrak X_q^{(\Upsilon)}(\Lambda)$, we rewrite\-~(\ref{Z_contour}) as
\begin{equation}
  \frac{Z^{(\Upsilon)}(\Lambda|\mathbf q)}{\widetilde{\mathfrak Z}^{(\Upsilon)}_{\mathbf q}(\Lambda)}=
  \sum_{\underline\xi\subset\mathfrak X^{(\Upsilon)}_q(\Lambda)}
  \left(\prod_{\xi\neq\xi'\in\underline\xi}\varphi_{\mathrm{ext}}(\xi,\xi')\right)
  \prod_{\xi\in\underline\xi}\left(
    \zeta^{(\Upsilon)}_{\mathbf q}(\xi)
    \prod_{\Gamma\in\underline\Gamma(\xi)}
    \prod_{l\in\Gamma}
    \frac{\widetilde Z^{(\Upsilon)}(\iota_{c(l),\geqslant \ell_0}^{(\Upsilon)}(l,\Gamma)|\mathbf q)}{\widetilde{\mathfrak Z}^{(\Upsilon)}_{\mathbf q,}(\iota_{c(l),\geqslant \ell_0}^{(\Upsilon)}(l,\Gamma))}
  \right)
  \label{Z_external_polymers}
\end{equation}
in which
\begin{itemize}
  \item $\varphi_{\mathrm{ext}}(\xi,\xi')\in\{0,1\}$ is equal to 1 if and only if $\xi$ and $\xi'$ are \define{compatible}, by which we mean that the support of $\xi_1\cup\xi_2\equiv(\underline\Gamma(\xi_1)\cup\underline\Gamma(\xi_2),\underline\sigma(\xi_1)\cup\underline\sigma(\xi_2))$ is {\it disconnected},
  \item $\zeta^{(\Upsilon)}_{\mathbf q}(\xi)$ is the {\it activity} of $\xi$:
  \begin{equation}
    \zeta^{(\Upsilon)}_{\mathbf q}(\xi):=
    \left(
      \prod_{\Gamma\in\underline\Gamma(\xi)}
      \eta^{(\Upsilon)}_{\mathbf q}(\Gamma)
    \right)
    \left(\prod_{\sigma\in\underline\sigma(\xi)}w_{\varpi_{\varrho,\Upsilon}(\sigma)}(|\sigma|)\right).
    \label{zeta_external}
  \end{equation}
\end{itemize}
\bigskip

\indent We will now define $\widetilde Z^{(\Upsilon)}(\iota_{c(l),\geqslant\ell_0}^{(\Upsilon)}(l,\Gamma)|\mathbf q)$, as the partition function of {\it non-trivial} external polymers. Trivial polymers are $q$-segments that go all the way through $\iota_{c(l),\geqslant\ell_0}^{(\Upsilon)}(l,\Gamma)$. By construction, every $c(l)$-segment in $\iota_{c(l),\geqslant\ell_0}^{(\Upsilon)}(l,\Gamma)$ is of length $\geqslant\ell_0$, but this is not necessarily true of the $q$-segments. Since short segments give a poor gain, we wish to avoid them, and, simply, define $\widetilde Z^{(\Upsilon)}(\iota_{c(l),\geqslant\ell_0}^{(\Upsilon)}(l,\Gamma)|\mathbf q)$ without them: for any finite $\Lambda'\subset\mathbb Z^2$,
\begin{equation}
  \frac{\widetilde Z^{(\Upsilon)}(\Lambda'|\mathbf q)}{\widetilde{\mathfrak Z}^{(\Upsilon)}_{\mathbf q}(\Lambda')}:=
  \sum_{\underline\xi\subset\widetilde{\mathfrak X}^{(\Upsilon)}_q(\Lambda')}
  \left(\prod_{\xi\neq\xi'\in\underline\xi}\varphi_{\mathrm{ext}}(\xi,\xi')\right)
  \prod_{\xi\in\underline\xi}\left(
    \zeta^{(\Upsilon)}_{\mathbf q}(\xi)
    \prod_{\Gamma\in\underline\Gamma(\xi)}
    \prod_{l\in\Gamma}
    \frac{\widetilde Z^{(\Upsilon)}(\iota_{c(l),\geqslant \ell_0}^{(\Upsilon)}(l,\Gamma)|\mathbf q)}{\widetilde{\mathfrak Z}^{(\Upsilon)}_{\mathbf q,}(\iota_{c(l),\geqslant \ell_0}^{(\Upsilon)}(l,\Gamma))}
  \right)
  \label{tildeZ_external_polymers}
\end{equation}
obtained from\-~(\ref{Z_external_polymers}) by replacing $\mathfrak X^{(\Upsilon)}_q(\Lambda')$ with $\widetilde{\mathfrak X}^{(\Upsilon)}_q(\Lambda')$, which is the set of \define{non-trivial polymers}:
\begin{equation}
  \widetilde{\mathfrak X}^{(\Upsilon)}_q(\Lambda):=\{\xi\in\mathfrak X_q^{(\Upsilon)}(\Lambda),\ \underline\Gamma(\xi)\neq\emptyset\}.
\end{equation}
\bigskip

{\bf Remark}: The reason that we can drop the trivial polymers comes from the inductive structure of the construction: we have $\widetilde Z(\iota|\mathbf q)$ instead of $Z(\iota|\mathbf c)$ in\-~(\ref{Z_external_polymers}) because we have multiplied and divided by $\widetilde Z(\iota|\mathbf q)$ and incorporated $Z(\iota|\mathbf c)$ into the flipping term $K_{\mathbf q,\mathbf c}$. Therefore, at this stage, $\widetilde Z(\iota|\mathbf q)$ could be, essentially, anything. Later on, we will need the fact that $K_{\mathbf q,\mathbf c}$ is, at most, exponentially large in the size of the boundary. This imposes constraints on $\widetilde Z(\iota|\mathbf q)$, which must not differ too much from $Z(\iota|\mathbf c)$. In this context, `not too much' means that they only differ by boundary terms. Trivial polymers, which go all the way through $\Lambda$, {\it are} boundary terms, which is why they can be dropped. This will be proved in lemma\-~\ref{lemma:boundK}.

\subsection{Polymer model}\label{subsec:polymer}
\indent In order to move from external polymers to polymers, we proceed recursively, by placing external polymers inside external polymers. Before defining the set of polymers, let us first introduce a few more definition: an external polymer $\xi$ is said to be \define{connected to $\partial\Lambda$} if the support of $(\partial\Lambda\cup\underline\Gamma(\xi),\underline\sigma(\xi))$ is connected. In addition, the \define{core} of $\xi$ is defined as
\begin{equation}
  \mathbb I_q^{(\Upsilon)}(\xi):=\bigcup_{\Gamma\in\underline\Gamma(\xi)}\bigcup_{l\in\Gamma}\mathbb I_{c(l)}^{(\Upsilon)}(l).
\end{equation}
Now, the set of \define{polymers} $\mathfrak P_q^{(\Upsilon)}(\Lambda)$ is defined recursively. Roughly (see below for a formal definition), a polymer consists of an external polymer with polymers inside it (here, the word `external' refers to the definition in section\-~\ref{subsec:external_polymers}: the external polymer is, obviously, not external to the polymers inside it). The polymers inside the external polymer are connected to it by segments.
\begin{itemize}
  \item An external polymer is, itself, a polymer: $\mathfrak P_q^{(\Upsilon)}(\Lambda)\supset\mathfrak X_q^{(\Upsilon)}(\Lambda)$.
  \item A polymer $\gamma\in\mathfrak P_q^{(\Upsilon)}(\Lambda)\setminus\mathfrak X_q^{(\Upsilon)}(\Lambda)$ that is not external, is the union of an external polymer $\xi(\gamma)\in\mathfrak X_q^{(\Upsilon)}(\Lambda)$ and of polymers $g_1,\cdots,g_n\in\mathfrak P_q^{(\Upsilon)}(\mathbb I_q^{(\emptyset)}(\xi(\gamma)))$ that are all $q$-connected to $\partial_q\mathbb I_q^{(\emptyset)}(\xi(\gamma))$, and compatible with $\xi(\gamma)$ and with each other. In this case, we define $\underline g(\gamma):=\{g_1,\cdots,g_n\}$.
  \item A polymer $\gamma\in\mathfrak P_q^{(\Upsilon)}(\Lambda)$ is said to be \define{connected to $\partial\Lambda$} if $\xi(\gamma)$ is connected to $\partial\Lambda$.
  \item Two polymers $\gamma,\gamma'\in\mathfrak P_q^{(\Upsilon)}(\Lambda)$ are said to be \define{compatible} if, for any $g\in\{\xi(\gamma)\}\cup\underline g(\gamma)$ and any $g'\in\{\xi(\gamma')\}\cup\underline g(\gamma')$, $g$ and $g'$ are compatible.
\end{itemize}
The \define{activity} of a polymer $\gamma$ is defined as
\begin{equation}
  \zeta_{\mathbf q}^{(\Upsilon)}(\gamma):=
  \zeta_{\mathbf q}^{(\Upsilon)}(\xi(\gamma))
  \prod_{g\in\underline g(\gamma)}\zeta_{\mathbf q}^{(\Upsilon)}(g).
  \label{zeta}
\end{equation}
\bigskip

\begin{figure}
  \hfil\includegraphics[width=8cm]{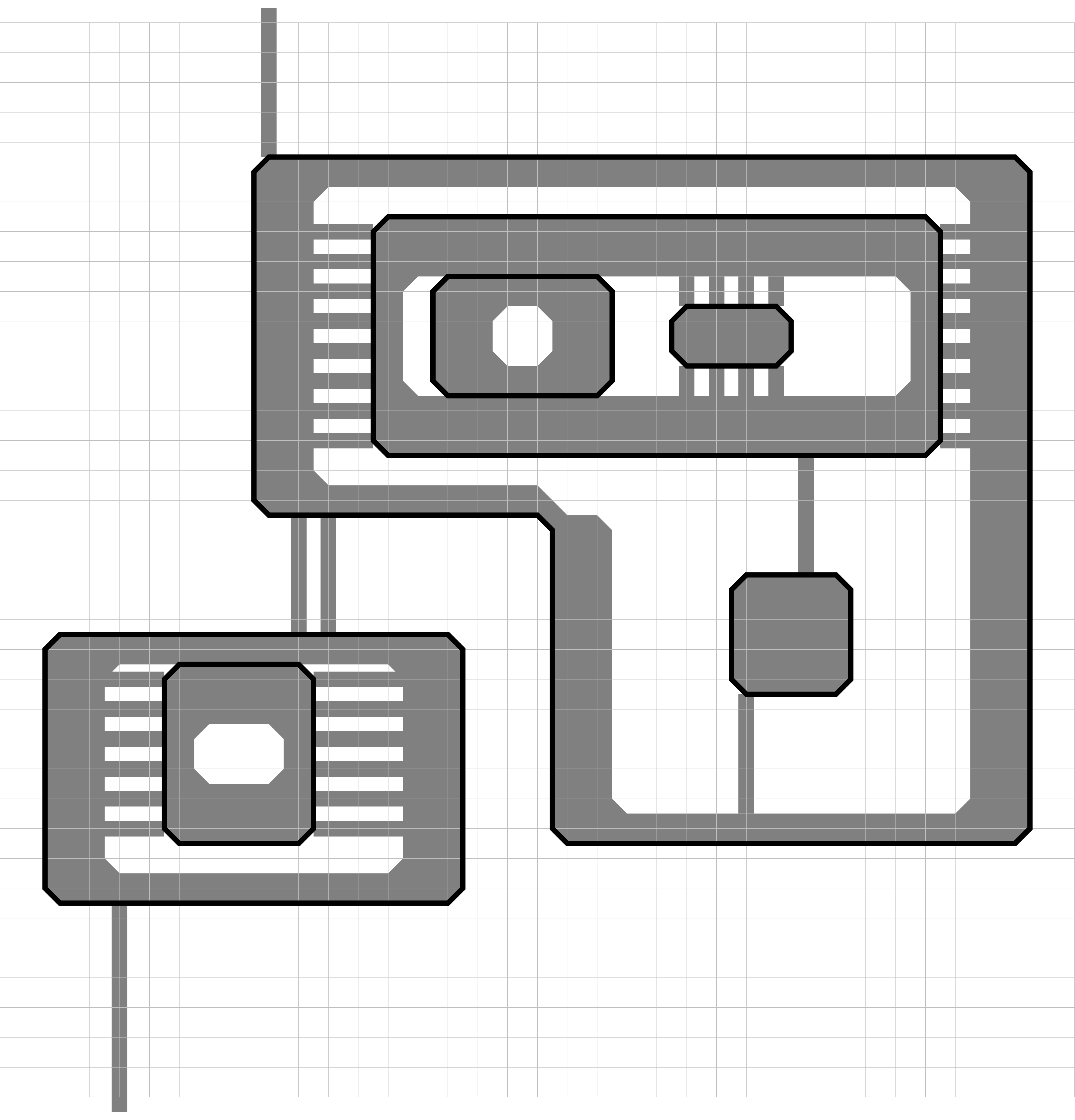}
  \caption{A polymer. It consists of an external polymer, which is the one drawn in figure\-~\ref{fig:polymers}, and a smaller polymer inside it. The smaller polymer consists of a single loop, and is connected to the external polymer by two vertical lines, whose length are $\geqslant\ell_0\equiv 4$. Note that this loop is an $\mathrm h$-bounding loop, even though it is directly inside an $\mathrm h$-bounding loop. Similarly, the smaller polymer is connected to the external one by vertical lines, instead of horizontal ones. This comes from the recursive structure of the construction of polymers.}
  \label{fig:polymer_inductive}
\end{figure}

\indent We are now ready to state the main result of this section, namely the mapping to the polymer model, stated in the following lemma.
\bigskip

\theoname{Lemma}{polymer model}\label{lemma:polymer}
  Consider a bounded subset $\Lambda\subset\mathbb Z^2$ such that $\mathbb Z^2\setminus\Lambda$ is connected, and the boundary $\partial\Lambda$ is a bounding loop. In addition, let $\mathbf q$ be a boundary condition. If every $q$-segment of $\Lambda$ is of length $\geqslant\ell_0$, that is, for every $\sigma\in\Sigma_q(\Lambda)$, $|\sigma|\geqslant\ell_0$, then
  \begin{equation}
    \frac{Z^{(\Upsilon)}(\Lambda|\mathbf q)}{\widetilde{\mathfrak Z}^{(\Upsilon)}_{\mathbf q}(\Lambda)}=
    \sum_{\underline\gamma\subset\mathfrak P_q(\Lambda)}
    \left(\prod_{\gamma\neq\gamma'\in\underline\gamma}\Phi(\gamma,\gamma')\right)
    \prod_{\gamma\in\underline\gamma}\zeta_{\mathbf q}^{(\Upsilon)}(\gamma)
    \label{Z_polymer}
  \end{equation}
  in which $\Phi(\gamma,\gamma')\in\{0,1\}$ is equal to 1 if and only if $\gamma$ and $\gamma'$ are {\it compatible}.
\endtheo
\bigskip

\indent\underline{Proof}:
  We will actually prove that\-~(\ref{tildeZ_external_polymers}) can be rewritten as
  \begin{equation}
    \frac{\widetilde Z^{(\Upsilon)}(\Lambda'|\mathbf q)}{\widetilde{\mathfrak Z}^{(\Upsilon)}_{\mathbf q}(\Lambda')}=
    \sum_{\underline\gamma\subset\widetilde{\mathfrak P}_q^{(\Upsilon)}(\Lambda')}
    \left(\prod_{\gamma\neq\gamma'\in\underline\gamma}\Phi(\gamma,\gamma')\right)
    \prod_{\gamma\in\underline\gamma}\zeta_{\mathbf q}^{(\Upsilon)}(\gamma)
    \label{tildeZ_polymer}
  \end{equation}
  in which  $\widetilde{\mathfrak P}_q^{(\Upsilon)}(\Lambda)$ is the set of {\it non-trivial} polymers, which we define as follows. The set of \define{trivial} polymers is the set of polymers that consist of a single trivial external polymer. The set of \define{non-trivial} polymers is the complement of the set of trivial polymers. By\-~(\ref{Z_external_polymers}), this implies\-~(\ref{Z_polymer}).
  \bigskip

  \indent Equation\-~(\ref{tildeZ_external_polymers}) states that we can deduce the expression of the right side of\-~(\ref{tildeZ_polymer}) from the same expression for smaller sets $\Lambda'\subsetneq\Lambda$. It follows from the principle of mathematical induction, that if we know\-~(\ref{tildeZ_polymer}) for the smallest possible sets, then we can compute the left side of\-~(\ref{tildeZ_polymer}) for sets of any size. If $\Lambda$ is so small that it cannot contain a contour, then\-~(\ref{tildeZ_polymer}) follows immediately from\-~(\ref{tildeZ_external_polymers}) (both sides of the equation are equal to 1). We now assume that\-~(\ref{Z_polymer}) holds for every strict subset of $\Lambda$. By inserting\-~(\ref{tildeZ_polymer}) into\-~(\ref{tildeZ_external_polymers}), we find
  \begin{equation}
    \begin{largearray}
      \frac{\widetilde Z^{(\Upsilon)}(\Lambda|\mathbf q)}{\widetilde{\mathfrak Z}^{(\Upsilon)}_{\mathbf q}(\Lambda)}=
      \sum_{\underline\gamma\subset\widetilde{\mathfrak X}^{(\Upsilon)}_q(\Lambda)}
      \left(\prod_{\gamma\neq\gamma'\in\underline\gamma}\varphi_{\mathrm{ext}}(\gamma,\gamma')\right)
      \cdot\\[0.5cm]\hfill\cdot
      \prod_{\gamma\in\underline\gamma}\left(
	\zeta^{(\Upsilon)}_{\mathbf q}(\gamma)
	\prod_{l\in\mathrm{supp}_{\mathcal L}(\gamma)}
	\left(
	  \sum_{\underline\gamma_l\subset\widetilde{\mathfrak P}^{(\Upsilon)}_q(\iota_{c(l),\geqslant\ell_0}^{(\Upsilon)}(l,\Gamma))}
	  \left(\prod_{\gamma_l\neq\gamma_l'\in\underline\gamma_l}\Phi(\gamma_l,\gamma_l')\right)
	  \prod_{\gamma_l\in\underline\gamma_l}\zeta_{\mathbf q}^{(\Upsilon)}(\gamma_l)
	\right)
      \right).
    \end{largearray}
    \label{induction}
  \end{equation}
  Following the recursive structure of the definition of the set of polymers, we group the unions of external polymers $\underline\gamma$ and polymers $\underline\gamma_l$ into a set of connected polymers that are pairwise compatible. We thus conclude the proof of\-~(\ref{tildeZ_polymer}) for $\Lambda$ from\-~(\ref{tildeZ_polymer}) for strict subsets of $\Lambda$.
\qed

\section{Cluster expansion of the polymer model}\label{sec:cluster}
\indent In this section, we will express the partition function of the polymer model\-~(\ref{Z_polymer}) as an absolutely convergent {\it cluster expansion}. To prove the convergence of the expansion, we will proceed by induction: assuming that the cluster expansion is absolutely convergent for strict subsets of $\Lambda$, we will prove that it converges for $\Lambda$. We split this result into three lemmas (see lemmas\-~\ref{lemma:boundzeta}, \ref{lemma:bound_entropy} and\-~\ref{lemma:boundK}). In the first, we prove a bound for the effective activity of polymers, in the second, we bound the entropy of the polymers, and in the third, prove the convergence of the cluster expansion.
\bigskip

\subsection{Cluster expansion}
\indent The cluster expansion allows us to compute the logarithm of the partition function\-~(\ref{Z_polymer}) in terms of an absolutely convergent series. This is a rather standard step in Pirogov-Sinai theory, and has been written about extensively. In this work, we will use a result of Bovier and Zahradn\'ik\-~\cite[Theorem\-~1]{BZ00}, which, using our notations, is summed up in the following lemma.
\bigskip

\theoname{Lemma}{cluster expansion}\label{lemma:bz}
  If there exist two functions $a,d$ that map polymers $\mathfrak P^{(\Upsilon)}_q(\Lambda)$ to $[0,\infty)$ and a number $\delta\geqslant0$, such that $\forall\gamma\in\mathfrak P^{(\Upsilon)}_q(\Lambda)$,
  \begin{equation}
    |\zeta^{(\Upsilon)}_{\mathbf q}(\gamma)|e^{a(\gamma)+d(\gamma)}\leqslant\delta<1
    ,\quad
    \sum_{\displaystyle\mathop{\scriptstyle\gamma'\in\mathfrak P^{(\Upsilon)}_q(\Lambda)}_{\gamma'\not\sim\gamma}}|\zeta^{(\Upsilon)}_{\mathbf q}(\gamma')|e^{a(\gamma')+d(\gamma')}\leqslant
    \frac{\delta}{|\log(1-\delta)|}
    a(\gamma)
    \label{cvcd}
  \end{equation}
  in which $\gamma'\not\sim\gamma$ means that $\gamma'$ and $\gamma$ are {\it not} compatible, then
  \begin{equation}
    \log\left(
      \frac{Z(\Lambda|\mathbf q)}{\widetilde{\mathfrak Z}_{\mathbf q}(\Lambda)}
    \right)=
    \sum_{\underline\gamma\sqsubset\mathfrak P^{(\Upsilon)}_q(\Lambda)}
    \Phi^T(\underline\gamma)
    \prod_{\gamma\in\underline\gamma}\zeta^{(\Upsilon)}_{\mathbf q}(\gamma)
    \label{ce}
  \end{equation}
  in which $\underline\gamma\sqsubset\mathfrak P_q^{(\Upsilon)}(\Lambda)$ (the symbol $\sqsubset$ is used instead of $\subset$) means that $\underline\gamma$ is a multiset (a multiset is similar to a set except for the fact that an element may appear several times in a multiset, in other words, a multiset is an unordered tuple) with elements in $\mathfrak P_q^{(\Upsilon)}(\Lambda)$, and $\Phi^T$ is the {\it Ursell function}, defined as (see \cite[(4.9)]{Ru99})
  \begin{equation}
    \Phi^T(\emptyset)=0
    ,\quad
    \Phi^T(\{\gamma\})=1
    ,\quad
    \Phi^T(\{\gamma_1,\cdots,\gamma_n\}):=
    \frac1{N_{\underline\gamma}!}
    \sum_{\mathfrak g\in\mathcal G^T(n)}\prod_{\{j,j'\}\in\mathcal E(\mathfrak g)}(\Phi(\gamma_j,\gamma_{j'})-1)
    \label{PhiT}
  \end{equation}
  in which $\mathcal G^T(n)$ is the set of connected graphs on $n$ vertices and $\mathcal E(\mathfrak g)$ is the set of edges of $\mathfrak g$. In addition, for every $\gamma\in\mathfrak P^{(\Upsilon)}_q(\Lambda)$, and, if $n_{\gamma_i}$ is the multiplicity of $\gamma_i$ in $(\gamma_1,\cdots,\gamma_n)$, then $N_{\underline\gamma}!\equiv\prod_{j=1}^n(n_{\gamma_j}!)^{\frac1{n_{\gamma_j}}}$. In addition, for every $\gamma\in\mathfrak C_\nu(\Lambda)$,
  \begin{equation}
    \sum_{\underline\gamma\sqsubset\mathfrak P^{(\Upsilon)}_q(\Lambda)}
    \left|
      \Phi^T(\{\gamma\}\sqcup\underline\gamma)
      \prod_{\gamma'\in\underline\gamma}
      \left(\zeta^{(\Upsilon)}_{\mathbf q}(\gamma')e^{d(\gamma')}\right)
    \right|
    \leqslant
    e^{a(\gamma)}
    \label{ce_remainder}
  \end{equation}
  where $\sqcup$ denotes the union operation in the sense of multisets.
\endtheo

\subsection{Bound on the polymer activity}
\indent We will now prove a bound on the activity $\zeta_{\mathbf q}^{(\Upsilon)}$ of a polymer. We will prove this bound under the assumption that $K_{\mathbf q,\mathbf c}^{(\Upsilon)}(\Lambda')$ is, at most, exponentially large in $|\partial\Lambda'|$, a fact which we will prove in lemma\-~\ref{lemma:boundK}. (That proof is based on a cluster expansion of $Z$ and $\widetilde Z$, in which the only clusters that contribute are those which interact with the boundary.)
\bigskip

\theoname{Lemma}{bound on the polymer activity}\label{lemma:boundzeta}
  If
  \begin{equation}
    e^J\gg z\gg1,
    \label{activityconditions}
  \end{equation}
  and, $\exists\cst C{cst:K}>0$ such that, for every $\Lambda'\subsetneq\Lambda$,
  \begin{equation}
    \left|K^{(\Upsilon)}_{\mathbf q,\mathbf c}(\Lambda')\right|\leqslant e^{\cst C{cst:K}(|\partial_c\Lambda'|+\ell_0e^{-3J}|\partial_q\Lambda'|)}.
    \label{assum_K}
  \end{equation}
  then, $\forall\gamma\in\mathfrak P^{(\Upsilon)}_q(\Lambda)$,
  \begin{equation}
    \left|\zeta^{(\Upsilon)}_{\mathbf q}(\gamma)\right|\leqslant
    e^{-\Xi^{(\Upsilon)}(\gamma)}
    \label{boundzeta}
  \end{equation}
  where
  \begin{equation}
    \Xi^{(\Upsilon)}(\gamma):=
    \frac12\bar J\mathfrak l(\gamma)
    +\bar\kappa\mathfrak s(\gamma)
    -(J+\log\bar\epsilon)\mathfrak m^{(\Upsilon)}(\gamma)
    -\cst C{cst:K}\ell_0^2e^{-3J}\mathfrak b(\gamma)
    -J\mathfrak v_1^{(\Upsilon)}(\gamma)
    -J\mathfrak v_2^{(\Upsilon)}(\gamma)
    -\cst C{cst:source}|\Upsilon|
    \label{Xi}
  \end{equation}
  for some constant $\cst C{cst:source}>0$, with
  \begin{equation}
    \bar J:=J-\cst C{cst:Jexp}
    ,\quad
    \bar\kappa:=\kappa\cst C{cst:sigma}
    ,\quad
    \bar\epsilon:=\epsilon\cst C{cst:m}
    \label{bars}
  \end{equation}
  in which $\cst C{cst:Jexp}>2\cst C{cst:K}$ and $\cst C{cst:sigma},\cst C{cst:m}>0$ are constants,
  \begin{equation}
    \epsilon:=\frac1{\sqrt{ze^J}}\ll1
    ,\quad
    \kappa:=\frac1{e^J\sqrt{ze^J}}\ll1.
  \end{equation}
  and (the following definitions are recursive)
  \begin{equation}
    \mathfrak l(\gamma):=\sum_{\Gamma\in\underline\Gamma(\xi(\gamma))}\sum_{l\in\Gamma}|l|+\sum_{g\in\underline g(\gamma)}\mathfrak l(g)
    ,\quad
    \mathfrak s(\gamma):=\sum_{\sigma\in\underline\sigma(\xi(\gamma))}|\sigma|+\sum_{g\in\underline g(\gamma)}\mathfrak s(g)
  \end{equation}
  \begin{equation}
    \mathfrak m^{(\Upsilon)}(\gamma):=\sum_{\Gamma\in\underline\Gamma(\xi(\gamma))}\sum_{\displaystyle\mathop{\scriptstyle\sigma\in\mathcal S_{\Lambda,<\ell_0}^{(\Upsilon)}(\Gamma)}_{|\sigma|=1}}1+\sum_{g\in\underline g(\gamma)}\mathfrak m^{(\Upsilon)}(g)
    ,\quad
    \mathfrak b(\gamma):=\mathds 1_{\mathfrak l(\gamma)-\mathfrak m^{(\Upsilon)}(\gamma)\geqslant\ell_0}(\mathfrak l(\gamma)-\mathfrak m^{(\Upsilon)}(\gamma))
    .
    \label{frakm}
  \end{equation}
  \begin{equation}
    \mathfrak v_1^{(\Upsilon)}(\gamma)
    :=
    \sum_{\Gamma\in\underline\Gamma(\xi(\gamma))}
    3(|\mathbb D_{-q}(\Upsilon_{-q})|-|\mathbb D_q(\Upsilon_{-q})|)
    +
    \sum_{g\in\underline g(\gamma)}\mathfrak v_1^{(\Upsilon)}(g)
    \label{frakv1}
  \end{equation}
  \begin{equation}
    \mathfrak v_2^{(\Upsilon)}(\gamma)
    :=
    \sum_{\Gamma\in\underline\Gamma(\xi(\gamma))}
    \sum_{\upsilon\in \mathbb D_q(\Upsilon_{-q})\cup \mathbb D_{-q}(\Upsilon_q)}
    |\mathfrak K^{(\Upsilon,\mathrm{ext})}(\upsilon,\Gamma)|
    +
    \sum_{g\in\underline g(\gamma)}\mathfrak v_2^{(\Upsilon)}(g)
    \label{frakv2}
  \end{equation}
  in which $\mathfrak K^{(\Upsilon,\mathrm{ext})}(\upsilon,\Gamma):=\mathfrak K^{(\Upsilon,\mathrm{ext})}_{\mathrm v}(\upsilon,\Gamma)\cup\mathfrak K^{(\Upsilon,\mathrm{ext})}_{\mathrm h}(\upsilon,\Gamma)$ (see\-~(\ref{frakK})).
\endtheo
\restorepagebreakaftereq
\bigskip

\indent\underline{Proof}:
  We recall that $\zeta^{(\Upsilon)}_{\mathbf q}(\gamma)$ was defined in\-~(\ref{zeta}). We proceed by first bounding $\zeta^{(\Upsilon)}_{\mathbf q}(\gamma)$ for $\gamma\equiv\xi\in\mathfrak X^{(\Upsilon)}_q(\Lambda)$, and conclude the proof by induction. To that end we bound the terms appearing in\-~(\ref{zeta_external}) one by one.
  \bigskip

  \point First of all, by\-~(\ref{eigenvalues}),
  \begin{equation}
    \lambda_+=\sqrt{ze^J}(1+\kappa+O(\epsilon^2\kappa))
    ,\quad
    \frac{\lambda_-}{\lambda_+}=-(1-\kappa)+O(\kappa\epsilon)
    ,\quad
    \frac{\lambda_0}{\lambda_+}=\frac1{\sqrt{ze^J}}(1+O(e^{-J}))
    \label{bound_lambda}
  \end{equation}
  by\-~(\ref{b}),
  \begin{equation}
    b_+=\frac1{2ze^{2J}}(1+O(\epsilon))
    ,\quad
    \frac{b_-}{b_+}=1-4\epsilon+8\epsilon^2+5\kappa+O(\epsilon^3)
    ,\quad
    \frac{b_0}{b_+}=2ze^{2J}(1+O(\epsilon))
    \label{bound_b}
  \end{equation}
  and by\-~(\ref{nu}) and\-~(\ref{omegabeta}),
  \begin{equation}
    \nu_+(\omega_0)=\lambda_+
    ,\quad
    \frac{\nu_-(\omega_0)}{\nu_+(\omega_0)}=\frac{\lambda_-}{\lambda_+}
    ,\quad
    \frac{\nu_0(\omega_0)}{\nu_+(\omega_0)}=\frac{\lambda_0}{\lambda_+}
    \label{bound_nu0}
  \end{equation}
  and
  \begin{equation}
    \nu_+(\omega_1)=e^J\sqrt{ze^J}(1+O(\epsilon))
    ,\quad
    \frac{\nu_-(\omega_1)}{\nu_+(\omega_1)}=-(1+2\epsilon+2\epsilon^2-3\kappa)+O(\epsilon^3)
    ,\quad
    \frac{\nu_0(\omega_1)}{\nu_+(\omega_1)}=-\frac{e^{-\frac52J}}{z^{\frac32}}(1+O(\epsilon)).
    \label{bound_nu1}
  \end{equation}
  \bigskip

  \point We bound $\mathfrak x^{(\Upsilon)}_c(\Gamma)$, which was defined in\-~(\ref{frakx}): by\-~(\ref{bound_lambda}) through\-~(\ref{bound_nu1}),
  \begin{equation}
    \sqrt{b_+}\nu_+(\omega_0)e^{\frac12J}=\frac1{\sqrt2}(1+O(\epsilon))
    ,\quad
    \sqrt{b_+}\nu_+(\omega_1)e^{-\frac12J}=\frac1{\sqrt2}(1+O(\epsilon))
    ,\quad
    \frac{\sqrt{ze^J}}{\lambda_+}<1.
    \label{ineqbnu}
  \end{equation}
  In addition, for any $\omega\equiv(\omega^{(0)},\omega^{(1)})\in\{\omega_0,\omega_1\}^2$,
  \begin{equation}
    \frac{\nu_-(\omega^{(0)})\nu_-(\omega^{(1)})b_-}{\nu_+(\omega^{(0)})\nu_+(\omega^{(1)})b_+}
    <1
    ,\quad
    \frac{\nu_0(\omega^{(0)})\nu_0(\omega^{(1)})b_0}{\nu_+(\omega^{(0)})\nu_+(\omega^{(1)})b_+}
    =O(e^J)
    \label{bound_nub}
  \end{equation}
  so, by\-~(\ref{eW}),
  \begin{equation}
    |e^{-W^{(\omega)}(|\sigma|)}|< 1+O(e^J\epsilon^{|\sigma|}).
  \end{equation}
  If $|\sigma|\geqslant 2$, then $O(e^J\epsilon^{|\sigma|})<1$, whereas if $|\sigma|=1$, it is of order $O(e^{J+\log\epsilon})$ and may be large. Therefore,
  \begin{equation}
    |\mathfrak x^{(\Upsilon)}_{\mathbf q}(\Gamma)|<\left(O(e^{J+\log\epsilon})\right)^{\mathfrak m^{(\Upsilon)}(\Gamma)}
    \label{bound_frakx}
  \end{equation}
  in which $\mathfrak m^{(\Upsilon)}$ was defined in\-~(\ref{frakm}) (Each segment of length $\geqslant 2$ and $<\ell_0$ contributes, at most, $1+O(e^J\epsilon^2)$, which we absorb into the $\frac1{\sqrt 2}$ factors in\-~(\ref{ineqbnu}), which we can do since we can bound the number of segments by the lengths of the loops (or rather, by the number of $-c(l)$-edges in the outer loop and the number of $c(l)$-edges on the boundary of its core).)
  \bigskip
  
  \point We now turn to $\mathfrak u_q^{(\Upsilon)}(\Gamma)$, defined in\-~(\ref{fraku}). By\-~(\ref{ineqbnu}),
  \begin{equation}
    |\mathfrak u_q^{(\Upsilon)}(\Gamma)|<O(1)^{|\Upsilon|}e^{J(\mathfrak v_1^{(\Upsilon)}(\Gamma)+\mathfrak v_2^{(\Upsilon)}(\Gamma))}
  \end{equation}
  in which $\mathfrak v_i$ was defined in\-~(\ref{frakv1}) and\-~(\ref{frakv2}).
  \bigskip

\begin{figure}
    \hfil\includegraphics[width=7cm]{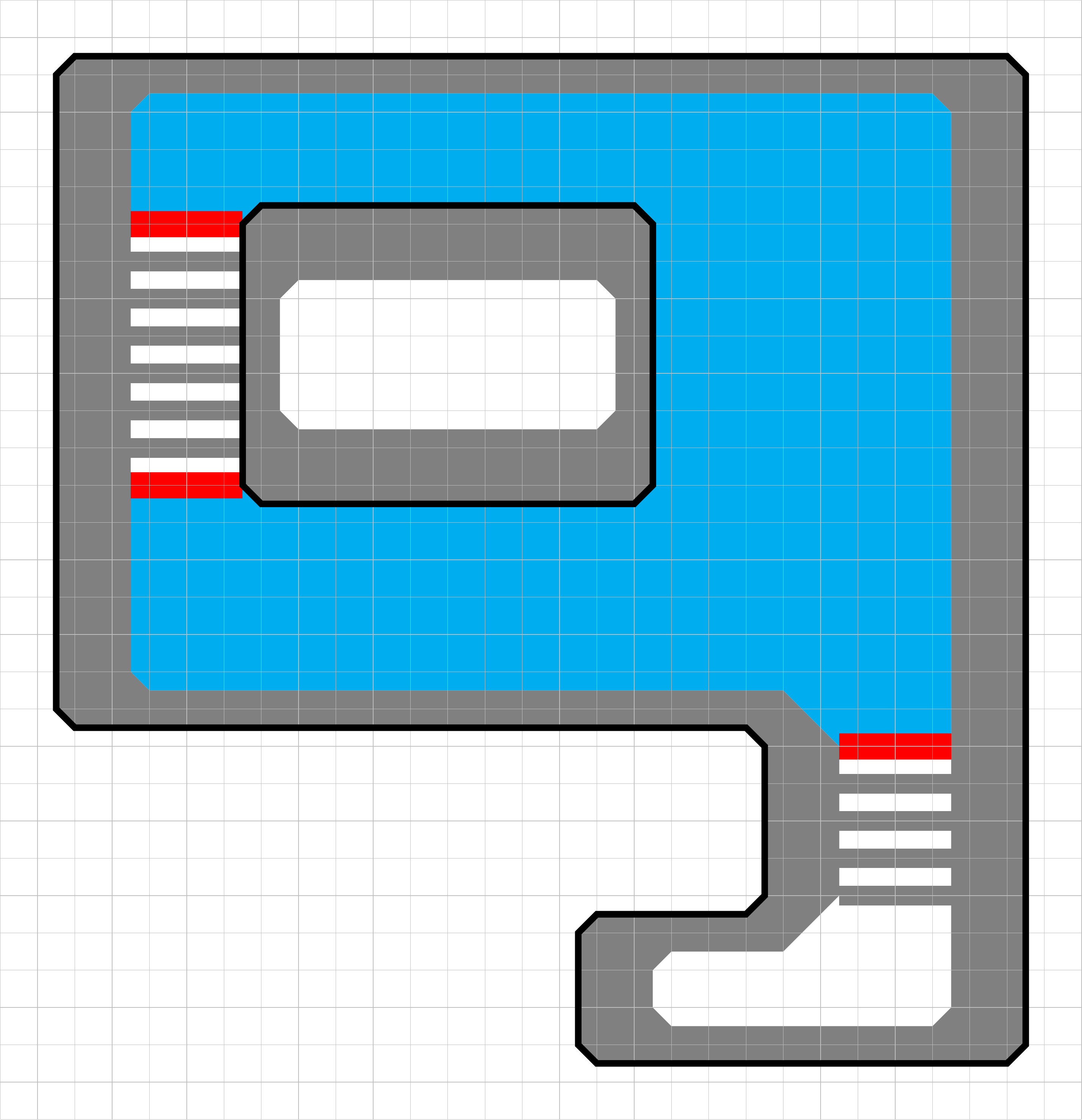}
    \caption{A polymer, in which we have highlighted one of the connected components of the padding (in cyan, color online). Whereas most of the boundary of the padding is part of a loop (or of the boundary of the core of a loop), some parts are segments of length $<\ell_0\equiv4$. These are depicted as thicker red (color online) lines.}
    \label{fig:iota_boundary}
  \end{figure}

  \point We will now bound $K^{(\Upsilon)}_{\mathbf q,\mathbf c(l)}$, which was defined in\-~(\ref{K}). By\-~(\ref{assum_K}),
  \begin{equation}
    \sum_{l\in\Gamma}
    \log K^{(\Upsilon)}_{\mathbf q,\mathbf c(l)}(\iota_{c(l),\geqslant\ell_0}^{(\Upsilon)}(l,\Gamma))
    \leqslant
    \cst C{cst:K}\sum_{l\in\Gamma}|\partial_{c(l)}\iota_{c(l),\geqslant\ell_0}^{(\Upsilon)}(l,\Gamma)|
    +
    \cst C{cst:K}\ell_0e^{-3J}\sum_{l\in\Gamma}|\partial_q\iota_{c(l),\geqslant\ell_0}^{(\Upsilon)}(l,\Gamma)|
    .
    \label{bound_K_prel}
  \end{equation}
  (If $q=c(l)$, then this inequality is slightly suboptimal: only the first term is needed. However, this bound is good enough for our purposes.) In addition, since every edge in $\partial_{c(l)}\iota_{c(l),\geqslant\ell_0}^{(\Upsilon)}(l,\Gamma)$ is necessarily part of a loop (or, rather, of the boundary of the core of a loop, but this distinction does not matter much since there exists an injective map from the boundary of the core to its loop) or a source (see figure\-~\ref{fig:polymers}),
  \begin{equation}
    \sum_{l\in\Gamma}|\partial_{c(l)}\iota_{c(l),\geqslant\ell_0}^{(\Upsilon)}(l,\Gamma)|
    \leqslant
    \mathfrak l(\Gamma)
    +4|\Upsilon|
    .
    \label{bound_count_pc}
  \end{equation}
  On the other hand, the edges in $\partial_q\iota_{c(l),\geqslant\ell_0}^{(\Upsilon)}(l,\Gamma)$ are not necessarily in a loop or a source: if $q\neq c(l)$, then portions of the $q$-boundary of $\iota_{c(l),\geqslant\ell_0}^{(\Upsilon)}(l,\Gamma)$ can consist of segments of length $<\ell_0$ (see figure\-~\ref{fig:iota_boundary}). However, we can bound the number of times this may happen, as follows. Consider a connected component $I$ of $\partial_{c(l)}\iota_{c(l)m\geqslant\ell_0}^{(\Upsilon)}(l,\Gamma)$ and a loop $l'$ in $\partial I$. We go through the edges in $l'$ in, say, clockwise order, which gives us an ordered list of edges. The edges that intersect a segment of length $<\ell_0$ are called `bad'. We then group consecutive bad edges together, and the edge immediately following a bad group is called `good'. By construction, there is are least as many good edges as there are groups of bad ones. In addition, since bad edges touch segments of length $<\ell_0$, each group of consecutive bad edges contains $<\ell_0$ elements. Therefore,
  \begin{equation}
    \#\{\mathrm{bad\ edges}\}<\ell_0\times\ \#\{\mathrm{good\ edges}\}
  \end{equation}
  We then construct an injective map from the set of good edges to the set of edges in $l$ that are not connected to a segment of length 1. The map is defined as follows: given a good edge $e$
  \begin{itemize}
    \item if $e$ is already part of a loop, then the map returns $e$ itself, which may not be connected to a segment of length 1 (otherwise, it would not be part of the boundary of $\iota_{c(l),\geqslant\ell_0}^{(\Upsilon)}(l,\Gamma)$),
    \item if $e$ is on the boundary of the mantle of a loop, then we use the mapping alluded to earlier to map the boundary of the mantle of a loop to the loop itself (we have not defined it formally, a task which we leave to the reader),
    \item if $e$ is on the boundary of a source, then that source must, itself, neighbor a loop or its mantle (if it did not, then the loop $l'$ would simply go around the source and there would be no bad edges), in which case, the map returns the edge at which the source is connected to the loop (if that edge is on the boundary of a mantle, then we use the map mentioned above).
  \end{itemize}
  All in all, this implies that
  \begin{equation}
    \#\{\mathrm{bad\ edges}\}<\ell_0(\mathfrak l(\Gamma)-\mathfrak m^{(\Upsilon)}(\Gamma))
    .
  \end{equation}
  Finally, this may only occur if $\Gamma$ is large enough to contain a non-empty $\iota_{c(l),\geqslant\ell_0}^{(\Upsilon)}$, that is, if
  \begin{equation}
    \mathfrak l(\Gamma)-\mathfrak m^{(\Upsilon)}(\Gamma)\geqslant \ell_0.
  \end{equation}
  Therefore,
  \begin{equation}
    \sum_{l\in\Gamma}|\partial_q\iota_{c(l),\geqslant\ell_0}^{(\Upsilon)}(l,\Gamma)|
    \leqslant
    \mathfrak l(\Gamma)
    +4|\Upsilon|
    +
    \mathds 1_{\mathfrak l(\Gamma)-\mathfrak m^{(\Upsilon)}(\Gamma)\geqslant\ell_0}\ell_0(\mathfrak l(\Gamma)-\mathfrak m^{(\Upsilon)}(\Gamma))
    .
    \label{bound_count_pq}
  \end{equation}
  Thus, inserting\-~(\ref{bound_count_pc}) and\-~(\ref{bound_count_pq}) into\-~(\ref{bound_K_prel}), we find
  \begin{equation}
    \prod_{l\in\Gamma}
    K^{(\Upsilon)}_{\mathbf q,\mathbf c(l)}(\iota_{c(l),\geqslant\ell_0}^{(\Upsilon)}(l,\Gamma))
    \leqslant
    e^{\cst C{cst:K}(2\mathfrak l(\Gamma)+\mathds 1_{\mathfrak l(\gamma)-\mathfrak m^{(\Upsilon)}(\gamma)\geqslant\ell_0}\ell_0^2e^{-3J}(\mathfrak l(\Gamma)-\mathfrak m^{(\Upsilon)}(\Gamma))}
    O(1)^{|\Upsilon|}
    .
  \end{equation}
  \bigskip

  \point We turn, now, to $w$, which was defined in\-~(\ref{w}). By\-~(\ref{bound_nub}), since $|\sigma|\geqslant \ell_0$,
  \begin{equation}
    |w_\varpi(|\sigma|)|=
    O(e^{-\kappa(1+O(\epsilon))|\sigma|}).
    \label{bound_w}
  \end{equation}
  \bigskip

  \point By injecting\-~(\ref{bound_frakx}), (\ref{bound_w}) and\-~(\ref{assum_K}) into\-~(\ref{zeta_external}), we find
  \begin{equation}
    \begin{largearray}
      \left|\zeta^{(\Upsilon)}_{\mathbf q}(\xi(\gamma))\right|\leqslant
      O(e^{J+\log\epsilon})^{\mathfrak m^{(\Upsilon)}(\xi(\gamma))}
      e^{-\frac12J\mathfrak l(\xi(\gamma))+\cst C{cst:K}(2\mathfrak l(\xi(\gamma))+\ell_0^2e^{-3J}\mathfrak b(\xi(\gamma))+J(\mathfrak v_1^{(\Upsilon)}(\xi(\gamma))+\mathfrak v_2^{(\Upsilon)}(\xi(\gamma)))}
      \cdot\\[0.5cm]\hfill\cdot
      O(1)^{|\Upsilon|}
      \left(\prod_{\sigma\in\underline\sigma(\xi)}O(e^{-\kappa(1+O(\epsilon))|\sigma|})\right).
    \end{largearray}
    \label{bound_zeta_inproof}
  \end{equation}
  We conclude the proof by induction.
\qed

\subsection{Bound on the polymer entropy}
\indent We now bound the number of possible polymers, weighted by their activity.
\bigskip

\theoname{Lemma}{bound on the polymer entropy}\label{lemma:bound_entropy}
  For $0<\alpha,\beta<1$ and $\gamma\in\mathfrak P^{(\Upsilon)}_q(\Lambda)$, let
  \begin{equation}
    a(\gamma):=\alpha\Xi^{(\Upsilon)}(\gamma)
    ,\quad
    d(\gamma):=\beta\Xi^{(\Upsilon)}(\gamma)
  \end{equation}
  which are both positive. If
  \begin{equation}
    J\gg z\gg 1
  \end{equation}
  and
  \begin{equation}
    \ell_0=\cst C{cst:ell0}\kappa^{-1}
    \label{ell0}
  \end{equation}
  for some constant $\cst C{cst:ell0}>\max\{1,(\alpha\cst C{cst:sigma})^{-1}\}$ (in which $\cst C{cst:sigma}$ is the constant appearing in\-~(\ref{bars})), and
  \begin{equation}
    \theta:=1-\alpha-\beta
    ,\quad
    \frac12\mathfrak t\leqslant \theta\leqslant 1
    ,\quad
    \mathfrak t=1+O(zJ^{-1})
    \label{bound_theta}
  \end{equation}
  then, for every $\gamma\in\mathfrak P^{(\Upsilon)}_q(\Lambda)$,
  \begin{equation}
    \sum_{\displaystyle\mathop{\scriptstyle\gamma'\in\mathfrak P^{(\Upsilon)}_q(\Lambda)}_{\gamma'\not\sim\gamma}}
    e^{-\theta\Xi^{(\Upsilon)}(\gamma')}
    \leqslant
    a(\gamma)
    \label{bound_entropy}
  \end{equation}
  in which $\gamma'\not\sim\gamma$ means that $\gamma'$ and $\gamma$ are incompatible, which implies that\-~(\ref{cvcd}), and, consequently, lemma\-~\ref{lemma:bz} hold.
\endtheo
\bigskip

\indent\underline{Proof}:
  We will first focus on the sum over non-trivial polymers, and then turn to the trivial ones.
  \bigskip

  \point Let us, for the moment, neglect the sources, and discuss their role later on. We will show that for every edge $e\in\mathbb Z^2$,
  \begin{equation}
    \sum_{\displaystyle\mathop{\scriptstyle\gamma'\in\widetilde{\mathfrak P}^{(\emptyset)}_q(\Lambda)}_{\partial\xi(\gamma')\ni e}}
    e^{-\theta\Xi^{(\emptyset)}(\gamma')}
    \leqslant
    e^{-3\theta\bar J}
    \label{bound_entropy1_nosources}
  \end{equation}
  in which $\partial\xi(\gamma'):=\bigcup_{\Gamma\in\underline\Gamma(\xi(\gamma'))}\Gamma$.
  \bigskip

\begin{figure}
    \hfil\includegraphics[height=8cm]{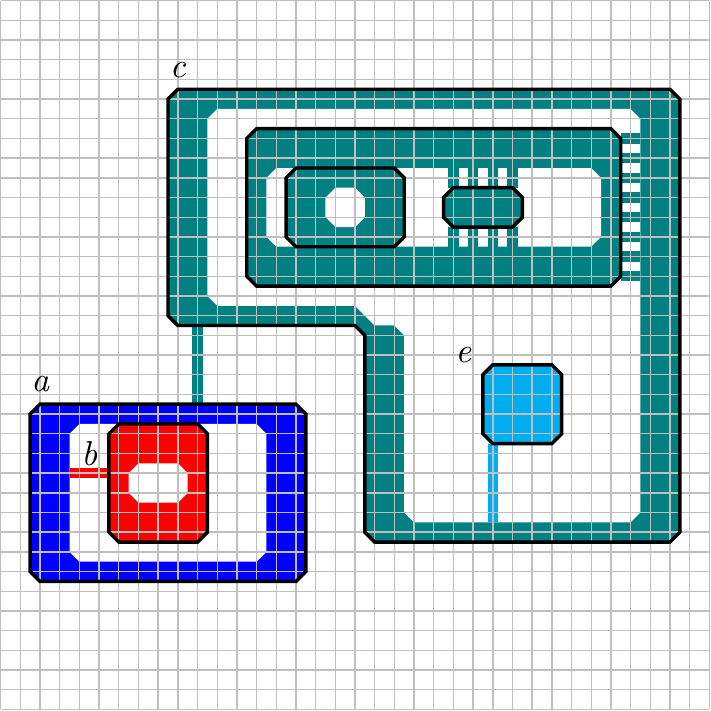}
    \hfil\raise2.5cm\hbox{\includegraphics[height=3cm]{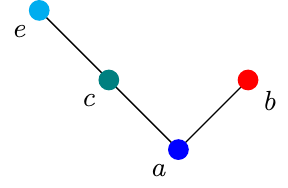}}
    \caption{A backbone of the polymer in figure\-~\ref{fig:polymers}. The blue ({\it a}) loop is the root. The lollipops are depicted in different colors and labeled by different letters. Each consists of a stem and a head, which consists of loops which are at distance $\leqslant 1$ from each other. The tree structure of the backbone is drawn as well.}
    \label{fig:lollipops}
  \end{figure}

  \subpoint A polymer $\gamma'\in\widetilde{\mathfrak P}^{(\emptyset)}_q(\Lambda)$ consists of {\it loops} and {\it segments} which are either of length $<\ell_0$, in which case they connect two loops in the same contour, or their length is $\geqslant\ell_0$. By lemma\-~\ref{lemma:boundzeta}, loops come with a gain factor $e^{-\frac12\bar J|l|}$, and segments which are $\geqslant\ell_0$ come with a gain factor $e^{-\bar\kappa|\sigma|}$. Shorter segments do not have such a gain, and segments of length\-~1 actually come with a loss factor $e^{J+\log\bar\epsilon}$. This loss is less dramatic than might seem at first glance: segments of length\-~1 necessarily connect two loops (since loops are at a distance $\geqslant\ell_0$ from the boundary), and, when taking the $e^{-\frac12J}$ factors coming from the endpoints of the segment, one finds that length\-~1 segments actually contribute $O(\epsilon)$, which is small. Nevertheless, this gain factor is much smaller than for longer segments, which is a fact we will have to deal with. The trick is to consider loops that are at distance\-~1 from each other as a single object, and introduce the notion of a \define{head}, which is a contour $\Gamma\in\mathfrak C_q^{(\emptyset)}(\Lambda)$ whose segments (if any) are all of length\-~1: $\forall\sigma\in\mathcal S_{\Lambda,<\ell_0}^{(\emptyset)}(\Gamma)$, $|\sigma|=1$. Thus, a polymer $\gamma'\in\widetilde{\mathfrak P}^{(\emptyset)}_q(\Lambda)$ consists of {\it heads} and {\it segments} (see figure\-~\ref{fig:lollipops}). For simplicity of exposition, we will consider loops that are not separated by a segment (see figure\-~\ref{fig:lollipops}) as belonging to the same head.
  \bigskip

  \indent We then define the set of {\it backbones} $\underline{\bar\gamma'}$ of $\gamma'$ as the set of polymers obtained from $\gamma'$ by removing segments of length $>1$ in such a way that, while  the support of $\bar\gamma'$ is still connected, it would not be if we removed any more segments of length $>1$. Among the backbones in $\underline{\bar\gamma'}$, we pick one arbitrarily, denote it by $\bar\gamma'$ and call it {\it the} \define{backbone} of $\gamma'$ (see figure\-~\ref{fig:lollipops}). The main idea is to bound the entropy of the backbone, after which we bound the entropy of the full polymer.
  \bigskip

  \subpoint The backbone $\bar\gamma'$ consists of $L\geqslant1$ heads and $n_{\mathrm s}$ segments of length $>1$, and has a natural tree structure (see figure\-~\ref{fig:lollipops}): if we associate a node to each head and a branch to every pair of nodes that corresponds to heads that are connected by a segment, then the resulting graph is a tree (a tree is a graph with no loops), denoted by $\mathbb T(\bar\gamma')$. We call the head containing the edge $e$ the \define{root} of the tree. Every other head has a unique \define{parent}, which is defined as the unique neighbor of the head that is closest to the root (using the natural graph distance on the tree). A head together with the segment that connects it to its parent is called a \define{lollipop}, and the segment is called the \define{stem} of the lollipop. The backbone is completely determined by the tree $\mathbb T(\bar\gamma')$, the shape of the root head and the lollipops, and the points on the heads to which the lollipops are attached.
  \begin{itemize}
  \item The number of rooted trees with $L$ nodes is bounded by
  \begin{equation}
    \#\ \mathrm{trees\ with\ }L\mathrm{\ nodes}\ \leqslant4^{L-1}
    \label{count_trees}
  \end{equation}
  (which can be proved rather easily by mapping the set of trees to 1-dimensional walks with $2(L-1)$ steps, see\-~\cite[lemma\-~A.1]{GM01}).
  \item The number of possible shapes of a lollipop is estimated as follows. Let us focus on the $i$-th lollipop $\Gamma_i$. It consists of a stem of length $\ell_i$, and a head, which is a union of bounding loops. The head is {\it connected} to the stem at an edge $e_i$ (in the sense discussed in section\-~\ref{subsec:external_contours}). By definition, every segment in $\mathcal S_{\Lambda,<\ell_0}^{(\emptyset)}(\Gamma_i)$ is of length 1, and every such segment is connected to exactly 2 edges of the head. Conversely, every edge in the head may be connected to 0 or 1 length-1 segments. We denote the number of edges in the head that are connected to a length-1 segment by $n_{\mathrm z,i}$, and fix the number of remaining edges to $n_{\mathrm l,i}$. Then, we estimate the number of possible heads with $n_{\mathrm z,i}$ and $n_{\mathrm l,i}$ fixed. A head can be seen as a connected subgraph of a finite-degree graph: for instance, consider the graph $\mathfrak G$ whose vertices correspond to the edges of $\mathbb Z^2$ and whose edges correspond to every pair of edges of $\mathbb Z^2$ that are at distance $\leqslant 2$ from each other. A head is a connected subgraph of this graph and has $n_{\mathrm l,i}+n_{\mathrm z,i}$ vertices. Therefore, the number of possible head shapes is bounded by
  \begin{equation}
    \#\ \mathrm{lollipop\ head\ shapes}\ \leqslant\cst c{cst:count_head}^{n_{\mathrm l,i}+n_{\mathrm z,i}}
    \label{count_heads}
  \end{equation}
  for some constant $\cst c{cst:count_head}>0$ (which depends only on the degree of the graph $\mathfrak G$).
  \item Once the tree structure and the shapes of the lollipops are fixed, we are left with positioning the lollipops. Given a lollipop $\Gamma_i$, the tree structure tells us to which other lollipop $\Gamma_j$ its stem is connected. Therefore, it suffices to bound the number ways $\Gamma_i$ can be connected to $\Gamma_j$ by $2^{n_{\mathrm l,j}}$. Thus,
  \begin{equation}
    \#\ \mathrm{lollipop\ positions}\ \leqslant\prod_{i=1}^L2^{n_{\mathrm l,i}}.
    \label{count_positions}
  \end{equation}
  \end{itemize}
  \bigskip

  \indent We now express the weight $e^{-\theta\Xi^{(\emptyset)}(\bar\gamma')}$ (see\-~(\ref{Xi})) of the backbone in terms of lollipops. We fix the lengths of the stems $\ell_2,\cdots,\ell_L\geqslant 2$ (we take the convention that the first head is the root, which does not have a stem), as well as the numbers $n_{\mathrm l,1},\cdots,n_{\mathrm l,L}$ of edges in each head that are not connected to a length-1 segment, and the numbers $n_{\mathrm z,1},\cdots,n_{\mathrm z,L}$ of edges in each head that are connected to a length-1 segment. We have
  \begin{equation}
    \mathfrak l(\gamma')=\sum_{i=1}^L(n_{\mathrm l,i}+n_{\mathrm z,i})
    ,\quad
    \mathfrak s(\bar\gamma')=\sum_{i=2}^L\ell_i
    ,\quad
    \mathfrak m^{(\emptyset)}(\bar\gamma')=\frac12\sum_{i=1}^Ln_{\mathrm z,i}.
    \label{backbone_as_walk_h}
  \end{equation}
  Therefore, by\-~(\ref{Xi})
  \begin{equation}
    \Xi^{(\emptyset)}(\bar\gamma')=
    \sum_{i=1}^L\left(\frac12\bar Jn_{\mathrm l,i}-\frac12\log\bar\epsilon\ n_{\mathrm z,i}\right)
    +\bar\kappa\sum_{i=2}^L\ell_i
    -\chi(n_{\mathrm l})
    \label{bound_Xi}
  \end{equation}
  with
  \begin{equation}
    \chi(n_{\mathrm l}):=
    \cst C{cst:K}\ell_0^2e^{-3J}\mathds 1_{\sum_i n_{\mathrm l,i}\geqslant\ell_0}\sum_{i=1}^L n_{\mathrm{l,i}}
    .
  \end{equation}
  Note that, since $z\ll J$, this shows that $\Xi^{(\emptyset)}(\gamma')\geqslant 0$.
  \bigskip
  
\begin{figure}
    \hfil\raise0.75cm\hbox{\includegraphics[height=1.5cm]{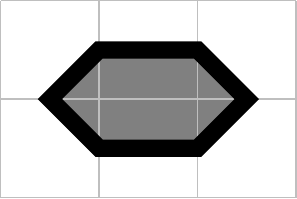}}
    \hfil\includegraphics[height=3cm]{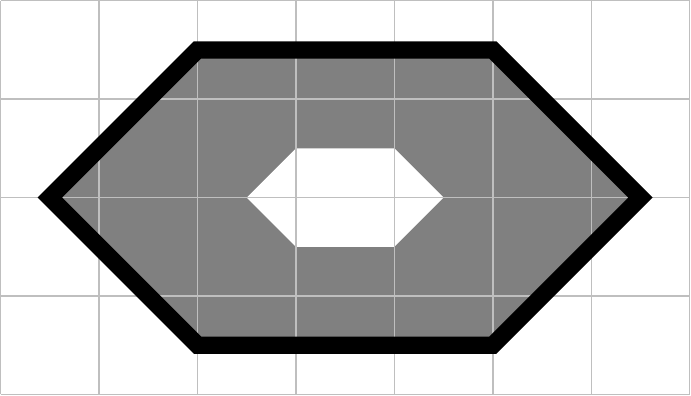}
    \caption{Left: the smallest possible head has 6 edges. Right: the smallest $\mathrm h$-loop that can contain an $\mathrm h$-dimer has 18 edges.}
    \label{fig:smallest}
  \end{figure}

  \indent In addition, by simple geometric considerations, for every $i\in\{1,\cdots,L\}$,
  \begin{equation}
    n_{\mathrm l,i}\geqslant 6
    \label{bound_nz}
  \end{equation}
  Indeed there are at least 6 edges in every head that can not be connected to a length-1 segment (see figure\-~\ref{fig:smallest}). Those edges are the $q$-edges with the largest $q$-component, the $q$-edges with the smallest $q$-component, and the $-q$-edges with the largest $q$-component. There are at least 2 of each, which adds up to 6.
  \bigskip

  \subpoint Given a backbone, we can construct a family of polymers by adding segments to it. The weight of an additional segment of length $\ell$ is $\leqslant 1$. In addition, the number of ways in which on can add segments is bounded by
  \begin{equation}
    \#\ \mathrm{polymers\ compatible\ with\ backbone}\ \leqslant\prod_{i=1}^L2^{n_{\mathrm l,i}}
    \label{count_polymers}
  \end{equation}
  (the estimate corresponds to allowing for segments to be added to any point of a loop, which is an over-counting).
  \bigskip
  
  \subsubpoint Thus, we bound, by\-~(\ref{count_trees}) through\-~(\ref{count_polymers}),
  \begin{equation}
    \begin{largearray}
      \sum_{\displaystyle\mathop{\scriptstyle\gamma'\in\widetilde{\mathfrak P}^{(\emptyset)}_q(\Lambda)}_{\partial\gamma'\ni e}}
      e^{-\theta\Xi^{(\emptyset)}(\gamma')}
      \leqslant
      \sum_{L=1}^\infty
      4^{L-1}
      \sum_{n_{\mathrm l,1},\cdots,n_{\mathrm l,L}\geqslant 6}
      \ \sum_{n_{\mathrm z,1},\cdots,n_{\mathrm z,L}\geqslant 0}
      \ \sum_{\ell_2,\cdots,\ell_L\geqslant 2}
      \cdot\\[-0.25cm]\hfill\cdot
      e^{\theta\chi(n_{\mathrm l})}
      \left(\prod_{i=1}^L\left(\cst c{cst:lolly}\ e^{-\theta\frac12\bar J}\right)^{n_{\mathrm l,i}}\right)
      \left(\prod_{i=1}^L\left(\cst c{cst:lolly}\ \bar\epsilon^{\frac12\theta}\right)^{n_{\mathrm z,i}}\right)
      \prod_{i=2}^{L}F(\ell_i)
    \end{largearray}
    \label{bound_entropy_ns1}
  \end{equation}
  where $\cst c{cst:lolly}>0$ is a constant, and
  \begin{equation}
    F(\ell):=
    \left\{\begin{array}{ll}
      e^{-\theta\bar\kappa\ell}&\mathrm{\ if\ }\ell\geqslant\ell_0\\
      1&\mathrm{\ otherwise}.
    \end{array}\right.
    \label{F}
  \end{equation}
  \bigskip

  \subsubpoint Let us now get rid of the $e^{\theta\chi}$ factor. If $\sum_i n_{\mathrm l,i}\geqslant\ell_0$, then the first factor can be rewritten as
  \begin{equation}
      e^{\theta\chi(n_{\mathrm l})}
      \prod_{i=1}^L\left(e^{-\theta\frac12\bar J}\right)^{n_{\mathrm l,i}}
      =
      \prod_{i=1}^L\left(e^{-\theta\frac12\bar J+\theta\cst C{cst:K}\ell_0^2e^{3J}}\right)^{n_{\mathrm l,i}}
  \end{equation}
  and, since $\ell_0=\cst C{cst:ell0}\kappa^{-1}$ and $\kappa^{-1}=e^{\frac32J}\sqrt z$,
  \begin{equation}
    \cst C{cst:K}\ell_0^2e^{-3J}=
    \cst C{cst:K}\cst C{cst:ell0}^2z
    .
  \end{equation}
  Therefore, we can get rid of $\chi$ by replacing $\theta$ with $\theta(1-2\cst C{cst:K}\cst C{cst:ell0}^2z\bar J^{-1})$:
  \begin{equation}
      e^{\theta\chi(n_{\mathrm l})}
      \prod_{i=1}^L\left(e^{-\theta\frac12\bar J}\right)^{n_{\mathrm l,i}}
      =
      \prod_{i=1}^L\left(e^{-\theta(1+O(zJ^{-1}))\frac12\bar J}\right)^{n_{\mathrm l,i}}.
  \end{equation}
  \bigskip

  \subsubpoint Therefore, if $z$ and $J$ are large enough, since $\ell_0=\cst C{cst:ell0}\kappa^{-1}$, $\sum_\ell F(\ell)=O(\theta\bar\kappa)^{-1}$ and
  \begin{equation}
    \sum_{\displaystyle\mathop{\scriptstyle\gamma'\in\widetilde{\mathfrak P}^{(\emptyset)}_q(\Lambda)}_{\partial\gamma'\ni e}}
    e^{-\theta\Xi^{(\emptyset)}(\gamma')}
    \leqslant
    e^{-3\theta\bar J}
    \sum_{L=1}^\infty\cst c{cst:entropy_ii}^L
    \left(e^{-3\bar\theta\bar J}(\theta\bar\kappa)^{-1}\right)^{L-1}
  \end{equation}
  for some constant $\cst c{cst:entropy_ii}>0$, and with $\bar\theta=\theta(1+O(zJ^{-1}))$. Note that the first factor is $e^{-3\theta\bar J}$ instead of $e^{-3\theta(1+O(zJ^{-1}))\bar J}$. This follows from the fact that,
  \begin{itemize}
    \item if $L=1$, then the $(1+O(zJ^{-1}))$ correction only arises if $n_{\mathrm l,1}\geqslant\ell_0$, in which case the factor would be $e^{-\frac12\theta(1+O(zJ^{-1}))\bar J\ell_0}\gg e^{-3\theta\bar J}$
    \item if $L>1$, then the correction can be absorbed in $e^{-3\theta\bar J(1+O(zJ^{-1}))(L-1)}$.
  \end{itemize}
  In addition, provided $z\ll e^{(6\bar\theta-3)J}$, which is true if $\bar\theta=\theta(1+O(zJ^{-1}))>\frac12$, we have $\bar\kappa\ll e^{3\bar\theta\bar J}$, from which\-~(\ref{bound_entropy1_nosources}) follows.
  \bigskip

  \point We now take the sources into account, and show that
  \begin{equation}
    \sum_{\displaystyle\mathop{\scriptstyle\gamma'\in\widetilde{\mathfrak P}^{(\Upsilon)}_q(\Lambda)}_{\partial\xi(\gamma')\ni e}}
    e^{-\theta\Xi^{(\Upsilon)}(\gamma')}
    \leqslant
    \left\{\begin{array}{ll}
      e^{-3\theta\bar J}&\mathrm{if\ }\mathrm{dist}_1(e,\Upsilon)>3
      \\
      e^{-\theta\bar J}&\mathrm{otherwise}
    \end{array}\right.
    \label{bound_entropy1}
  \end{equation}
  in which $\mathrm{dist}_1$ is the distance induced by the $1$-norm: $|(x,y)|_1=|x|+|y|$. For simplicity of exposition, we will only consider the case in which there are two sources. This is enough for the purpose of computing two-point correlations, and the argument can easily be generalized to an arbitrary number of sources.
  \bigskip

  \subpoint We will first deal with $\mathfrak v_1$ (see\-~(\ref{frakv1})). The key observation is that, in order for $|\mathbb D_{-q}(\Upsilon_{-q})|>0$, the size of one of the heads of $\Gamma$ must be large enough:
  \begin{equation}
    \mathfrak l(\Gamma)-2\mathfrak m^{(\emptyset)}(\Gamma)\geqslant 18|\mathbb D_{-q}(\Upsilon_{-q})|
    .
  \end{equation}
  Indeed, $\Upsilon_{-q}$ is empty unless it is contained inside at least one $-q$-loop, and the smallest $-q$-loop that can contain a $-q$-dimer is of length 18 (see figure\-~\ref{fig:smallest}). Furthermore, since distinct sources are at distance $\geqslant\ell_0$ from each other, if a loop contains two sources, then it is much larger than $2\times 18$. Now, since $\mathfrak l(\Gamma)-2\mathfrak m^{(\emptyset)}(\Gamma)\geqslant 6$ (which is the size of the smallest possible loop, see the discussion above), we have
  \begin{equation}
    \frac12\mathfrak l(\Gamma)-\mathfrak m^{(\emptyset)}(\Gamma)-\mathfrak v_1^{(\Upsilon)}(\Gamma)
    \geqslant
    \frac12\left(\frac12\mathfrak l(\Gamma)-\mathfrak m^{(\emptyset)}(\Gamma)\right)+\frac32
    .
    \label{absorb_frakv1}
  \end{equation}
  We can thus absorb $\mathfrak v_1$ by replacing $e^{-\theta\frac12\bar Jn_{\mathrm l,i}}$ in\-~(\ref{bound_entropy_ns1}) by $e^{-\theta\frac14\bar Jn_{\mathrm l,i}-\theta\frac32\bar J}$.
  \bigskip

  \subpoint After having thus absorbed $\mathfrak v_1$, the remaining contribution of sources comes from loops that are in contact with a source (through $\mathfrak v_2$, see\-~(\ref{frakv2})), or at distance 1 from a source (from the $\Upsilon$-dependence of $\mathfrak m^{(\Upsilon)}$, see\-~(\ref{frakm})). When such an event occurs, $\mathfrak m^{(\Upsilon)}$ and $\mathfrak v_2$ give rise to a large factor, which is counter-balanced by the gain in the entropy coming from the constraint that the loop in question is pinned down by the source.

  \indent The large factor produced by $\mathfrak v_2$ is $e^{J|\mathfrak K^{(\mathrm ext)}|}(\upsilon,\Gamma)$ (see\-~(\ref{frakK})), that is, it is exponentially large in the number of external contact points of each head. Consider a head which is in contact with exactly one source, and does not encircle another source. Denoting its length by $n_{\mathrm l}$ (using the notation introduced above), we note that it can have, at most, $\min(\frac12n_{\mathrm l}-1,6)$ external contact points (since the head has to wind around the source in order to have many contact points). A similar argument holds for the factor produced by $\mathfrak m^{(\Upsilon)}$, which implies that the overall contribution of a head neighboring a source is bounded by $e^{\frac12J\min(n_{\mathrm l}-2,12)}$. Since the head does not encircle another source, there is no need to absorb the contribution of $\mathfrak v_1$ as we did above, and $\frac12\mathfrak l(\Gamma)-\mathfrak m^{(\emptyset)}(\Gamma)$ will contribute $\frac12n_{\mathrm l}$, instead of $\frac14n_{\mathrm l}+\frac32$ as per\-~(\ref{absorb_frakv1}). Thus, the overall contribution of this loop to\-~(\ref{bound_entropy_ns1}) is
  \begin{equation}
    e^{\theta\bar J\min(-1,-\frac12n_{\mathrm l}+6)}
    .
  \end{equation}
  
  \indent Now, consider a head that either is in contact with both sources, or touches one and encircles the other. Such a head must, therefore, be quite large: since sources are separated by at least $\ell_0$, $n_{\mathrm l}\geqslant 2\ell_0$. This time, we must absorb $\mathfrak v_1$ as explained above, and, by\-~(\ref{absorb_frakv1}), find that the head will contribute
  \begin{equation}
    e^{\theta\bar J(-\frac14n_{\mathrm l}+\frac92)}
    \mathds 1_{n_{\mathrm l}\geqslant 2\ell_0}
    .
  \end{equation}

  \indent When a head that is not the root (recall that, when counting the number of possible backbones, we identified the head containing the edge $e$ as the {\it root}) is in contact with a source, then there is no need to sum over the length of its stem. Equivalently, since the sum over the length of a lollipop stem produces a factor proportional to $\kappa^{-1}$, we can sum over the length of the stem, and correct the weight of the loop by a factor proportional to $\kappa$.
  \bigskip

  \indent Finally, we turn to the root head. The number of points at which it comes within a distance 1 of a source $\upsilon$ is bounded by $\frac12n_{\mathrm l}+1-\mathrm{dist}_1(e,\upsilon)$. Therefore, provided it only comes in contact with a single source, and does not encircle another, it contributes
  \begin{equation}
    e^{\theta\bar J\min(1-\mathrm{dist}_1(e,\upsilon),-1,-\frac12n_{\mathrm l}+6)}
    .
  \end{equation}
  Note that, by a very similar argument, one checks that $\Xi^{(\Upsilon)}(\gamma')\geqslant 0$ even in the presence of sources.

  \subpoint All in all, in the presence of sources, (\ref{bound_entropy_ns1}) becomes
  \begin{equation}
    \begin{largearray}
      \sum_{\displaystyle\mathop{\scriptstyle\gamma'\in\widetilde{\mathfrak P}^{(\Upsilon)}_q(\Lambda)}_{\partial\gamma'\ni e}}
      e^{-\theta\Xi^{(\Upsilon)}(\gamma')}
      \leqslant
      \cst c{cst:sources_in_entropy}
      \sum_{L=1}^\infty
      4^{L-1}
      \sum_{n_{\mathrm l,1},\cdots,n_{\mathrm l,L}\geqslant 6}
      \ \sum_{n_{\mathrm z,1},\cdots,n_{\mathrm z,L}\geqslant 0}
      \ \sum_{\ell_2,\cdots,\ell_L\geqslant 2}
      e^{\theta\chi(n_{\mathrm l})}
      \left(\prod_{i=1}^L\left(\cst c{cst:lolly}\ \bar\epsilon^{\frac12\theta}\right)^{n_{\mathrm z,i}}\right)
      \cdot\\[-0.25cm]\hfill\cdot
      \left(\cst c{cst:lolly}^{n_{\mathrm l,1}}\left(e^{-\theta\frac14\bar Jn_{\mathrm l,1}-\theta\frac32\bar J}\left(1+\mathds 1_{n_{\mathrm l,1}\geqslant 2\ell_0}e^{\theta 6\bar J}\right)+e^{-\theta\frac12\bar J\max(2\mathrm{dist}_1(e,\Upsilon)-2,2,n_{\mathrm l,1}-12)}\right)\right)
      \cdot\\[0.5cm]\hfill\cdot
      \left(\prod_{i=2}^L\left(\cst c{cst:lolly}^{n_{\mathrm l,i}}\left(e^{-\theta\frac14\bar Jn_{\mathrm l,i}-\theta\frac32\bar J}\left(1+\kappa\mathds 1_{n_{\mathrm l,i}\geqslant 2\ell_0}e^{\theta 6\bar J}\right)+\kappa e^{-\theta\frac12\bar J\max(2,n_{\mathrm l,i}-12)}\right)\right)\right)
      \prod_{i=2}^{L}F(\ell_i)
      \label{bound_withsources}
    \end{largearray}
  \end{equation}
  for some constant $\cst c{cst:sources_in_entropy}>0$. The rest of the computation is identical to the case without sources, and yields\-~(\ref{bound_entropy1}).
  \bigskip

  \point We can now estimate the sum over non-trivial $\gamma'$ that intersect a given $\gamma$ by summing over the position of $e$ in such a way that $\gamma'$ is incompatible with $\gamma$. Such an incompatibility arises only if a loop of $\gamma'$ is at a distance $<\ell_0$ from a loop of $\gamma$, or if a segment or loop of $\gamma'$ intersects a segment or loop of $\gamma$. This yields a factor $O(\ell_0\mathfrak l(\gamma)+\mathfrak s(\gamma))$. However, if $\gamma'$ is at a 1-distance that is $\leqslant 3$ from a source, then the sum over its position yields a constant rather than $O(\ell_0\mathfrak l(\gamma)+\mathfrak s(\gamma))$. Therefore,
  \begin{equation}
    \sum_{\displaystyle\mathop{\scriptstyle\gamma'\in\widetilde{\mathfrak P}^{(\Upsilon)}_q(\Lambda)}_{\gamma'\not\sim\gamma}}
    e^{-\theta\Xi^{(\Upsilon)}(\gamma')}
    \leqslant
    \kappa^{-1}e^{-3\theta\bar J}O(\mathfrak l(\gamma)+\kappa\mathfrak s(\gamma))
    +O(e^{-\theta\bar J})
    \ll a(\gamma).
    \label{bound_cvcd_loop}
  \end{equation}
  \bigskip

  \point Let us now turn to the contribution of trivial polymers $\gamma'\in\mathfrak P^{(\Upsilon)}_q(\Lambda)\setminus\widetilde{\mathfrak P}^{(\Upsilon)}_q(\Lambda)$. The activity of such polymers is bounded by
  \begin{equation}
    e^{-\theta\Xi^{(\Upsilon)}(\gamma')}\leqslant e^{-\theta\bar\kappa\ell_0}=
    e^{-\theta\cst C{cst:ell0}\cst C{cst:sigma}}
  \end{equation}
  where $\cst C{cst:sigma}$ was introduced in\-~(\ref{bars}), and $\cst C{cst:ell0}$ in\-~(\ref{ell0}).
  \bigskip
  
  \subpoint If $\gamma$ is non-trivial, then $\gamma'\not\sim\gamma$ only if $\gamma'$ intersects a loop of $\gamma$. Indeed, $\gamma'$ is a $q$-segment, and, in order for it to intersect a $q$-segment of $\gamma$, it will have to intersect the loops at its endpoints, and $-q$-segments of $\gamma$ must lie {\it inside} a loop of $\gamma$. Therefore,
  \begin{equation}
    \sum_{\displaystyle\mathop{\scriptstyle\gamma'\in\mathfrak P^{(\Upsilon)}_q(\Lambda)\setminus\widetilde{\mathfrak P}^{(\Upsilon)}_q(\Lambda)}_{\gamma'\not\sim\gamma}}
    e^{-\theta\Xi^{(\Upsilon)}(\gamma')}
    =
    O(\mathfrak l(\gamma))
    \ll a(\gamma)
    \label{bound_cvcd_noloop}
  \end{equation}
  \bigskip
  
  \subpoint If $\gamma$ is trivial, then there is only one position for $\gamma'$ that will intersect $\gamma$, and
  \begin{equation}
    \sum_{\displaystyle\mathop{\scriptstyle\gamma'\in\mathfrak P^{(\Upsilon)}_q(\Lambda)\setminus\widetilde{\mathfrak P}^{(\Upsilon)}_q(\Lambda)}_{\gamma'\not\sim\gamma}}
    e^{-\theta\Xi^{(\Upsilon)}(\gamma')}
    \leqslant
    e^{-\theta\cst C{cst:ell0}\cst C{cst:sigma}}
    \label{bound_cvcd_noloop2}
  \end{equation}
  whereas
  \begin{equation}
    a(\gamma)\geqslant\alpha\bar\kappa\ell_0=\alpha\cst C{cst:ell0}\cst C{cst:sigma}>1.
  \end{equation}
  \bigskip
\qed

\subsection{Polymer-source interaction}
\indent In this section we introduce the notion of a polymer {\it interacting} with a source, which will be useful in the following to compute observables from the cluster expansion.
\bigskip

\theoname{Definition}{polymer-source interaction}\label{def:polymer_source_interaction}
  First of all, we generalize the definition of the polymer activity $\zeta_{\mathbf q}^{(\Upsilon)}(\gamma)$ in\-~(\ref{zeta}), which, so far, has only been defined for polymers $\gamma\in\mathfrak P^{(\Upsilon)}_q(\Lambda)$. We extend this definition to polymers with a different family of sources $\Upsilon'$: if $\gamma\in\mathfrak P_q^{(\Upsilon')}(\Lambda)\setminus\mathfrak P_q^{(\Upsilon)}(\Lambda)$, then we set $\zeta_{\mathbf q}^{(\Upsilon)}(\gamma)\equiv0$.
  \bigskip

  Given a polymer $\gamma\in\mathfrak P_q^{(\Upsilon)}(\Lambda)$ and a source $E\in\mathcal E(\Lambda)$, we say that $\gamma$ \define{interacts} with $E$ if $\zeta_{\mathbf q}^{(\{E\})}(\gamma)\neq\zeta_{\mathbf q}^{(\emptyset)}(\gamma)$. In this case, we write $\gamma\& E$. Note that, in order for $\gamma$ to interact with $E$, it must either come within a distance $\ell_0$ of it or encircle it.
\endtheo
\bigskip

\theoname{Lemma}{entropy of a polymer interacting with a source}\label{lemma:polymer_source_interaction}
  There exists a constant $\cst C{cst:polymer_source}>0$ such that, for any $E\in\mathcal E(\Lambda)$ and a family of sources $\Upsilon'$,
  \begin{equation}
    \sum_{\displaystyle\mathop{\scriptstyle\gamma\in\mathfrak P_q^{(\Upsilon')}(\Lambda)}_{\gamma\& E}}
    |\zeta_{\mathbf q}^{(\Upsilon)}(\gamma)|\ e^{\alpha\Xi^{(\Upsilon)}(\gamma)}
    \leqslant
    \cst C{cst:polymer_source}e^{-(1-\alpha)\bar J}
    +4\max_{\displaystyle\mathop{\scriptstyle\sigma\in\Sigma_q(\Lambda^{(\Upsilon')})}_{\mathfrak d_q(\sigma,E)\leqslant 1}}e^{-(1-\alpha)\bar\kappa|\sigma|}
    \label{polymer_source}
  \end{equation}
  for $\alpha<\frac12\mathfrak t$ (see\-~(\ref{bound_theta})).
\endtheo
\bigskip

\indent\underline{Proof}:
  As noted in definition\-~\ref{def:polymer_source_interaction}, $\gamma$ may only interact with $E$ if it surrounds it, or comes within a distance $\ell_0$ of it. If $\gamma$ is trivial, then its activity is bounded by
  \begin{equation}
    \max_{\displaystyle\mathop{\scriptstyle\sigma\in\Sigma_q(\Lambda^{(\Upsilon')})}_{\mathfrak d_q(\sigma,E)\leqslant 1}}e^{-(1-\alpha)\bar\kappa|\sigma|}
  \end{equation}
  and there are fewer than 4 trivial polymers that interact with $E$. If $\gamma$ is non-trivial, let us fix one edge $e\in\partial\xi(\gamma)$ of one of its external loops. Furthermore, if $\mathrm{dist}_1(e,\Upsilon)>3$, then, since $\gamma\& E$, $\mathfrak l(\gamma)+\mathfrak s(\gamma)\geqslant\mathrm{dist}_1(e,E)-\ell_0$, which implies that,
  \begin{equation}
    \Xi^{(\Upsilon)}(\gamma)\geqslant\bar\kappa(\mathrm{dist}_1(e,E)-\ell_0)
    .
  \end{equation}
  Therefore, for $\beta\in(0,\frac12\mathfrak t-\alpha)$ (see\-~(\ref{bound_theta})),
  \begin{equation}
    \begin{largearray}
      \sum_{\displaystyle\mathop{\scriptstyle\gamma\in\mathfrak P_q^{(\Upsilon')}(\Lambda)}_{\gamma\& E}}
      |\zeta_{\mathbf q}^{(\Upsilon)}(\gamma)|\ e^{a(\gamma)}
      \leqslant
      \sum_{\displaystyle\mathop{\scriptstyle e\in\mathcal E(\Lambda)}_{\mathrm{dist}_1(e,\Upsilon)>3}}
      e^{-\beta\bar\kappa(\mathrm{dist}_1(e,E)-\ell_0)}
      \sum_{\displaystyle\mathop{\scriptstyle\gamma\in\widetilde{\mathfrak P}_q^{(\Upsilon)}(\Lambda)}_{\scriptstyle\partial\xi(\gamma)\ni e}}
      e^{-(1-\alpha-\beta)\Xi^{(\Upsilon)}(\gamma)}
      \\[1.5cm]\hfill
      +
      \sum_{\displaystyle\mathop{\scriptstyle e\in\mathcal E(\Lambda)}_{\mathrm{dist}_1(e,\Upsilon)\leqslant 3}}
      \sum_{\displaystyle\mathop{\scriptstyle\gamma\in\widetilde{\mathfrak P}_q^{(\Upsilon)}(\Lambda)}_{\scriptstyle\partial\xi(\gamma)\ni e}}
      e^{-(1-\alpha)\Xi^{(\Upsilon)}(\gamma)}
      +
      4\max_{\displaystyle\mathop{\scriptstyle\sigma\in\Sigma_q(\Lambda^{(\Upsilon')})}_{\mathfrak d_q(\sigma,E)\leqslant 1}}e^{-(1-\alpha)\bar\kappa|\sigma|}
      .
    \end{largearray}
    \label{polymer_source1}
  \end{equation}
  Thus, by\-~(\ref{bound_entropy1}),
  \begin{equation}
    \begin{largearray}
      \sum_{\displaystyle\mathop{\scriptstyle\gamma\in\mathfrak P_q^{(\Upsilon')}(\Lambda)}_{\gamma\& E}}
      |\zeta_{\mathbf q}^{(\Upsilon)}(\gamma)|\ e^{a(\gamma)}
      \leqslant
      e^{-(1-\alpha-\beta)3\bar J}
      \sum_{\displaystyle\mathop{\scriptstyle e\in\mathcal E(\Lambda)}_{\mathrm{dist}_1(e,\Upsilon)>3}}
      e^{-\beta\bar\kappa(\mathrm{dist}_1(e,E)-\ell_0)}
      \\[0.5cm]\hfill
      +
      e^{-(1-\alpha)\bar J}
      \sum_{\displaystyle\mathop{\scriptstyle e\in\mathcal E(\Lambda)}_{\mathrm{dist}_1(e,\Upsilon)\leqslant 3}}1
      +4\max_{\displaystyle\mathop{\scriptstyle\sigma\in\Sigma_q(\Lambda^{(\Upsilon')})}_{\mathfrak d_q(\sigma,E)\leqslant 1}}e^{-(1-\alpha)\bar\kappa|\sigma|}
      .
    \end{largearray}
    \label{polymer_source2}
  \end{equation}
  Finally, using\-~(\ref{ell0}),
  \begin{equation}
    \sum_{\displaystyle\mathop{\scriptstyle e\in\mathcal E(\Lambda)}_{\mathrm{dist}_1(e,\Upsilon)>3}}
      e^{-\beta\bar\kappa(\mathrm{dist}_1(e,E)-\ell_0)}
      =
      O(\beta\kappa)^{-1}
  \end{equation}
  from which\-~(\ref{polymer_source}) follows, using $\kappa<e^{2J}$ and taking $3\beta\leqslant 1-2\alpha$.
\qed

\subsection{Flipping term}
\indent We will now conclude the proof of the convergence of the cluster expansion, by proving\-~(\ref{assum_K}).
\bigskip

\theoname{Lemma}{bound on the flipping term}\label{lemma:boundK}
  There exists a constant $\cst C{cst:K}>0$ such that, for every boundary condition $\mathbf c\equiv(c,\varsigma,\ell_0)$,
  \nopagebreakaftereq
  \begin{equation}
    \left|K^{(\Upsilon)}_{\mathbf q,\mathbf c}(\Lambda)\right|\leqslant e^{\cst C{cst:K}(|\partial_c\Lambda|+\ell_0e^{-3J}|\partial_q\Lambda|)}.
    \label{assum_K_lemma}
  \end{equation}
\endtheo
\restorepagebreakaftereq
\bigskip

\indent\underline{Proof}:
  The main idea of the proof is to compute $Z^{(\Upsilon)}(\Lambda'|\mathbf \mathbf c)/\widetilde{\mathfrak Z}^{(\Upsilon)}_{\mathbf c}(\Lambda')$ and $\widetilde Z^{(\Upsilon)}(\Lambda'|\mathbf \mathbf q)/\widetilde{\mathfrak Z}^{(\Upsilon)}_{\mathbf q}(\Lambda')$ using the cluster expansion presented in lemma\-~\ref{lemma:bz} whose convergence is ensured by lemmas\-~\ref{lemma:boundzeta} and\-~\ref{lemma:bound_entropy}. We then isolate the {\it bulk} terms, which cancel out, and the {\it boundary} terms, which yield\-~(\ref{assum_K_lemma}). As we will see, it suffices to consider only the first term in\-~(\ref{ce}) and bound the remainder according to\-~(\ref{ce_remainder}).
  \bigskip

  \point{\bf Sources.} The first step is to eliminate the sources. We will focus on $Z^{(\Upsilon)}(\Lambda'|\mathbf \mathbf c)/\widetilde{\mathfrak Z}^{(\Upsilon)}_{\mathbf c}(\Lambda')$, the argument for the other ratio is very similar. Let, for $t\in[0,1]$,
  \begin{equation}
    \bar\zeta_{\mathbf c}^{(\Upsilon,\emptyset)}(\gamma|t):=
    t\zeta_{\mathbf c}^{(\Upsilon)}(\gamma)
    +
    (1-t)\zeta_{\mathbf c}^{(\emptyset)}(\gamma)
  \end{equation}
  and
  \begin{equation}
    \overline{\mathfrak P}_c^{(\Upsilon,\emptyset)}(\Lambda)
    :=
    \mathfrak P_c^{(\Upsilon)}(\Lambda)
    \cup
    \mathfrak P_c^{(\emptyset)}(\Lambda)
  \end{equation}
  in terms of which
  \begin{equation}
    \begin{largearray}
      \log\left(
	\frac{Z^{(\Upsilon)}(\Lambda|\mathbf c)}{\widetilde{\mathfrak Z}^{(\Upsilon)}_{\mathbf c}(\Lambda)}
      \right)
      -
      \log\left(
	\frac{Z^{(\emptyset)}(\Lambda|\mathbf c)}{\widetilde{\mathfrak Z}^{(\emptyset)}_{\mathbf c}(\Lambda)}
      \right)
      =
      \int_0^1dt
      \sum_{m=1}^\infty
      \sum_{\gamma\in\overline{\mathfrak P}_c^{(\Upsilon,\emptyset)}(\Lambda)}
      \sum_{\underline\gamma\sqsubset\overline{\mathfrak P}_c^{(\Upsilon,\emptyset)}(\Lambda)\setminus\{\gamma\}}
      \Phi^T(\{\gamma\}^m\sqcup\underline\gamma)
      \cdot\\[0.75cm]\hfill\cdot
      \left(\zeta^{(\Upsilon)}_{\mathbf c}(\gamma)-\zeta^{(\emptyset)}_{\mathbf c}(\gamma)\right)
      m(\bar\zeta_{\mathbf c}^{(\Upsilon,\emptyset)}(\gamma|t))^{m-1}
      \prod_{\gamma'\in\underline\gamma}\bar\zeta^{(\Upsilon,\emptyset)}_{\mathbf c}(\gamma'|t)
    \end{largearray}
    \label{logZ_cluster}
  \end{equation}
  where $\{\gamma\}^m$ is the multiset with $m$ elements, all of which are $\gamma$. Furthermore, by\-~(\ref{boundzeta}),
  \begin{equation}
    m|\bar\zeta_{\mathbf c}^{(\Upsilon,\emptyset)}(\gamma|t)|^{m-1}\leqslant 1
  \end{equation}
  so, by\-~(\ref{ce_remainder}),
  \begin{equation}
    \left|
      \log\left(
	\frac{Z^{(\Upsilon)}(\Lambda|\mathbf c)}{\widetilde{\mathfrak Z}^{(\Upsilon)}_{\mathbf c}(\Lambda)}
      \right)
      -
      \log\left(
	\frac{Z^{(\emptyset)}(\Lambda|\mathbf c)}{\widetilde{\mathfrak Z}^{(\emptyset)}_{\mathbf c}(\Lambda)}
      \right)
    \right|
    \leqslant
    \sum_{\gamma\in\overline{\mathfrak P}_c^{(\emptyset)}(\Lambda)}
    \left|\zeta^{(\Upsilon)}_{\mathbf c}(\gamma)-\zeta^{(\emptyset)}_{\mathbf c}(\gamma)\right|
    e^{a(\gamma)}
    .
  \end{equation}
  Furthermore, $\zeta_{\mathbf c}^{(\Upsilon)}(\gamma)-\zeta_{\mathbf c}^{(\emptyset)}(\gamma)$ differs from 0 only if $\gamma$ interacts with at least one source in $\Upsilon$ (see definition\-~\ref{def:polymer_source_interaction}). Therefore, by lemma\-~\ref{lemma:polymer_source_interaction},
  \begin{equation}
    \left|
      \log\left(
	\frac{Z^{(\Upsilon)}(\Lambda|\mathbf c)}{\widetilde{\mathfrak Z}^{(\Upsilon)}_{\mathbf c}(\Lambda)}
      \right)
      -
      \log\left(
	\frac{Z^{(\emptyset)}(\Lambda|\mathbf c)}{\widetilde{\mathfrak Z}^{(\emptyset)}_{\mathbf c}(\Lambda)}
      \right)
    \right|
    =O(1)
    .
    \label{bound_ratio_sources}
  \end{equation}
  The same bound holds for $\log(\widetilde Z^{(\Upsilon)}(\Lambda'|\mathbf \mathbf q)/\widetilde{\mathfrak Z}^{(\Upsilon)}_{\mathbf q}(\Lambda'))$. We are thus left with estimating $K_{\mathbf q,\mathbf c}^{(\emptyset)}(\Lambda)$.
  \bigskip

  \point{\bf Bulk terms.} Some of the terms in the cluster expansion\-~(\ref{ce}) are independent of the boundary, and cannot contribute to $K^{(\emptyset)}_{\mathbf q,\mathbf c}(\Lambda)$ since it only involves boundary terms (since the two ratios in\-~(\ref{K}) only differ through their boundary conditions). Let us now make this idea more precise.
  \bigskip
 
  \subpoint Among the polymers, some are connected to the boundary, which we call {\it boundary polymers}, while the others are not, and are called {\it bulk polymers}. Boundary polymers depend on the boundary, since they are connected to it, so we partition
  \begin{equation}
    \mathfrak P^{(\emptyset)}_c(\Lambda)=\mathfrak P_c^{(\partial)}(\Lambda)\cup\mathfrak P_c^{(\circ)}(\Lambda)
    ,\quad
    \widetilde{\mathfrak P}^{(\emptyset)}_q(\Lambda)=\widetilde{\mathfrak P}_q^{(\partial)}(\Lambda)\cup\mathfrak P_q^{(\circ)}(\Lambda)
  \end{equation}
  in which $\mathfrak P_c^{(\partial)}(\Lambda)$ and $\widetilde{\mathfrak P}_q^{(\partial)}(\Lambda)$ are sets of boundary polymers, and rewrite\-~(\ref{ce}) as
  \begin{equation}
    \log\left(
      \frac{Z^{(\emptyset)}(\Lambda|\mathbf c)}{\widetilde{\mathfrak Z}^{(\emptyset)}_{\mathbf c}(\Lambda)}
    \right)=
    \mathfrak B_{\mathbf c}(\mathfrak P^{(\circ)}_c(\Lambda))
    +\overline{\mathfrak B}_{\mathbf c}(\mathfrak P^{(\partial)}_c(\Lambda),\mathfrak P^{(\circ)}_c(\Lambda))
  \end{equation}
  and
  \begin{equation}
    \log\left(
      \frac{\widetilde Z^{(\emptyset)}(\Lambda|\mathbf q)}{\widetilde{\mathfrak Z}^{(\emptyset)}_{\mathbf q}(\Lambda)}
    \right)=
    \mathfrak B_{\mathbf q}(\mathfrak P^{(\circ)}_q(\Lambda))
    +\overline{\mathfrak B}_{\mathbf q}(\widetilde{\mathfrak P}^{(\partial)}_q(\Lambda),\mathfrak P^{(\circ)}_q(\Lambda))
  \end{equation}
  where $\mathfrak B$ is the contribution of clusters involving only bulk polymers, and $\overline{\mathfrak B}$ is the contribution of clusters that contain at least one boundary polymer:
  \begin{equation}
    \mathfrak B_{\mathbf q}(\mathfrak P):=
    \sum_{\underline\gamma\sqsubset\mathfrak P}
    \Phi^T(\underline\gamma)
    \prod_{\gamma\in\underline\gamma}\zeta^{(\emptyset)}_{\mathbf q}(\gamma).
    \label{frakB}
  \end{equation}
  and
  \begin{equation}
    \overline{\mathfrak B}_{\mathbf q}(\mathfrak P,\mathfrak Q):=
    \sum_{m=1}^\infty
    \sum_{\gamma\in\mathfrak P}
    \sum_{\underline\gamma\sqsubset(\mathfrak P\cup\mathfrak Q)\setminus\{\gamma\}}
    \Phi^T(\{\gamma\}^m\sqcup\underline\gamma)
    (\zeta^{(\emptyset)}_{\mathbf q}(\gamma))^m
    \prod_{\gamma'\in\underline\gamma}\zeta^{(\emptyset)}_{\mathbf q}(\gamma').
    \label{frakB_overline}
  \end{equation}
  \bigskip
  
  {\bf Remark}: When defining $\overline{\mathfrak B}$, we separate out one of the polymers, $\gamma$, and ask that it be a boundary polymer. When doing so, we must sum over the multiplicity of $\gamma$ {\it separately} (the sum over $m$ in\-~(\ref{frakB_overline})). If we did not do so, we would be overcounting polymer configurations (this can easily be seen on a simple example where the set of polymers consists of only two objects). We do this because we are writing identities, but, in the following, we will want bounds, for which we do not need to split the sum over the multiplicity from the sum over $\underline\gamma$.
  \bigskip

  \subpoint Bulk polymers still depend on the boundary, because it restricts the polymers to be inside $\Lambda$. To remove this dependence, we introduce the set of {\it infinite-volume} polymers: $\mathfrak P_q^{(\circ)}(\infty)$ is the set of finite polymers, defined as in section\-~\ref{subsec:polymer}, except that, instead of requiring that the polymers be inside $\Lambda$, they are merely required to be finite. In addition, given a vertex $\mathbf v\in\mathbb Z^2$, we define $\mathfrak P_q^{(\circ)}(\infty|\mathbf v)$ as the set of polymers whose upper-leftmost vertex is $\mathbf v$. We then rewrite\-~(\ref{ce}) as the sum over clusters of infinite-volume polymers minus the sum over clusters of infinite-volume polymers which are not contained within $\Lambda$:
  \begin{equation}
    \mathfrak B_{\mathbf q}(\mathfrak P_q^{(\circ)}(\Lambda))=
    \mathfrak F_q(|\Lambda|)
    -\mathfrak G_q(\Lambda)
    ,\quad
    \mathfrak B_{\mathbf c}(\mathfrak P_c^{(\circ)}(\Lambda))=
    \mathfrak F_c(|\Lambda|)
    -\mathfrak G_c(\Lambda)
  \end{equation}
  with
  \begin{equation}
    \mathfrak F_q(|\Lambda|):=
    \sum_{\mathbf v\in \Lambda}
    \sum_{m=1}^\infty
    \sum_{\gamma\in\mathfrak P_q^{(\circ)}(\infty|\mathbf v)}
    (\zeta^{(\emptyset)}_{\mathbf q}(\gamma))^m
    \sum_{\underline\gamma\sqsubset\mathfrak P_q^{(\circ)}(\infty)\setminus\{\gamma\}}
    \Phi^T(\{\gamma\}^m\sqcup\underline\gamma)
    \prod_{\gamma'\in\underline\gamma}\zeta^{(\emptyset)}_{\mathbf q}(\gamma')
  \end{equation}
  which, by translation invariance, only depends on $\Lambda$ through $|\Lambda|$, and
  \begin{equation}
    \mathfrak G_q(\Lambda):=
    \sum_{\mathbf v\in\Lambda}
    \sum_{m=1}^\infty
    \sum_{\gamma\in\mathfrak P_q^{(\circ)}(\infty|\mathbf v)}
    (\zeta^{(\emptyset)}_{\mathbf q}(\gamma))^m
    \sum_{\displaystyle\mathop{\scriptstyle\underline\gamma\sqsubset\mathfrak P_q^{(\circ)}(\infty)\setminus\{\gamma\}}_{\{\gamma\}^m\sqcup\underline\gamma\not\sqsubset\mathfrak P^{(\circ)}_q(\Lambda)}}
    \Phi^T(\{\gamma\}^m\sqcup\underline\gamma)
    \prod_{\gamma'\in\underline\gamma}\zeta^{(\emptyset)}_{\mathbf q}(\gamma').
  \end{equation}
  Thus,
  \begin{equation}
    \begin{largearray}
      \log K^{(\emptyset)}_{\mathbf q,\mathbf c}(\Lambda)
      \\[0.5cm]\hfill
      =
      \left(
	\overline{\mathfrak B}_{\mathbf c}(\mathfrak P^{(\partial)}_c(\Lambda),\mathfrak P_c^{(\circ)}(\Lambda))
	-
	\overline{\mathfrak B}_{\mathbf q}(\widetilde{\mathfrak P}^{(\partial)}_q(\Lambda),\mathfrak P_q^{(\circ)}(\Lambda))
      \right)
      +\left(
	\mathfrak F_c(|\Lambda|)
	-
	\mathfrak F_q(|\Lambda|)
      \right)
      +\left(
	\mathfrak G_c(\Lambda)
	-
	\mathfrak G_q(\Lambda)
      \right).
    \end{largearray}
    \label{K_BFG}
  \end{equation}
  \bigskip

  \subpoint The cancellation of the bulk terms follows from the observation that 
  \begin{equation}
    \mathfrak F_q(|\Lambda|)=\mathfrak F_c(|\Lambda|)
    \label{symmetry_frakF}
  \end{equation}
  which is obvious if $c=q$, and follows from the invariance of the system under $\frac\pi2$ rotations if $c\neq q$.
  \bigskip

  \point{\bf Boundary terms.} We now bound $\overline{\mathfrak B}(\mathfrak P^{(\partial)},\mathfrak P^{(\circ)})$ and $\mathfrak G$.
  \bigskip

  \subpoint We isolate the dominant term in\-~(\ref{frakB_overline}):
  \begin{equation}
    \overline{\mathfrak B}_{\mathbf q}(\mathfrak P,\mathfrak Q)=
    \sum_{\gamma\in\mathfrak P}
    \zeta^{(\emptyset)}_{\mathbf q}(\gamma)
    +
    \sum_{m=1}^\infty
    \sum_{\gamma\in\mathfrak P}
    (\zeta^{(\emptyset)}_{\mathbf q}(\gamma))^m
    \sum_{\underline\gamma\sqsubset(\mathfrak P\cup\mathfrak Q)\setminus\{\gamma\}}
    \Phi^T(\{\gamma\}^m\sqcup\underline\gamma)
    \prod_{\gamma'\in\underline\gamma}\left(e^{-d(\gamma')+d(\gamma')}\zeta^{(\emptyset)}_{\mathbf q}(\gamma')\right)
    \label{barfrakB_dominant}
  \end{equation}
  in which we set $d(\gamma):=\beta\Xi^{(\Upsilon)}(\gamma)$ as in lemma\-~\ref{lemma:bound_entropy}.
  \bigskip
  
  \subsubpoint Let us consider $\overline{\mathfrak B}_{\mathbf q}(\widetilde{\mathfrak P}_q^{(\partial)}(\Lambda),\mathfrak P_q^{(\circ)}(\Lambda))$. By lemma\-~\ref{lemma:bound_entropy} and, more precisely, (\ref{bound_entropy1_nosources}), for any $\theta\in(\frac12\mathfrak t,1]$,
  \begin{equation}
    \sum_{\gamma\in\widetilde{\mathfrak P}_q^{(\partial)}(\Lambda)}e^{-\theta\Xi^{(\Upsilon)}(\gamma)}
    \leqslant
    \sum_{e\in\Lambda}
    \sum_{\displaystyle\mathop{\scriptstyle\gamma\in\widetilde{\mathfrak P}_q^{(\partial)}(\Lambda)}_{\xi(\gamma)\ni e}}
    F(\mathfrak d_q(e,\partial_q\Lambda))
    e^{-\theta\Xi^{(\Upsilon)}(\gamma)}
    \leqslant
    \cst c{cst:kappa}\kappa^{-1}e^{-3\theta\bar J}|\partial_q\Lambda|
    \label{bound_1part_q}
  \end{equation}
  for some constant $\cst c{cst:kappa}>0$, where $F$ was defined in\-~(\ref{F}). In addition using the fact that, by\-~(\ref{Xi}), for $\gamma\in\widetilde{\mathfrak P}_q^{(\partial)}(\Lambda)$,
  \begin{equation}
    d(\gamma)\equiv\beta\Xi^{(\Upsilon)}(\gamma)\geqslant 3\beta\bar J
  \end{equation}
  so that, by\-~(\ref{ce_remainder}),
  \begin{equation}
    \sum_{\underline\gamma\sqsubset\widetilde{\mathfrak P}_q^{(\emptyset)}(\Lambda)}
    \left|
      \Phi^T(\{\gamma\}\sqcup\underline\gamma)
      \prod_{\gamma'\in\underline\gamma}\left(e^{-d(\gamma')+d(\gamma')}\zeta^{(\emptyset)}_{\mathbf q}(\gamma')\right)
    \right|
    \leqslant e^{a(\gamma)}e^{-3\beta\bar J}
  \end{equation}
  in which we recall that $a(\gamma)=\alpha\Xi^{(\Upsilon)}(\gamma)$ (since we are interested in an upper bound, we can reabsorb the sum over $m$ into the sum over $\underline\gamma$), and, by\-~(\ref{bound_1part_q}),
  \begin{equation}
    \sum_{\gamma\in\widetilde{\mathfrak P}_q^{(\partial)}(\Lambda)}e^{a(\gamma)}|\zeta^{(\emptyset)}_{\mathbf q}(\gamma)|
    \leqslant \cst c{cst:kappa}\kappa^{-1}e^{-3(1-\alpha)\bar J}|\partial_q\Lambda|.
  \end{equation}
  Thus
  \begin{equation}
    |\overline{\mathfrak B}_q(\widetilde{\mathfrak P}_q^{(\partial)}(\Lambda),\mathfrak P_q^{(\circ)}(\Lambda))|\leqslant
    \cst c{cst:kappa}\kappa^{-1}e^{-3\bar J}|\partial_q\Lambda|\left(1+e^{-3(\beta-\alpha)\bar J}\right).
    \label{bound_frakB_q}
  \end{equation}
  \bigskip

  \subsubpoint We now turn to $\overline{\mathfrak B}_{\mathbf c}(\mathfrak P_c^{(\partial)}(\Lambda)\mathfrak P_c^{(\circ)}(\Lambda))$. By lemma\-~\ref{lemma:bound_entropy},
  \begin{equation}
    \begin{array}{>\displaystyle r@{\ }>\displaystyle l}
      \sum_{\gamma\in\mathfrak P_c^{(\partial)}(\Lambda)}e^{-\theta\Xi^{(\Upsilon)}(\gamma)}
      =&
      \sum_{\gamma\in\widetilde{\mathfrak P}_c^{(\partial)}(\Lambda)}e^{-\theta\Xi^{(\Upsilon)}(\gamma)}
      +
      \sum_{\gamma\in\mathfrak P_c^{(\partial)}(\Lambda)\setminus\widetilde{\mathfrak P}_c^{(\partial)}(\Lambda)}e^{-\theta\Xi^{(\Upsilon)}(\gamma)}
      \\[0.75cm]
      \leqslant&
      \cst c{cst:kappa}\kappa^{-1}e^{-3\theta\bar J}|\partial_c\Lambda|
      +
      e^{-\theta\bar\kappa\ell_0}|\partial_c\Lambda|
      \label{bound_frakB_c_clusters}
    \end{array}
  \end{equation}
  so, by a similar reasoning as in the previous paragraph,
  \begin{equation}
    |\overline{\mathfrak B}_c(\mathfrak P_c^{(\partial)}(\Lambda),\mathfrak P_c^{(\circ)}(\Lambda))|\leqslant
    |\partial_c\Lambda|
    \left(
      \left(e^{-\bar\kappa\ell_0}+\cst c{cst:kappa}\kappa^{-1}e^{-3\bar J}\right)
      +
      e^{-\beta\bar\kappa\ell_0}\left(e^{-\bar\kappa\ell_0(1-\alpha)}+\cst c{cst:kappa}\kappa^{-1}e^{-3\bar J(1-\alpha)}\right)
    \right).
    \label{bound_frakB_c}
  \end{equation}
  \bigskip

  \subpoint We now turn to $\mathfrak G$. We fix a vertex $\mathbf v\in\Lambda$. If $\mathfrak d_q(\mathbf v,\partial_q\Lambda)<\ell_0$, then, proceeding in the same way as for $\mathfrak B$, we bound
  \begin{equation}
    \sum_{\gamma\in\mathfrak P_q^{(\circ)}(\infty|\mathbf v)}
    |\zeta^{(\emptyset)}_{\mathbf q}(\gamma)|
    \sum_{\displaystyle\mathop{\scriptstyle\underline\gamma\sqsubset\mathfrak P_q^{(\circ)}(\infty)}_{\{\gamma\}\sqcup\underline\gamma\not\sqsubset\mathfrak P^{(\circ)}_q(\Lambda)}}
    \Phi^T(\{\gamma\}\sqcup\underline\gamma)
    \prod_{\gamma'\in\underline\gamma}|\zeta^{(\emptyset)}_{\mathbf q}(\gamma')|
    \leqslant
    e^{-3\bar J}
    \left(
      1+e^{-3(1-\alpha)\bar J}
    \right).
    \label{bound_frakG1}
  \end{equation}
  If $\mathfrak d_q(\mathbf v,\partial_q\Lambda)\geqslant\ell_0$, then the condition $\gamma\sqcup\underline\gamma\not\sqsubset(\mathfrak P^{(\circ)}_q(\Lambda)$ implies that
  \begin{equation}
    \sum_{\gamma'\in\gamma\cup\underline\gamma}(\mathfrak l(\gamma')+\mathfrak s(\gamma'))\geqslant \mathfrak d_q(\mathbf v,\partial_q\Lambda)-\ell_0.
  \end{equation}
  In addition, $\mathfrak l(\gamma)\geqslant 6$. Therefore
  \begin{equation}
    \sum_{\gamma'\in\underline\gamma}\Xi^{(\Upsilon)}(\gamma)
    \geqslant
    3\bar J|\underline\gamma|
    +
    \bar\kappa(\mathfrak d_q(\mathbf v,\partial_q\Lambda)-\ell_0-4)
    .
  \end{equation}
  Therefore, proceeding as for $\mathfrak B$,
  \begin{equation}
    \begin{largearray}
      \sum_{\gamma\in\mathfrak P_q^{(\circ)}(\infty|\mathbf v)}
      |\zeta^{(\emptyset)}_{\mathbf q}(\gamma)|
      \sum_{\displaystyle\mathop{\scriptstyle\underline\gamma\sqsubset\mathfrak P_q^{(\circ)}(\infty)}_{\{\gamma\}\sqcup\underline\gamma\not\sqsubset\mathfrak P^{(\circ)}_q(\Lambda)}}
      \Phi^T(\{\gamma\}\sqcup\underline\gamma)
      \prod_{\gamma'\in\underline\gamma}|\zeta^{(\emptyset)}_{\mathbf q}(\gamma')|
      \\[-0.5cm]\hfill
      \leqslant
      e^{-\beta\bar\kappa(\mathfrak d_q(\mathbf v,\partial_q\Lambda)-\ell_0)}
      e^{-3\bar J}
      \left(
	1+e^{-3(1-\alpha)\bar J}
      \right).
    \end{largearray}
    \label{bound_frakG_clusters}
  \end{equation}
  Therefore,
  \begin{equation}
    |\mathfrak G_q(\Lambda)|\leqslant\cst c{cst:kappa2}\ell_0e^{-3\bar J}|\partial_q\Lambda|\left(1+e^{-3(1-\alpha)\bar J}\right)
    \label{bound_frakG}
  \end{equation}
  for some constant $\cst c{cst:kappa2}>0$.
  \bigskip

  \point We conclude the proof by injecting\-~(\ref{symmetry_frakF}), (\ref{bound_frakB_q}), (\ref{bound_frakB_c}) and\-~(\ref{bound_frakG}) into\-~(\ref{K_BFG}), and using the fact that $\kappa^{-1}<\ell_0$ (\ref{ell0}).
\qed

\section{Nematic phase}\label{sec:nematic}
\indent We are now ready to prove theorem\-~\ref{theo:nematic}. Let $\Lambda_L$ by a square box of side-length $L$.
\bigskip

\point Given an edge $e\in\mathcal E(\Lambda_L)$, we have
\begin{equation}
  \left<\mathds 1_e\right>_{\Lambda_L,\mathbf q}=
  \frac{\widetilde{\mathfrak Z}_{\mathbf q}^{(\{e\})}(\Lambda_L)}{\widetilde{\mathfrak Z}_{\mathbf q}^{(\emptyset)}(\Lambda_L)}
  \exp\left(
    \log\left(\frac{Z^{(\{e\})}(\Lambda_L|\mathbf q)}{\widetilde{\mathfrak Z}_{\mathbf q}^{(\{e\})}(\Lambda_L)}\right)
    -
    \log\left(\frac{Z^{(\emptyset)}(\Lambda_L|\mathbf q)}{\widetilde{\mathfrak Z}_{\mathbf q}^{(\emptyset)}(\Lambda_L)}\right)
  \right)
\end{equation}
\bigskip

\subpoint Let us first bound the exponent in the thermodynamic limit. As in lemma\-~\ref{lemma:boundK}, we define
\begin{equation}
  \bar\zeta_{\mathbf c}^{(\{e\},\emptyset)}(\gamma|t):=
  t\zeta_{\mathbf c}^{(\{e\})}(\gamma_0)
  +
  (1-t)\zeta_{\mathbf c}^{(\emptyset)}(\gamma_0)
\end{equation}
and
\begin{equation}
  \overline{\mathfrak P}_q^{(\{e\},\emptyset)}(\Lambda_L)
  :=
  \mathfrak P_q^{(\{e\})}(\Lambda_L)
  \cup
  \mathfrak P_q^{(\emptyset)}(\Lambda_L)
\end{equation}
and, using the cluster expansion\-~(\ref{ce}), we write
\begin{equation}
  \begin{largearray}
    \log\left(
      \frac{Z^{(\{e\})}(\Lambda_L|\mathbf q)}{\widetilde{\mathfrak Z}^{(\{e\})}_{\mathbf q}(\Lambda_L)}
    \right)
    -
    \log\left(
      \frac{Z^{(\emptyset)}(\Lambda_L|\mathbf q)}{\widetilde{\mathfrak Z}^{(\emptyset)}_{\mathbf q}(\Lambda_L)}
    \right)
    =
    \int_0^1dt
    \sum_{m=1}^\infty
    \sum_{\gamma\in\overline{\mathfrak P}_q^{(\{e\},\emptyset)}(\Lambda_L)}
    \sum_{\underline\gamma\sqsubset\overline{\mathfrak P}_q^{(\{e\},\emptyset)}(\Lambda_L)\setminus\{\gamma\}}
    \\[1.0cm]\hfill
    \Phi^T(\{\gamma\}^m\sqcup\underline\gamma)
    \left(\zeta^{(\{e\})}_{\mathbf q}(\gamma)-\zeta^{(\emptyset)}_{\mathbf q}(\gamma)\right)
    m(\bar\zeta^{(\{e\},\emptyset)}_{\mathbf q}(\gamma|t))^{m-1}
    \prod_{\gamma'\in\underline\gamma}\bar\zeta^{(\{e\},\emptyset)}_{\mathbf q}(\gamma'|t)
    .
  \end{largearray}
  \label{1v_cluster}
\end{equation}
We split the sum into a {\it bulk} and a {\it boundary} contribution, similarly to lemma\-~\ref{lemma:boundK}. We define $\overline{\mathfrak P}_q^{(\{e\},\emptyset)}(\infty)$ as the set of polymers $\gamma$ for which $\exists L$ such that $\gamma$ is in $\overline{\mathfrak P}_q^{(\{e\},\emptyset)}(\Lambda_L)$ while not being connected to the boundary. We then split
\begin{equation}
  \log\left(
    \frac{Z^{(\{e\})}(\Lambda_L|\mathbf q)}{\widetilde{\mathfrak Z}^{(\{e\})}_{\mathbf q}(\Lambda_L)}
  \right)
  -
  \log\left(
    \frac{Z^{(\emptyset)}(\Lambda_L|\mathbf q)}{\widetilde{\mathfrak Z}^{(\emptyset)}_{\mathbf q}(\Lambda_L)}
  \right)
  =
  I^{(e)}-R_1^{(e)}(\Lambda_L)+R_2^{(e)}(\Lambda_L)
\end{equation}
where
\begin{equation}
  \begin{largearray}
    I^{(e)}
    :=
    \int_0^1dt
    \sum_{m=1}^\infty
    \sum_{\gamma\in\overline{\mathfrak P}_q^{(\{e\},\emptyset)}(\infty)}
    \sum_{\underline\gamma\sqsubset\overline{\mathfrak P}_q^{(\{e\},\emptyset)}(\infty)\setminus\{\gamma\}}
    \\[1cm]\hfill
    \Phi^T(\{\gamma\}^m\sqcup\underline\gamma)
    \left(\zeta^{(\{e\})}_{\mathbf q}(\gamma)-\zeta^{(\emptyset)}_{\mathbf q}(\gamma)\right)
    m(\bar\zeta^{(\{e\},\emptyset)}_{\mathbf q}(\gamma|t))^{m-1}
    \prod_{\gamma'\in\underline\gamma}\bar\zeta^{(\{e\},\emptyset)}_{\mathbf q}(\gamma'|t)
  \end{largearray}
  \label{I_cluster}
\end{equation}
\begin{equation}
  \begin{largearray}
    R_1^{(e)}(\Lambda_L)
    :=
    \int_0^1dt
    \sum_{m=1}^\infty
    \sum_{\gamma\in\overline{\mathfrak P}_q^{(\{e\},\emptyset)}(\infty)}
    \sum_{\displaystyle\mathop{\scriptstyle\underline\gamma\sqsubset\overline{\mathfrak P}_q^{(\{e\},\emptyset)}(\infty)\setminus\{\gamma\}}_{\{\gamma\}^m\sqcup\underline\gamma\not\sqsubset\overline{\mathfrak P}_q^{(\{e\},\emptyset)}(\Lambda_L)}}
    \\[1cm]\hfill
    \Phi^T(\{\gamma\}^m\sqcup\underline\gamma)
    \left(\zeta^{(\{e\})}_{\mathbf q}(\gamma)-\zeta^{(\emptyset)}_{\mathbf q}(\gamma)\right)
    m(\bar\zeta^{(\{e\},\emptyset)}_{\mathbf q}(\gamma|t))^{m-1}
    \prod_{\gamma'\in\underline\gamma}\bar\zeta^{(\{e\},\emptyset)}_{\mathbf q}(\gamma'|t)
  \end{largearray}
  \label{R_cluster}
\end{equation}
and
\begin{equation}
  \begin{largearray}
    R_2^{(e)}(\Lambda_L)
    :=
    \int_0^1dt
    \sum_{m=1}^\infty
    \sum_{\gamma\in\overline{\mathfrak P}_q^{(\{e\},\emptyset)}(\Lambda_L)}
    \sum_{\displaystyle\mathop{\scriptstyle\underline\gamma\sqsubset\overline{\mathfrak P}_q^{(\{e\},\emptyset)}(\Lambda_L)\setminus\{\gamma\}}_{\{\gamma\}^m\sqcup\underline\gamma\not\sqsubset\overline{\mathfrak P}_q^{(\{e\},\emptyset)}(\infty)}}
    \\[1cm]\hfill
    \Phi^T(\{\gamma\}^m\sqcup\underline\gamma)
    \left(\zeta^{(\{e\})}_{\mathbf q}(\gamma)-\zeta^{(\emptyset)}_{\mathbf q}(\gamma)\right)
    m(\bar\zeta^{(\{e\},\emptyset)}_{\mathbf q}(\gamma|t))^{m-1}
    \prod_{\gamma'\in\underline\gamma}\bar\zeta^{(\{e\},\emptyset)}_{\mathbf q}(\gamma'|t)
    .
  \end{largearray}
  \label{R2_cluster}
\end{equation}
By\-~(\ref{boundzeta}),
\begin{equation}
  m|\bar\zeta_{\mathbf q}^{(\{e\},\emptyset)}(\gamma|t)|^{m-1}\leqslant 1
\end{equation}
so, by\-~(\ref{ce_remainder}),
\begin{equation}
  |I^{(e)}|
  \leqslant
  \sum_{\gamma\in\overline{\mathfrak P}_q^{(\{e\},\emptyset)}(\infty)}
  \left|\zeta^{(\{e\})}_{\mathbf q}(\gamma)-\zeta^{(\emptyset)}_{\mathbf q}(\gamma)\right|
  e^{a(\gamma)}
  .
\end{equation}
Furthermore, $\zeta_{\mathbf q}^{(\{e\})}(\gamma)-\zeta_{\mathbf q}^{(\emptyset)}(\gamma)$ differs from 0 only if $\gamma$ interacts with $e$ (see definition\-~\ref{def:polymer_source_interaction}), so, by lemma\-~\ref{lemma:polymer_source_interaction},
\begin{equation}
  |I^{(e)}|
  \leqslant
  2\cst C{cst:polymer_source}e^{-(1-\alpha)\bar J}
  .
  \label{bound_I}
\end{equation}
In addition, the clusters $\gamma_0,\cdots,\gamma_n$ that contribute to $R_1$ or $R_2$ interact with $e$ as well as with the boundary of $\Lambda_L$. Therefore, for such clusters,
\begin{equation}
  \sum_{i=0}^n\Xi(\gamma_i)\geqslant\bar\kappa\left(\mathrm{dist}_1(e,\partial\Lambda)-2\ell_0\right)
\end{equation}
which goes to $\infty$ as $L\to\infty$. Therefore,
\begin{equation}
  \prod_{j=0}^n|\zeta_{\mathbf q}^{(\Upsilon)}(\gamma_j)|
  \leqslant
  e^{-\beta\bar\kappa(\mathrm{dist}_1(e,\partial\Lambda)-2\ell_0)}
  \prod_{j=0}^n
  e^{-(1-\beta)\Xi^{(\Upsilon)}(\gamma_j)}
\end{equation}
for $\beta\in(0,\frac12\mathfrak t-\alpha)$. We then use lemma\-~\ref{lemma:bz} and\-~\ref{lemma:polymer_source_interaction} to bound
\begin{equation}
  |R_i^{(e)}(\Lambda_L)|=
  O\left(e^{-\beta\bar\kappa(\mathrm{dist}_1(e,\partial\Lambda)-2\ell_0)}\right)
\end{equation}
so
\begin{equation}
  |R_1^{(e)}(\Lambda_L)|+|R_2^{(e)}(\Lambda_L)|\mathop{\longrightarrow}_{L\to\infty}0
  .
\end{equation}
Therefore,
\begin{equation}
  \log\left(\frac{Z^{(\{e\})}(\Lambda_L|\mathbf q)}{\widetilde{\mathfrak Z}_{\mathbf q}^{(\{e\})}(\Lambda_L)}\right)
  -
  \log\left(\frac{Z^{(\emptyset)}(\Lambda_L|\mathbf q)}{\widetilde{\mathfrak Z}_{\mathbf q}^{(\emptyset)}(\Lambda_L)}\right)
  \mathop{\longrightarrow}_{L\to\infty}
  I^{(e)}
\end{equation}
which is independent of the position of $e$, and is bounded as per\-~(\ref{bound_I}).
\bigskip

\subpoint We turn, now, to the ratio of $\widetilde{\mathfrak Z}$'s, which can be computed explicitly from\-~(\ref{tildefrakZ}) and\-~(\ref{idfrakZ}): if $e$ is vertical and $L$ is large enough, then
\begin{equation}
  \frac{\widetilde{\mathfrak Z}_{\mathbf v}^{(\{e\})}(\Lambda_L)}{\widetilde{\mathfrak Z}_{\mathbf v}^{(\emptyset)}(\Lambda_L)}
  =
  z\nu_+^2(\omega_1)b_+\lambda_+^{-2}
  =
  \frac12(1+O(\epsilon))
\end{equation}
in which we used\-~(\ref{bound_lambda}) through\-~(\ref{bound_nu1}). If $e$ is horizontal, then
\begin{equation}
  \frac{\widetilde{\mathfrak Z}_{\mathbf v}^{(\{e\})}(\Lambda_L)}{\widetilde{\mathfrak Z}_{\mathbf v}^{(\emptyset)}(\Lambda_L)}
  =
  z\nu_+^4(\omega_0)b_+^2\lambda_+^{-2}
  =
  O(e^{-3J})
  .
\end{equation}
This proves that $\left<\mathds 1_e\right>_{\mathbf v}$ is independent of the position of $e$, as well as\-~(\ref{bound_1v}) and\-~(\ref{bound_1h}).
\bigskip

\point We now consider two edges $e,e'\in\mathcal E(\Lambda_L)$ which are at a distance of at least $\ell_0$. We have
\begin{equation}
  \left<\mathds 1_e\mathds 1_{e'}\right>_{\Lambda_L,\mathbf q}
  -
  \left<\mathds 1_e\right>_{\Lambda_L,\mathbf q}
  \left<\mathds 1_{e'}\right>_{\Lambda_L,\mathbf q}
  =
  \left<\mathds 1_e\right>_{\Lambda_L,\mathbf q}
  \left<\mathds 1_{e'}\right>_{\Lambda_L,\mathbf q}
  \left(
    \frac{\widetilde{\mathfrak Z}^{(\{e,e'\})}_{\mathbf q}(\Lambda_L)\widetilde{\mathfrak Z}^{(\emptyset)}_{\mathbf q}(\Lambda_L)}{\widetilde{\mathfrak Z}^{(\{e\})}_{\mathbf q}(\Lambda_L)\widetilde{\mathfrak Z}^{(\{e'\})}_{\mathbf q}(\Lambda_L)}
    e^{\mathbb K^{(e,e')}_{\mathbf q}(\Lambda_L)}
    -1
  \right)
  \label{1vv_cluster}
\end{equation}
with
\begin{equation}
  \begin{largearray}
    \mathbb K_{\mathbf q}^{(e,e')}(\Lambda_L)
    :=
    \log\left(\frac{Z^{(\{e,e'\})}(\Lambda_L|\mathbf q)}{\widetilde{\mathfrak Z}_{\mathbf q}^{(\{e,e'\})}(\Lambda_L)}\right)
    +
    \log\left(\frac{Z^{(\emptyset)}(\Lambda_L|\mathbf q)}{\widetilde{\mathfrak Z}_{\mathbf q}^{(\emptyset)}(\Lambda_L)}\right)
    \\[0.5cm]\hfill
    -
    \log\left(\frac{Z^{(\{e\})}(\Lambda_L|\mathbf q)}{\widetilde{\mathfrak Z}_{\mathbf q}^{(\{e\})}(\Lambda_L)}\right)
    -
    \log\left(\frac{Z^{(\{e'\})}(\Lambda_L|\mathbf q)}{\widetilde{\mathfrak Z}_{\mathbf q}^{(\{e'\})}(\Lambda_L)}\right)
    .
  \end{largearray}
  \label{bbK_cluster}
\end{equation}
First of all, by\-~(\ref{tildefrakZ}) and\-~(\ref{idfrakZ}),
\begin{equation}
  \frac{\widetilde{\mathfrak Z}^{(\{e,e'\})}_{\mathbf q}(\Lambda_L)\widetilde{\mathfrak Z}^{(\emptyset)}_{\mathbf q}(\Lambda_L)}{\widetilde{\mathfrak Z}^{(\{e\})}_{\mathbf q}(\Lambda_L)\widetilde{\mathfrak Z}^{(\{e'\})}_{\mathbf q}(\Lambda_L)}=1
  .
\end{equation}
We then write $\mathbb K_{\mathbf q}^{(e,e')}(\Lambda_L)$ using the cluster expansion\-~(\ref{ce}), and note that the only clusters $\gamma_0,\cdots,\gamma_n$ that contribute are those that interact with {\it both} $e$ and $e'$. For such clusters, denoting the vertical and horizontal distances by $\mathrm{dist}_{\mathrm v}$ and $\mathrm{dist}_{\mathrm h}$ (these are the induced by the semi-norms $\|(x,y)\|_{\mathrm v}:=|y|$ and $\|(x,y)\|_{\mathrm h}:=|x|$),
\begin{equation}
  \sum_{i=0}^n\Xi(\gamma_i)
  \geqslant
  \bar\kappa\left(\mathrm{dist}_{\mathrm v}(e,e')-2\ell_0\right)
  +
  \bar J\mathrm{dist}_{\mathrm h}(e,e')
\end{equation}
so, by once again estimating
\begin{equation}
  \prod_{j=0}^n|\zeta_{\mathbf q}^{(\Upsilon)}(\gamma_j)|
  \leqslant
  e^{-\beta(\bar\kappa(\mathrm{dist}_1(e,e')-2\ell_0)+\bar J\mathrm{dist}_{\mathrm h}(e,e'))}
  \prod_{j=0}^n
  e^{-(1-\beta)\Xi^{(\Upsilon)}(\gamma_j)}
\end{equation}
for $\beta\in(0,\frac12\mathfrak t-\alpha)$, and using lemma\-~\ref{lemma:bz} and\-~\ref{lemma:polymer_source_interaction} to bound
\begin{equation}
  |\mathbb K_{\mathbf q}^{(e,e')}(\Lambda_L)|
  =
  O\left(e^{-\beta(\bar\kappa(\mathrm{dist}_{\mathrm v}(e,e')-2\ell_0)+\bar J\mathrm{dist}_{\mathrm h}(e,e'))}\right)
\end{equation}
from which\-~(\ref{bound_1vv}) follows.
\qed

\vfill
\eject


\begin{thebibliography}{WWW99}
\small
\bibitem[AH80]{AH80}D.B. Abraham, O.J. Heilmann - {\it Interacting dimers on the simple cubic lattice as a model for liquid crystals}, Journal of Physics A: Mathematical and General, volume~\-13, number~\-3, pages~\-1051-1062, 1980,\par\penalty10000
doi:{\tt\color{blue}\href{http://dx.doi.org/10.1088/0305-4470/13/3/038}{10.1088/0305-4470/13/3/038}}.\par\medskip
 
\bibitem[ACM14]{ACM14}D. Alberici, P. Contucci, E. Mingione - {\it A mean-field monomer-dimer model with attractive interaction: Exact solution and rigorous results}, Journal of Mathematical Physics, volume~\-55, page~\-063301, 2014,\par\penalty10000
doi:{\tt\color{blue}\href{http://dx.doi.org/10.1063/1.4881725}{10.1063/1.4881725}}, arxiv:{\tt\color{blue}\href{http://arxiv.org/abs/1311.6551}{1311.6551}}.\par\medskip
 
\bibitem[Al16]{Al16}D. Alberici - {\it A Cluster Expansion Approach to the Heilmann-Lieb Liquid Crystal Model}, Journal of Statistical Physics, volume~\-162, issue~\-3, pages~\-761-791, 2016,\par\penalty10000
doi:{\tt\color{blue}\href{http://dx.doi.org/10.1007/s10955-015-1421-8}{10.1007/s10955-015-1421-8}}, arxiv:{\tt\color{blue}\href{http://arxiv.org/abs/1506.02255}{1506.02255}}.\par\medskip
 
\bibitem[AZ82]{AZ82}N. Angelescu, V.A. Zagrebnov - {\it A lattice model of liquid crystals with matrix order parameter}, Journal of Physics A: Mathematical and General, volume~\-15, issue~\-11, pages L639-L643, 1982,\par\penalty10000
doi:{\tt\color{blue}\href{http://dx.doi.org/10.1088/0305-4470/15/11/012}{10.1088/0305-4470/15/11/012}}.\par\medskip
 
\bibitem[BZ00]{BZ00}A. Bovier, M. Zahradn\'\i k - {\it A Simple Inductive Approach to the Problem of Convergence of Cluster Expansions of Polymer Models}, Journal of Statistical Physics, volume~\-100, issue~\-3-4, pages~\-765-778, 2000,\par\penalty10000
doi:{\tt\color{blue}\href{http://dx.doi.org/10.1023/A:1018631710626}{10.1023/A:1018631710626}}.\par\medskip
 
\bibitem[BKL84]{BKL84}J. Bricmont, K. Kuroda, J.L. Lebowitz - {\it The struture of Gibbs states and phase coexistence for non-symmetric continuum Widom-Rowlinson models}, Zeitschrift f\"ur Wahrscheinlichkeitstheorie und Verwandte Gebiete, volume~\-67, issue~\-2, pages~\-121-138, 1984,\par\penalty10000
doi:{\tt\color{blue}\href{http://dx.doi.org/10.1007/BF00535264}{10.1007/BF00535264}}.\par\medskip
 
\bibitem[BKL85]{BKL85}J. Bricmont, K. Kuroda, J.L. Lebowitz - {\it First order phase transitions in lattice and continuous systems: extension of Pirogov-Sinai theory}, Communications in Mathematical Physics, volume~\-101, issue~\-4, pages~\-501-538, 1985,\par\penalty10000
doi:{\tt\color{blue}\href{http://dx.doi.org/10.1007/BF01210743}{10.1007/BF01210743}}.\par\medskip
 
\bibitem[DG13]{DG13}M. Disertori, A. Giuliani - {\it The nematic phase of a system of long hard rods}, Communications in Mathematical Physics, volume~\-323, pages~\-143-175, 2013,\par\penalty10000
doi:{\tt\color{blue}\href{http://dx.doi.org/10.1007/s00220-013-1767-1}{10.1007/s00220-013-1767-1}}, arxiv:{\tt\color{blue}\href{http://arxiv.org/abs/1112.5564}{1112.5564}}.\par\medskip
 
\bibitem[GBG04]{GBG04}G. Gallavotti, F. Bonetto, G. Gentile - {\it Aspects of Ergodic, Qualitative and Statistical Theory of Motion}, Springer, 2004.\par\medskip
 
\bibitem[GM01]{GM01}G. Gentile, V. Mastropietro - {\it Renormalization group for one-dimensional fermions - a review on mathematical results}, Physics Reports, volume~\-352, pages~\-273-437, 2001,\par\penalty10000
doi:{\tt\color{blue}\href{http://dx.doi.org/10.1016/S0370-1573(01)00041-2}{10.1016/S0370-1573(01)00041-2}}.\par\medskip
 
\bibitem[HL72]{HL72}O.J. Heilmann, E.H. Lieb - {\it Theory of monomer-dimer systems}, Communications in Mathematical Physics, volume~\-25, issue~\-3, pages~\-190-232, 1972,\par\penalty10000
doi:{\tt\color{blue}\href{http://dx.doi.org/10.1007/BF01877590}{10.1007/BF01877590}}.\par\medskip
 
\bibitem[HL79]{HL79}O.J. Heilmann, E.H. Lieb - {\it Lattice models for liquid crystals}, Journal of Statistical Physics, volume~\-20, issue~\-6, pages~\-679-693, 1979,\par\penalty10000
doi:{\tt\color{blue}\href{http://dx.doi.org/10.1007/BF01009518}{10.1007/BF01009518}}.\par\medskip
 
\bibitem[IVZ06]{IVZ06}D. Ioffe, Y. Velenik, M. Zahradn\'\i k - {\it Entropy-Driven Phase Transition in a Polydisperse Hard-Rods Lattice System}, Journal of Statistical Physics, volume~\-122, issue~\-4, pages~\-761-786, 2006,\par\penalty10000
doi:{\tt\color{blue}\href{http://dx.doi.org/10.1007/s10955-005-8085-8}{10.1007/s10955-005-8085-8}}, arxiv:{\tt\color{blue}\href{http://arxiv.org/abs/math/0503222}{math/0503222}}.\par\medskip
 
\bibitem[KP84]{KP84}R. Koteck\'y, D. Preiss - {\it An inductive approach to the Pirogov-Sinai theory}, Proceedings of the~\-11th Winter School on Abstract Analysis, Rendiconti del Circolo Matematico di Palermo, Serie II, supplemento~\-3, pages~\-161-164, 1984.\par\medskip
 
\bibitem[KP86]{KP86}R. Koteck\'y, D. Preiss - {\it Cluster expansion for abstract polymer models}, Communications in Mathematical Physics, volume~\-103, issue~\-3, pages~\-491-498, 1986,\par\penalty10000
doi:{\tt\color{blue}\href{http://dx.doi.org/10.1007/BF01211762}{10.1007/BF01211762}}.\par\medskip
 
\bibitem[Ma37]{Ma37}J.E. Mayer - {\it The Statistical Mechanics of Condensing Systems. I}, The Journal of Chemical Physics, volume~\-5, issue~\-67, pages~\-67-73, 1937,\par\penalty10000
doi:{\tt\color{blue}\href{http://dx.doi.org/10.1063/1.1749933}{10.1063/1.1749933}}.\par\medskip
 
\bibitem[On49]{On49}L. Onsager - {\it The effects of shape on the interaction of colloidal particles}, Annals of the New York Academy of Sciences, volume~\-51, pages~\-627-659, 1949,\par\penalty10000
doi:{\tt\color{blue}\href{http://dx.doi.org/10.1111/j.1749-6632.1949.tb27296.x}{10.1111/j.1749-6632.1949.tb27296.x}}.\par\medskip
 
\bibitem[PCF14]{PCF14}S. Papanikolaou, D. Charrier, E. Fradkin - {\it Ising nematic fluid phase of hard-core dimers on the square lattice}, Physical Review B, volume~\-89, page~\-035128, 2014,\par\penalty10000
doi:{\tt\color{blue}\href{http://dx.doi.org/10.1103/PhysRevB.89.035128}{10.1103/PhysRevB.89.035128}}, arxiv:{\tt\color{blue}\href{http://arxiv.org/abs/1310.4173}{1310.4173}}.\par\medskip
 
\bibitem[PS75]{PS75}S.A. Pirogov, Y.G. Sinai - {\it Phase diagrams of classical lattice systems}, Theoretical and Mathematical Physics, volume~\-25, pages~\-1185-1192, 1975,\par\penalty10000
doi:{\tt\color{blue}\href{http://dx.doi.org/10.1007/BF01040127}{10.1007/BF01040127}}.\par\medskip
 
\bibitem[Ru99]{Ru99}D. Ruelle - {\it Statistical mechanics: rigorous results}, Imperial College Press, World Scientific, (first edition: Benjamin, 1969), 1999.\par\medskip
 
\bibitem[Ur27]{Ur27}H.D. Ursell - {\it The evaluation of Gibbs' phase-integral for imperfect gases}, Mathematical Proceedings of the Cambridge Philosophical Society, volume~\-23, issue~\-6, pages~\-685-697, 1927,\par\penalty10000
doi:{\tt\color{blue}\href{http://dx.doi.org/10.1017/S0305004100011191}{10.1017/S0305004100011191}}.\par\medskip
 
\bibitem[Za96]{Za96}V.A. Zagrebnov - {\it Long-range order in a lattice-gas model of nematic liquid crystals}, Physica A: Statistical Mechanics and its Applications, volume~\-232, issues~\-3-4, pages~\-737-746, 1996,\par\penalty10000
doi:{\tt\color{blue}\href{http://dx.doi.org/10.1016/0378-4371(96)00181-1}{10.1016/0378-4371(96)00181-1}}.\par\medskip
 
\end{thebibliography}
\end{document}